\documentclass[
  aps,prx,twocolumn,
  superscriptaddress,
  longbibliography
]{revtex4-2}
\usepackage{empheq}
\usepackage{amsmath,amssymb,amsfonts}
\usepackage{graphicx}
\usepackage{bm}
\usepackage{hyperref}
\usepackage{tikz}
\usepackage{amsthm}
\newtheorem{remark}{Remark}
\begin{document}

\title{Nonperturbative Resummation of Divergent Time-Local Generators:\\
Disentangling Non-Markovian Dynamics}

\author{Dragomir Davidovic}
\email{dragomir.davidovic@physics.gatech.edu}
\affiliation{School of Physics, Georgia Institute of Technology, USA}

\date{} 

\begin{abstract}
Perturbative van Kampen cumulant expansions of time-local generators of open quantum systems generically diverge at long times, while the reduced dynamics remains regular. We show that these divergences encode information sufficient to reconstruct an approximate nonperturbative dynamical map. Although their resummation remains divergent, disentangling the dynamics according to its van Hove limit resolves the divergence and exposes its dynamical origin: the map approaches isolated times at which it becomes noninvertible. For the weak-coupling unbiased spin--boson model, the reconstructed non-Markovian dynamics takes the disentangled form
\begin{equation*}
\Phi=\Phi_{\mathrm{GAD}}+\Phi_{\mathrm{NS}}+\Phi_{\mathrm{Khal}},
\end{equation*}
separating generalized amplitude damping, exponentially damped nonsecular coherence mixing, and a projective, phase-selective Khalfin map. Thus, the onset of recurrent noninvertibility is the reduced-dynamical-map manifestation of the Khalfin transition from exponential to algebraic relaxation. Although the coherences acquire algebraic Khalfin tails, the distinguishability of quantum superposition states remains governed by an exponential decay law well after the Khalfin transition. Comparison with numerically exact TEMPO dynamics validates the procedure and shows that apparently complex non-Markovian dynamics can resolve into a small number of distinct, physically transparent dynamical processes.
\end{abstract}

\keywords{Open quantum systems, Non-Markovian dynamics, Quantum measurement,
Spin--boson model, Time--convolutionless methods}

\maketitle

\section{Introduction\label{Sec:Intro}}

Reduced open quantum system evolution is described by a completely positive
trace-preserving (CPTP) dynamical map $\Phi(t)$ acting on the density
matrix~\cite{BreuerPetruccione2002}. A time-local master equation
represents the same evolution through a generator $L(t)$ defined by
\begin{equation}
L(t)=\dot{\Phi}(t)\Phi^{-1}(t).
\label{eq:TCLgenerator}
\end{equation}
Thus the time-local generator is the logarithmic derivative of the dynamical map and exists only while the map remains invertible. 

Microscopically, time-local master equations are  derived from the time-convolutionless (TCL)
projection-operator expansion~\cite{ChaturvediShibata1979}. This provides one of two exact projection-operator representations of reduced dynamics,  the
other being the Nakajima--Zwanzig memory-kernel integro-differential equation~\cite{BreuerPetruccione2002}.

The time-local generator is obtained through the van Kampen cumulant expansion for stochastic linear differential equations, derived from the Feynman disentanglement theorem~\cite{VANKAMPEN1,VANKAMPEN2}. In this method, the dynamical map $\Phi(t)$ is reorganized into an exponential of cumulants $K_n(t)$ that encode progressively higher-order system--bath interactions. A remarkable feature of the expansion is that it remains valid for noncommuting interaction operators at different times.  When bath correlations decay sufficiently rapidly, the cumulants remove the secular growth present in the Dyson series order by order.

However, when environmental correlations decay algebraically rather than exponentially, the cumulant expansion diverges at long time scales, with divergences appearing already at fourth order~\cite{Crowder} and not removed by partial resummations of the leading cumulants~\cite{Lampert2025}. This occurs even though the state propagator remains well defined, consistent with the convergence of the Dyson expansion~\cite{RivasHuelga2012}. 

Rather than signaling a breakdown of the reduced dynamics, the divergence reflects the approach to a rank change of the dynamical map. As a singular value vanishes, the corresponding time-local generator becomes unbounded~\cite{hou2012singularity}. Although singular generators have previously been introduced through constructions~\cite{Chru2010,hou2012singularity,hegde2021open,jagadish2023noninvertibility,chru2026}, their microscopic origin and generality have remained largely unexplored. The partial resummations of the van Kampen cumulant expansion developed here show that such singularities arise generically from microscopic Hamiltonians whenever bath correlations decay algebraically rather than exponentially.

The resulting dynamics approaches a noninvertible quantum channel. At the singular time, two distinct initial states evolve to the same reduced state. Consequently, no measurement performed on the reduced system at that instant can determine which of the two states was initially prepared. The reduced dynamics therefore undergoes a complete instantaneous loss of distinguishability between the two initial states, even though this distinguishability is recovered once the dynamics becomes invertible again. 

Existing constructions~\cite{Chru2010,hou2012singularity,hegde2021open,jagadish2023noninvertibility,chru2026}, together with recent tensor-network simulations~\cite{strachan2024extracting}, indicate that such noninvertibility occurs at isolated times. Instead of remaining singular, the dynamics subsequently becomes invertible again, leading to repeated isolated noninvertibility events. The partial cumulant resummations presented here exhibit the same behavior: after the first singularity, invertibility is restored, and the cycle repeats indefinitely. Nevertheless, the key finding is that the residual distinguishability of the initial states, here characterized by the smallest singular value of the coherence block, continues to decay exponentially, as in Markovian dynamics. Long-lived memory therefore does not preserve the distinguishability of
quantum superpositions.

The analysis begins from the observation that the exponentially growing generator obtained from the partially resummed van Kampen expansion is not an artifact of the resummation, but the early-time manifestation of a true singular time-local generator. Although its growth is initially weak, it already contains sufficient information to reconstruct the singular behavior through an asymptotic reduction procedure.

The key idea is that, despite the emergence of long-lived non-Markovian memory, the dynamical map remains close to the reference semigroup generated by the Davies Markovian master equation~\cite{davies1974}. The semigroup continuously contracts the evolving state toward the stationary state. As this region shrinks, increasingly less information about the singular generator is required to correct the reference dynamics. The asymptotic reduction therefore requires only the contraction of the generator onto the exponentially shrinking neighborhood explored by the dynamics.

The balance between exponential relaxation generated by the Davies semigroup and the exponential growth of the time-local generator leaves a finite reduced evolution while exposing the singularities of the generator. The dynamical map is then reconstructed using the Feynman disentanglement theorem, with the Davies semigroup providing the reference evolution for the asymptotic reduction.

In continuum environments with slowly decaying correlations, the survival amplitude of a quantum state crosses over from exponential Markovian decay to the universal nonexponential decay law discovered by Khalfin~\cite{Khalfin1958}. Microscopically, this behavior reflects the pole--branch structure of the bath correlation function. Equivalently, it follows from the Fourier-transform constraints underlying the Khalfin effect: a continuum spectrum with an algebraically decaying correlation function necessarily produces a nonexponential long-time tail.

The Fourier-transform constraints underlying the Khalfin effect generalize
from survival amplitudes to reduced dynamical maps. The secular
($\Phi_{12,12}(t)$) and nonsecular ($\Phi_{12,21}(t)$) matrix elements
share the same algebraic tail but enter that regime at parametrically
different times because they differ in their weak-coupling scaling.
Their inevitable crossing produces a finite-time loss of invertibility
of the dynamical map, followed by an infinite sequence of isolated
noninvertibility points whose spacing is asymptotically determined by
the oscillation period and slowly modulated by the bath-correlation
function. Within the classification of
Ref.~\cite{jagadish2023noninvertibility}, this mechanism corresponds to
{\bf Type II}, where invertibility is lost while the dynamics remains
CP indivisible, rather than {\bf Type I}, where the loss of
invertibility is CP divisible.

In this sense, the loss of invertibility of the reduced dynamical map
is the reduced density-matrix counterpart of the Khalfin effect. The
transition from exponential to algebraic relaxation, originally
formulated for survival amplitudes, reappears in open-system dynamics
as the onset of singular time-local generators. Although the
coherences acquire algebraic Khalfin tails, the distinguishability of
quantum superposition states remains governed by the exponential
contraction inherited from the Davies Markovian master equation.

This paper’s central result is a simple structure of the quantum dynamical map in the unbiased weakly coupled spin-boson model (SBM). The map disentangles into three physically distinct dynamical processes,
\begin{equation}
\Phi
=
\Phi_{\mathrm{GAD}}
+
\Phi_{\mathrm{NS}}
+
\Phi_{\mathrm{Khal}},
\end{equation}
corresponding to generalized amplitude damping, exponentially damped
nonsecular coherence mixing, and a projective, phase-selective Khalfin
map. 

\paragraph*{Outline.}
The remainder of the paper develops the dynamical-map reconstruction and establishes its connection to singular time-local generators. We first formulate the reduced-dynamics framework and introduce
the disentangled representation based on a time-local generator.
We then analyze the reconstruction in the exactly solvable
rotating-wave approximation before applying the method to the full
spin--boson model. Readers primarily interested in the physical
content may proceed directly to Sec.~\ref{Sec:Disent}, where the
disentangled dynamical map is obtained and its constituent processes
are identified. The preceding sections provide the technical
derivation.

\section{The Standard Open System Setup\label{Sec:StandardSetup}}

We consider a finite--level quantum system weakly coupled to a bosonic
environment. The total Hamiltonian is
\begin{equation}
\label{Eq:Htotal}
    H_T = H_S + H_B + H_I .
\end{equation}

The system Hamiltonian is diagonal in its energy eigenbasis,
\begin{equation}
    H_S = \sum_{n=1}^{N} E_n \vert n\rangle \langle n \vert,
\end{equation}
with Bohr frequencies $\omega_{nm}=E_n-E_m$.
The environment consists of baths of harmonic modes
\begin{equation}
\label{Eq:Hbath}
    H_B = \sum_{k,\alpha} \omega_{k,\alpha} b_{k,\alpha}^{\dagger} b_{k,\alpha},
\end{equation}
where $\alpha$ and $k$ denote the bath and mode indices, respectively.
Different baths are assumed to be uncorrelated. 
and the coupling is taken in separable form
\begin{equation}
    H_I = \sum_\alpha A^\alpha \otimes F_\alpha,
\end{equation}
where $A^\alpha$ is a dimensionless Hermitian system operator that
couples the system to bath $\alpha$, normalized to have operator norm
of order unity and 
\begin{equation}
\label{Eq:Bath}
    F_\alpha = \sum_k g_{k,\alpha} (b_{k,\alpha} + b_{k,\alpha}^\dagger)
\end{equation}
is a Hermitian bath operator with coupling constants
$g_{k,\alpha}\propto\lambda\ll 1$.

\subsection*{Bath correlations}

The reduced dynamics are governed by the
bath correlation function
\begin{equation}
    C_\alpha(t)=\langle F_\alpha(t)F_\alpha(0)\rangle_\beta ,
\end{equation}
defined with respect to a thermal equilibrium state at inverse
temperature $\beta$.
The bath is characterized by the spectral density
\begin{equation}
\label{Eq:SDexpcut}
    \tilde{J}_{\omega}^\alpha = \pi \sum_k g_{k,\alpha}^2 \delta(\omega-\omega_{k,\alpha}), \qquad \omega>0,
\end{equation}
where the tilde denotes zero temperature. We employ the canonical continuum environments used in the spin--boson model~\cite{Leggett1987},
\begin{equation}
\label{Eq:SD0}
    \tilde{J}_\omega^\alpha = 2\pi\lambda^2
    \frac{\omega^s}{\omega_c^{\,s-1}}e^{-\omega/\omega_c}\Theta(\omega),
\end{equation}
where $\omega_c$ is the ultraviolet cutoff frequency and $s$ is the spectral exponent (Ohmic $s=1$, sub--Ohmic $0<s<1$, super--Ohmic $s>1$).

The thermal spectral density is
\begin{equation}
\label{Eq:Jbeta}
    J_{\omega}^\alpha=\frac{\tilde{J}_{\omega}^\alpha}{1-e^{-\beta\omega}}, \qquad \omega>0,
\end{equation}
with the extension to negative frequencies fixed by detailed balance,
\begin{equation}
    J_{-\omega}^\alpha=e^{-\beta\omega}J_{\omega}^\alpha.
\end{equation}

The bath correlation function admits the Fourier representation
\begin{equation}
\label{eq:BCF3}
    C_\alpha(t)=\frac{1}{\pi}\int_{-\infty}^{\infty} d\omega\,
    J_{\omega}^\alpha\, e^{-i\omega t},
\end{equation}
which at zero temperature reduces to
\begin{equation}
\label{Eq:BCF}
\tilde{C}_\alpha(t)=\frac{2\lambda^2\omega_c^2\Gamma(s+1)}{(1+i\omega_c t)^{s+1}}.
\end{equation}

\subsection*{Half--sided transform}

A central quantity is the half--sided Fourier transform
\begin{equation}
\label{Eq:Gamma}
\Gamma_{\omega}^\alpha=\int_{0}^{\infty}dt\, C_\alpha(t)e^{i\omega t}
= J_{\omega}^\alpha+iS_{\omega}^\alpha,
\end{equation}
which is analytic for $\mathrm{Im}\,\omega>0$.
Its finite-time counterpart defines the memory kernel
\begin{equation}
\label{Eq:timedSD}
    \Gamma_{\omega}^\alpha(t)=\int_{0}^{t} d\tau\, C_\alpha(\tau)e^{i\omega\tau},
\end{equation}
which is entire in $\omega$ and converges to $\Gamma_{\omega}^\alpha$ as $t\to\infty$.

\subsection*{Assumptions}

The analysis assumes a controlled weak--coupling open--system limit with
well separated dynamical scales.

\begin{enumerate}

\item \textit{Weak coupling.}
\[
\lambda^2\ll1,
\]
which is necessary for the Davies generator $L_0$ to provide the leading reduced dynamics.

\item \textit{Resolved system frequencies.}
The bath-induced relaxation rate remains small compared with the Bohr
frequencies,
\[
\gamma \sim J(\omega_{nm}) \ll |\omega_{nm}|.
\]
which in the spin--boson model reduces to $\gamma\ll\Delta$.

These scale separations are the necessary and sufficient conditions underlying the van Hove weak-coupling limit and ensure the existence of a well-defined pole contribution to the dynamics.

\item \textit{Continuum environment.}
The bath is taken in the thermodynamic limit with a continuous spectral
density~\eqref{Eq:SD0} and finite cutoff $\omega_c$, producing long--time
branch--cut contributions responsible for nonexponential (Khalfin)
decay.

\item \textit{Separated timescales.}
There exists an intermediate time window
\[
T_1,T_2 \ll t \ll t_{\mathrm{rec}},
\]
with relaxation times $T_{1,2}$ and recurrence time $t_{\mathrm{rec}}$.
The reconstruction and interpretation of generator singularities apply
in this regime.

\end{enumerate}

These conditions define the parameter domain in which divergences of the
time--local generator are interpreted.

\section{Time-Local Generator \label{Sec:PartialTCL}}

To uncover the structure of the divergent generator, we first examine the cumulant (van Kampen) expansion. In this formulation the propagator can be written as a time--ordered exponential,
\begin{equation}
\Phi(t)=\mathcal T \exp\left[\int_0^t d\tau\,L(\tau)\right],
\qquad
L(t)=\sum_{n=0}^\infty L_n(t),
\label{eq:cumulantgenerator}
\end{equation}
where the cumulants $L_n(t)$ encode progressively higher-order system--bath interactions~\cite{VANKAMPEN1,VANKAMPEN2}. 
Combining Eqs.~\eqref{eq:TCLgenerator} and ~\eqref{eq:cumulantgenerator}, the reduced density matrix satisfies the master equation
\begin{equation}
    \frac{d\rho(t)}{dt}=L(t)\rho(t).\label{Eq:TCLmeq}
\end{equation}
From a quantum field theory perspective, master equations such as Redfdield or Lindblad can  be viewed as partial resummations of an infinite perturbation series~\cite{kaplanek2020hot,brahma2025time}.
Since each cumulant $L_{n}(t)$ already resums an infinite subset of Dyson terms, the cumulant expansion is not merely a perturbative series but a hierarchy of nested resummations. Viewed in this way, selected successive cumulants act as filters on the underlying Dyson expansion. 

Central theme of this work is that such nested partial resummations  expose the processes associated with noninvertible dynamics. The first indication of this mechanism emerged in the analysis of slowly decaying environments in Ref.~\cite{Lampert2025}. 
There, the cumulant hierarchy was filtered to isolate the fastest-growing contributions, corresponding to the slowest decaying contributions at low perturbative order. Among these, a distinguished subset was found to be embeddable within the TCL2 generator. Resummation of the entire embeddable subsequence yielded an effective TCL2 generator with renormalized transition frequencies.

Using the partially resummed TCL generator obtained in Ref.~\cite{Lampert2025} as a starting point, we consider the case of multiple uncorrelated baths. The resulting generator is
\begin{equation}
\label{Eq:resumationGen}
    L(t) = -i[H_S,\,\cdot\,] + K(t),
\end{equation}
with the resummed dissipator
\begin{equation}
\label{Eq:renogen}
\begin{aligned}
    K_{nm,ij}(t) &= \sum_\alpha A^\alpha_{ni}A^\alpha_{jm}
    \left[\Gamma^\alpha_{\omega_{in}^{(j)}(t)}(t)
    +\big[\Gamma^\alpha_{\omega_{jm}^{(i)}(t)}(t)\big]^\star\right] \\
    &-\sum_{\alpha,k}\Big[
    \delta_{jm}A^\alpha_{nk}A^\alpha_{ki}\Gamma^\alpha_{\omega_{ik}^{(j)}(t)}(t)\\
    &+\delta_{ni}A_{jk}^\alpha A^\alpha_{km}\big[\Gamma^\alpha_{\omega_{jk}^{(i)}(t)}(t)\big]^\star
    \Big],
\end{aligned}
\end{equation}
where the renormalized complex frequencies are
\begin{align}
\nonumber
\omega_{in}^{(j)}(t)
&= \omega_{in}
- i\Big[
  \sum_{c,\alpha}\!\Big(\lvert A^\alpha_{ic}\rvert^{2}\,\Gamma^\alpha_{\omega_{ic}}(t)
  - \lvert A^\alpha_{nc}\rvert^{2}\,\Gamma^\alpha_{\omega_{nc}}(t)\Big)
  \\&+ \sum_\alpha 2 J^\alpha_{0}(t)\,A^\alpha_{jj}\big(A^\alpha_{nn}-A^\alpha_{ii}\big)
\Big].
\label{Eq:renofreqs}
\end{align}

These frequencies inherit the antisymmetry property of the initial Bohr frequencies, 
\begin{equation}
\omega_{in}^{(j)}(t)=-\omega_{ni}^{(j)}(t)\label{Eq{antisymmetry}}.
\end{equation}
 The imaginary
parts generated by the bath correlation function therefore
enter with opposite signs for the two transitions. If the
component associated with $\omega^{(j)}_{\mathrm{in}}$ acquires a decay
rate $\gamma^{(j)}_{\mathrm{in}}(t)>0$, the conjugate component
necessarily acquires the rate $-\gamma^{(j)}_{\mathrm{in}}(t)$.
When the bath correlation function cannot suppress the growing component, the cumulant expansion diverges and the corresponding time-local generator becomes unbounded. Vice versa, exponentially decaying bath corelation functions admit asymptotic limit of the generator if the coupling is sufficiently weak~\cite{Lampert2025}.

The first term in Eq.~\eqref{Eq:renofreqs} is the Fermi--Golden--Rule contribution,
while the second term $\Omega^\alpha_{jni}(t)=2 J_{0}^\alpha(t)\,A^\alpha_{jj}\big(A^\alpha_{nn}-A^\alpha_{ii}\big)$
is the spectral--overlap correction due to bath $\alpha$, that couples populations and
coherences beyond Born--Markov theory. For reference, and to clarify these expressions, the explicit
matrix elements for the spin--boson model coupled to a single bath are given in
Appendix~\ref{App:TCLSBM}. 

\begin{remark}
\normalfont
[Resemblance to the Redfield Equation]
The time-local generator in Eq.~\eqref{Eq:renogen} may superficially resemble a low-order TCL generator. This resemblance, however, should not be interpreted as indicating a second-order perturbative approximation. The renormalized transition frequencies appear as arguments of the memory kernels, and expanding the kernels in powers of the system--environment coupling generates contributions at arbitrarily high perturbative orders. Thus, the compact form of Eq.~\eqref{Eq:renogen} represents a partial resummation of the cumulant hierarchy rather than a second-order truncation. Similar compact generators arise in exact Gaussian TCL resummations~\cite{DAbbruzzoPRL}.
\end{remark}

\begin{remark}
\normalfont
[Accuracy of Eq.~\eqref{Eq:TCLmeq} relative to the Redfield Equation]
The \textit{}improved accuracy of Eq.~\eqref{Eq:TCLmeq} relative to the Redfield master equation has already been demonstrated by direct comparison with numerically exact TEMPO dynamics in Ref.~\cite{Lampert2025}. In particular, Fig.~6 of Ref.~\cite{Lampert2025} directly compares standard TCL2 and the resummed dynamics with TEMPO and demonstrates the improved accuracy obtained from the resummation. Moreover, Appendix~H of Ref.~\cite{Lampert2025} demonstrates analytically that, within the Markovian approximation, the asymptotic state obtained from the resummed dynamics is accurate through $O(\lambda^2)$,  while the Redfield (TCL2) master equation
asymptotic state is accurate only through $O(\lambda^0)$.
\end{remark}

\subsection{Generator validity range.} The generator $L(t)$  obtained from Ref.~\cite{Lampert2025}
establishes an internally consistent description up to a finite time
scale $t_L$ 
\[
t_L \approx \frac{s+1}{\nu_2},
\qquad
\nu_2=\frac{1}{T_2},
\]
where $T_2$ is the decoherence time. 
Although Ref.~\cite{Lampert2025} introduced the time scale $t_L$ 
using the auxiliary reference generator $L_M$, the same time scale is obtained when the Davies generator $L_0$ 
 is used instead. Since $\|L_M-L_0\|=O(\lambda^2)$ is time-independent, both references identify the same onset of growth-mode dominance up to perturbative corrections. We therefore adopt $L_0$
 as the reference generator throughout.
The time scale $t_L$
 therefore marks the crossover from a regime where the growth modes constitute a perturbative correction to one where they dominate the generator.

The reference generator $L_0$ is taken to be the Davies
weak-coupling generator~\cite{davies1974}, i.e. the
Markovian semigroup selected by the van Hove
weak-coupling limit of the microscopic Hamiltonian:
\begin{equation}
L_0=
-i[H_S+H_{\rm LS},\,\cdot\,]+K_0,
\end{equation}
where $K_0$ is the dissipator with matrix elements
(see, e.g., Ref.~\cite{davies1974,Davidovic2020})
\begin{align}
[K_0]_{nm,ij}
&= \sum_\alpha 2\,A^\alpha_{ni}A^\alpha_{jm}\,\delta_{\omega_{in},\omega_{jm}}\,J^\alpha_{\omega_{in}}
\nonumber\\
&\quad - \delta_{ni}\delta_{jm}\sum_{\alpha,k}\Big(
|A^\alpha_{nk}|^2\,J^\alpha_{\omega_{nk}}
+|A^\alpha_{jk}|^2\,J^\alpha_{\omega_{jk}}
\Big).
\label{Eq:DaviesGeneral}
\end{align}
The Lamb-shift Hamiltonian $H_{\rm LS}$ has matrix elements
\begin{equation}
[H_{\rm LS}]_{nm}=\delta_{nm}\sum_{\alpha,k} |A_{nk}|^2\,S^\alpha_{\omega_{nk}}.
\end{equation}
The specialization of $L_0$ to the spin--boson model is given
in Sec.~\ref{Sec:baseline}.

In the van Hove scaling $\tau=t\lambda^2$ the reduced dynamics converges to
this semigroup,
\begin{align}
\lim_{\lambda\to 0}\,
\big\|\rho_{\mathrm{ex}}(\tau/\lambda^2)-e^{L_0\,\tau/\lambda^2}\rho(0)\big\|=0,
\label{Eq:vanHovelimit}
\end{align}
so $e^{L_0 t}$ provides provides the leading reduced dynamics in this weak-coupling sense.

\section{Feynman Disentanglement Theorem\label{Sec:DisTheo}}

Since the exact dynamics remains asymptotically close to the contractive semigroup generated by the Davies generator, Eq.~\eqref{Eq:vanHovelimit}, it is natural to express the evolution relative to this reference semigroup.
\begin{equation}
\label{Eq:PeresC}
\rho_{\rm ex}(t)
=
\Phi_{\rm ex}(t)\rho(0)
=
\bigl(e^{L_0 t}
+
\mathbf C_{\rm ex}(t)\bigr)\rho(0).
\end{equation}

Here $\mathbf C_{\text{ex}}(t)$ is a  correction superoperator describing
deviations from the reference evolution, with
$\mathbf C_{\text{ex}}(0)=0$ so that $\Phi_{\rm ex}(0)=\mathbb I$. 
Similar use of a contractive semigroup as a leading reference evolution
appears in weak-coupling approaches
\cite{de2013approach,merkli2022dynamics}. 

The exact quantum dynamical map satisfies the master equation
\begin{equation}
\dot{\Phi}_{\text{ex}}(t)=L_{\text{ex}}(t)\Phi_{\text{ex}}(t),
\qquad
\Phi_{\text{ex}}(0)=1,
\end{equation}
provided that the dynamical map remains invertible.
Split the generator as
\begin{equation}
L_{\text{ex}}(t)=L_0+\delta L(t).
\end{equation}
The Feynman disentangling theorem then gives
\begin{equation}
\Phi_{\text{ex}}(t)
=
e^{L_0 t}\,
\mathcal T
\exp\!\left[
\int_0^t d\tau\,
e^{-L_0 \tau}\,\delta L(\tau)\,e^{L_0 \tau}
\right].
\end{equation}
Equivalently,
\begin{equation}
\Phi_\text{ex}(t)
=
e^{L_0 t}\,
\mathcal T
\exp\!\left[
\int_0^t d\tau\,
e^{-L_0 \tau}\,
\bigl(L_{\text{ex}}(\tau)-L_0\bigr)
e^{L_0 \tau}
\right].
\end{equation}

Since $\delta L(t)=O(\lambda^2)$, expansion to $O(\lambda^2)$ gives
\begin{equation}
\Phi_\text{ex}(t)
=
e^{L_0 t}
\left[
1+
\int_0^t d\tau\,
e^{-L_0 \tau}\,\delta L(\tau)\,e^{L_0 \tau}
\right]
+O(\lambda^4).
\end{equation}

Inserting the factor $e^{L_0 t}$ into the integral and comparing with
Eq.~\eqref{Eq:PeresC}, we identify the correction
$\mathbf C_{\text{ex}}(t)$ to accuracy $O(\lambda^2)$ as
\begin{equation}
\mathbf C_{\text{ex}}(t)
=
\int_0^t d\tau\,
e^{L_0(t-\tau)}
\bigl(L_{\text{ex}}(\tau)-L_0\bigr)
e^{L_0\tau}
+O(\lambda^4).
\label{Eq:CsolutionEX}
\end{equation}

The loss of invertibility of the dynamical map can make the norm
\(\|L_{\rm ex}(t)\|\) arbitrarily large, but does not produce a
corresponding growth of the dynamical map, which remains perfectly regular~\cite{Chru2010}. Consequently,
\(\mathbf C_{\rm ex}(t)\) remains finite. The time integral in
Eq.~\eqref{Eq:CsolutionEX} suppresses the divergence of
\(L_{\rm ex}(\tau)\) because its action is weighted by the contractive
Davies semigroup $e^{L_0\tau}$. The forward semigroup action restricts the late-time dynamics
to a reduced domain near the stationary state. Within this domain,
only those components of the divergent generator that affect the
approach to the stationary state contribute to the evolution, while
components that leave this behavior unchanged become redundant.

We now make the crucial step. The preceding discussion establishes  the exact generator need only be characterized through its action on an asymptotically shrinking neighborhood of stationary state. This observation motivates the following asymptotic reduction.

We assume that the same contractive mechanism remains
operative when the exact generator $L_{\rm ex}(t)$ is replaced
by the partially resummed generator $L(t)$. 
This motivates defining an approximate  map through
\begin{equation}
\label{Eq:Csolution}
\mathbf{C}(t)=\int_0^t d\tau\,
e^{L_0(t-\tau)}\big[L(\tau)-L_0\big]e^{L_0\tau}.
\end{equation}
The net map is then written as $e^{L_0 t}+\mathbf{C}(t)$, where the integral disentangles the map from the divergent
behavior of the approximate generator. To avoid repeated terminology, we refer to this map
interchangeably as the \emph{reconstructed} or
\emph{disentangled} map. The latter name reflects its
derivation from the disentangling theorem discussed above.

Note that the result in Eq.~\eqref{Eq:Csolution} is nonperturbative because the truncation applies only to the disentanglement step, not to the generator \(L(t)\), which is partially resummed and contains nested contributions from all orders of the interaction. Numerical results in this paper were obtained by solving the reconstruction equation
\begin{equation}
\label{Eq:Cdot}
\dot{\mathbf{C}}(t)-L_0\mathbf{C}(t)
=\big[L(t)-L_0\big]e^{L_0 t},
\end{equation}
which is equivalent to Eq.~\eqref{Eq:Csolution} but provides improved numerical stability. The differential formulation also makes clear that the growing modes of \(L(t)\) are preceded by the reference contraction $e^{L_0 t}$.

Also note that separating the integrand into \(L(\tau)\) and \(-L_0\) parts gives
\begin{align}
\mathbf C(t)
&=\int_0^t d\tau\, e^{L_0(t-\tau)}L(\tau)e^{L_0\tau}
-\int_0^t d\tau\, e^{L_0(t-\tau)}L_0e^{L_0\tau},
\end{align}
since \(L_0\) commutes with its exponential,
\begin{equation}
\int_0^t d\tau\, e^{L_0(t-\tau)}L_0e^{L_0\tau}
=t\,e^{L_0 t}L_0.
\label{Eq:linear_term}
\end{equation}
Hence
\begin{equation}
\begin{split}
\Phi(t)
&= e^{L_0 t}
- t\,e^{L_0 t}L_0
+ \int_0^t d\tau\, e^{L_0(t-\tau)}L(\tau)e^{L_0\tau}
\\
&= e^{L_0 t}\!\left[
I
- tL_0
+ \int_0^t d\tau\, e^{-L_0\tau}L(\tau)e^{L_0\tau}
\right].
\end{split}
\label{Eq:Phi_constraint}
\end{equation}

The last pair of equations show that the disentanglement procedure
necessarily produces a linear-in-time contribution proportional to
\(tL_0\). A purely exponential decay therefore requires that the
contribution generated by \(L(\tau)\) provide an equal and opposite
linear term. 

\section{RWA model and exact dynamics\label{Sec:RWAsetup}}

We consider the spin-boson model
\begin{equation}
H = H_S+H_B+H_I,\qquad
H_S=\tfrac{\Delta}{2}\sigma_z,\quad
H_I=A\otimes F .
\label{Eq:SBM_RWA}
\end{equation}
We specialize to unbiased transverse coupling,
\begin{equation}
A = \frac{1}{2}\sigma_x .
\end{equation}

Throughout the main text we further restrict to zero temperature. Excitation processes are then absent, so any loss of invertibility of the reduced dynamics cannot be attributed to thermal mixing and must instead originate from coherent system--bath interaction. In several expressions we write the matrix element $A_{12}^2$ explicitly as $1/4$; this factor should be understood as $A_{12}^2$, $A_{21}^2$, or $|A_{12}|^2$, depending on context, and thus the formalism extends trivially to an arbitrary transverse coupling direction.

In the rotating-wave approximation we retain only number-conserving terms,
\begin{align}
H_I &\approx \tfrac{1}{2}\left(\sigma_+\otimes F_-+\sigma_-\otimes F_+\right),
\\
F_+&=\sum_k g_k b_k^\dagger,\;\; F_-=\sum_k g_k^\ast b_k ,
\end{align}
so that the total excitation number
$N=\sigma_+\sigma_-+\sum_k b_k^\dagger b_k$
is conserved. At $T=0$ the dynamics from a single-excitation preparation is
therefore confined to the $N=1$ sector.

For an initial superposition
\begin{equation}
\vert\Psi(0)\rangle=c_1\vert 1\rangle+c_2\vert 2\rangle,\qquad
|c_1|^2+|c_2|^2=1,
\label{Eq:InitSuperposition}
\end{equation}
the exact reduced state has the form~\cite{zhang2012general}
\begin{equation}
\label{Eq:rhoExact}
\rho_{\rm exact}(t)=
\begin{bmatrix}
|c_1 f(t)|^2 & c_1 c_2^\ast f(t)\\
c_1^\ast c_2 f^\ast(t) & 1-|c_1 f(t)|^2
\end{bmatrix},
\end{equation}
corresponding to the quantum dynamical map
\begin{equation}
\label{Eq:RWAmap}
\Phi_{f}(t)=
    \begin{bmatrix}
       \vert f(t)\vert^2 & 0 & 0 & 0\\
       1-\vert f(t)\vert^2 & 1 & 0 & 0\\
        0   & 0& f^\star(t) & 0\\
        0   & 0 & 0 & f(t)
    \end{bmatrix},
\end{equation}
in the ordering $(\rho_{11},\rho_{22},\rho_{21},\rho_{12})$. So all reduced dynamics is determined by the
survival amplitude $f(t)$. The coherence sector contributes two singular values equal to $\vert f\vert$, while the population block yields one singular value of order  $\vert f\vert^2$,  and one of order unity. Hence, as $f(t)\to 0$ three singular values simultaneously vanish and the map approaches a rank-one projection.

The time--local generator is obtained by inserting Eq.~\eqref{Eq:RWAmap} into Eq.~\eqref{eq:TCLgenerator}:
\begin{equation}
\label{Eq:RWAgen}
 L_{f}(t)=
    \begin{bmatrix}
       \frac{\dot f}{f}+\frac{\dot f^\star}{f^\star} & 0 & 0 & 0\\
        -\frac{\dot f}{f}-\frac{\dot f^\star}{f^\star} & 0 & 0 & 0\\
        0  & 0  & \frac{\dot f^\star}{f\star} & 0 \\
        0   & 0 & 0 & \frac{\dot f}{f}
    \end{bmatrix}.
\end{equation}
Consequently the time-local generator diverges at the zeros of the survival amplitude. 

In the RWA model, the transverse Bloch component satisfies
\[
x(t)-iy(t)=2\rho_{12}(t)=f(t)\,[x(0)-iy(0)] .
\]
Thus the evolution acts on the $(x,y)$ plane as an isotropic contraction
combined with a rotation.

In the interaction picture, $f_I(t)=e^{-i\Delta t}f(t)$ obeys the exact
Volterra equation
\begin{equation}
\dot f_I(t)= -\frac{1}{4}\int_0^t d\tau\,
C(t-\tau)\,e^{i\Delta(t-\tau)} f_I(\tau).
\label{Eq:IDeq}
\end{equation}
Equation~\eqref{Eq:IDeq} is identical to the survival--amplitude
equation derived by Peres for quantum decay into a continuum
spectrum~\cite{Peres1980}. In such systems the survival amplitude
contains two distinct components: an exponential term associated
with a pole of the resolvent, and a delayed correlation term arising
from the continuum spectrum. As first shown by Khalfin~\cite{Khalfin1958},
the continuum part necessarily produces non-exponential long-time
behavior. In the present model the same pole and branch--cut structure
appears through the bath correlation function $C(t)$, which generates
an algebraic tail that interferes with the exponential pole
contribution.

The explicit solution in the Schrödinger picture is~\cite{zhang2012general} 
\begin{equation}
\label{Eq:BCintegral}
f(t)=\frac{1}{4\pi}\int_0^\infty d\omega\;
\frac{e^{-i\omega t}J_\omega}
{(\Delta-\omega+S_\omega /4)^2+(J_\omega/4)^2}.
\end{equation}
The integral representation confirms that the coherence amplitude remains finite and smooth for all times. Thus, a divergence of the logarithmic derivative $\dot f(t)/f(t)$ can occur only when the denominator becomes arbitrarily small. This realizes the mechanism anticipated in the Introduction: the reduced state remains regular while the dynamical map approaches loss of invertibility, and the singularity appears only in the logarithmic generator representation.

Equation~\eqref{Eq:BCintegral} makes explicit that the survival amplitude
is a Fourier transform of a bounded spectral density determined by the
bath. Consequently the late-time behavior cannot remain purely
exponential and must cross over to the algebraic decay.

\begin{figure}[h] \centering \includegraphics[width=0.45\textwidth]{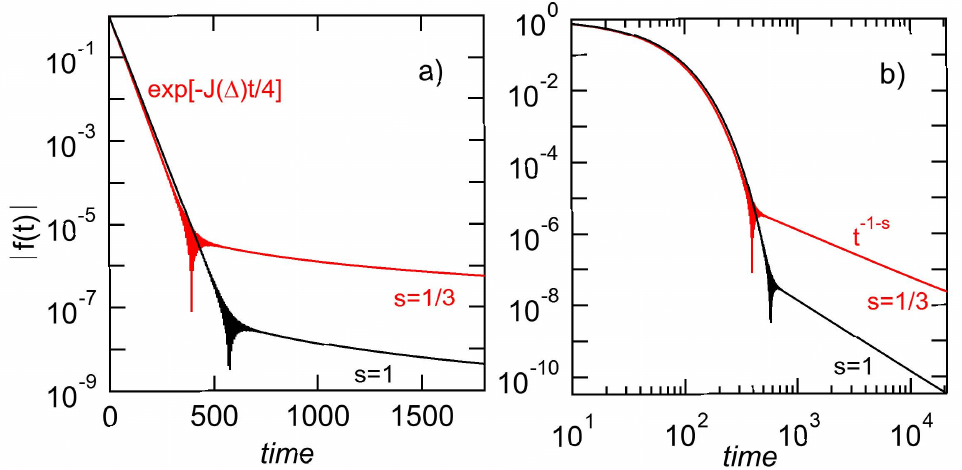} \caption{\label{Fig:Example1} Representative long-time dynamics of the coherence amplitude $|f(t)|$. 
(a,b) Log--linear and log--log plots showing the crossover from exponential (Markovian) to power-law (non-Markovian) decay. 
Near the phase lock-in time $t_P$, interference produces oscillations. 
Black: Ohmic bath ($s=1$, $\lambda^2=0.025$); red: sub-Ohmic bath ($s=1/3$, $\lambda^2=0.01$). 
In both cases $\omega_c=4$, $\Delta=1$.} \end{figure}

Representative exact solutions for several spectral exponents are shown in Fig.~\ref{Fig:Example1}. The coherence magnitude $|f(t)|$ crosses from an initial exponential decay to a late-time algebraic tail at the time scale $t_P$, defined by Eq.~\eqref{Eq:tp}. Near $t_P$, destructive interference between these two contributions produces deep minima in $|f(t)|$.

\subsection{Pole--plus--tail interference}\label{Sec:RWAlockin}

In the weak--coupling regime the integrand in Eq.~\eqref{Eq:BCintegral} is is sharply peaked near $\tilde{\Delta}=\Delta+S(\Delta)/4$. Expanding about this peak yields a Lorentzian lineshape and the Wigner--Weisskopf (Markovian) exponential contribution
\begin{equation}
\label{Eq:Markov}
f_M(t)=e^{-[i\tilde{\Delta}+J_\Delta/4]t},
\end{equation}
which accurately describes the coherence for $t\lesssim t_P$.

At late times the dominant frequencies scale as $\omega\sim 1/t$, and the lineshape in Eq.~\eqref{Eq:BCintegral} is controlled by its low--frequency behavior. Rapidly oscillating contributions average out and only the low-frequency part of the spectrum survives. Approximating the denominator by its $\omega=0$ value then gives the tail
\begin{equation}
\label{Eq:KhalfinExact}
f_C(t)\approx
\frac{1}{4[\Delta+S_0/4]^2}C(t),
\end{equation}
which captures the algebraic decay for $t\gtrsim t_P$.
The exact coherence is therefore well approximated by the two--component form
\begin{equation}
\label{Eq:RWAtwoComponent}
    f(t)\approx f_M(t)+f_C(t).
\end{equation}
Destructive interference between $f_M$ and $f_C$ produces
times at which $f(t)$ becomes anomalously small, approaching the kernel
of the dynamical map.

The time $t_P$ follows from the balance between the
Markovian pole contribution and the long--time tail,
$|f_M(t_P)|=|f_C(t_P)|$.
From Eq.~\eqref{Eq:Markov} one has
$|f_M(t)|=e^{-t/T_2}$, while the bath correlator for a spectral density
$J(\omega)\propto\omega^s e^{-\omega/\omega_c}$ yields the asymptotic
tail $|f_C(t)|\propto(\omega_c t)^{-(s+1)}$.
Equating the two gives
\begin{equation}
e^{-t_P/T_2}\sim(\omega_c t_P)^{-(s+1)},
\label{Eq:balance}
\end{equation}
so that
\[
\frac{t_P}{T_2}\sim (s+1)\ln(\omega_c t_P).
\]
In the weak--coupling regime $\omega_c T_2\gg1$, replacing
$t_P$ inside the logarithm by $T_2$ up to subleading $\ln\!\ln$
corrections yields
\begin{equation}
\begin{split}
t_P &\sim (s+1)\,T_2\,\ln(\omega_c T_2)\\
    &\sim (s+1)\,T_2\,\ln\frac{1}{\lambda^2}.
\end{split}
\label{Eq:tp}
\end{equation}

\begin{figure}[h] \centering \includegraphics[width=0.45\textwidth]{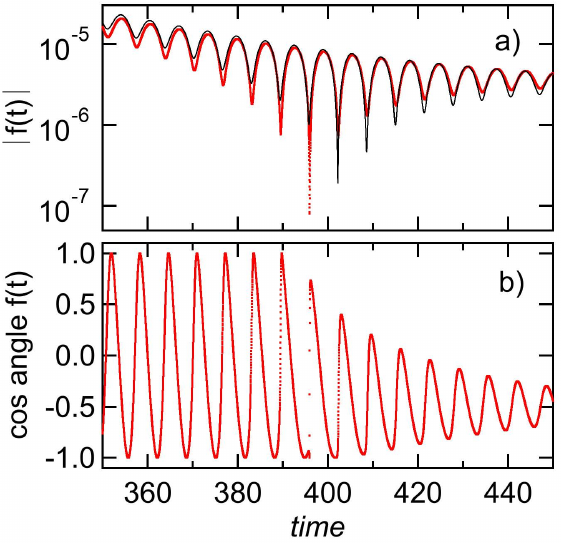} \caption{\label{Fig:PhaseSlip} Coherence magnitude (a) and phase (b) near $t_P$ in the RWA model. 
Constructive interference produces bursts, while destructive interference leads to near-extinction. 
The phase undergoes a rapid rotation approaching a $\pi$ slip, consistent with the map remaining invertible. 
Red dotted: exact solution; black solid: two-component model. 
Parameters: $s=1/3$, $\lambda^2=0.01$, $\omega_c=4$, $\Delta=1$.} \end{figure}

Near $t_P$, interference between $f_M$ and $f_C$ produces narrow modulation
windows in $|f(t)|$, ranging from transient bursts to near--extinctions and
accompanied by near-$\pi$ phase slips at destructive interference
[Fig.~\ref{Fig:PhaseSlip}(a,b)].

The dynamical map becomes strictly non-invertible only if $f(t)=0$ at a
finite time. For a continuum environment this requires exact destructive
interference between the pole contribution and the bath-correlation
tail. Such cancellation requires fine tuning of spectral phases and is therefore
non-generic for continuum RWA-Hamiltonians. The interference produces very small
but nonzero values of $f(t)$, so the map remains invertible at all finite
times while approaching a non-invertible channel asymptotically.

\section{\label{Sec:vanKampen}RWA Cumulant Expansion}

\subsection{Reference and TCL Generators}
\label{Sec:resumationresult}

We now apply the general cumulant analysis to the RWA Volterra
equation~\eqref{Eq:IDeq} using the van~Kampen ordered-cumulant
expansion. As in the general case, for algebraically decaying bath
correlation functions, cumulants beyond a critical order $n_{\max}(s)$
display late-time growth~\cite{Lampert2025}. Resumming the leading
growing cumulants reorganizes this algebraic growth into an exponential
time dependence. This leading resummation is a special case of the complex
renormalization introduced in Eq.~\eqref{Eq:renofreqs} and defines
the inflated TCL generator. Details are presented in
Appendix~\ref{Sec:nesting}.

The time-dependent renormalized frequency is given by
\begin{equation}
\label{eq:renoSELF}
    u(t)=\omega-\frac{i}{4}\Gamma_\omega(t),
\end{equation}
while the survival amplitude in the interaction picture satisfies
\begin{equation}
\label{Eq:genU}
    \left.\frac{\dot f_I}{f_I}\right|_{\rm resum}
    =-\frac{1}{4}\Gamma_{u(t)}(t)+O(\lambda^4).
\end{equation}

The Schr\"odinger-picture partially resummed generator is then
obtained from Eq.~\eqref{Eq:RWAgen} by replacing $\dot f/f$ with its
resummed counterpart, yielding
\begin{equation}
L(t)=
\begin{pmatrix}
L_{\mathrm{pop}}(t) & 0\\
0 & L_{\mathrm{coh}}(t)
\end{pmatrix},
\label{Eq:LSBM_blockRWA}
\end{equation}
in the vectorization order $\rho_{11},\rho_{22},\rho_{21},\rho_{12}$. Here
\begin{align}
L_{\mathrm{pop}}(t)
&=\frac{1}{2}
\begin{pmatrix}
-\mathrm{Re}\,\Gamma_{u(t)}(t) & 0\\
\mathrm{Re}\,\Gamma_{u(t)}(t) & 0
\end{pmatrix},
\label{Eq:RWAresumPO}
\end{align}
and
\begin{align}
L_{\mathrm{coh}}(t)
&=
\begin{pmatrix}
i\Delta-\tfrac{1}{4}\Gamma_{u(t)}^\star(t) & 0\\
0 & -i\Delta-\tfrac{1}{4}\Gamma_{u(t)}(t)
\end{pmatrix}.
\label{Eq:RWAresumCO}
\end{align}
The generator exhibits exponential growth at long times, since
$\mathrm{Im}\,u(t)<0$ as $t\to\infty$.

\paragraph*{Remark}
The full ordered-cumulant resummation contains nested renormalizations
(Appendix~\ref{Sec:nesting}) that can shift the renormalized frequency
onto a different analytic branch,
\begin{equation}
    u=\omega-\tfrac{i}{4}\,
      \Gamma_{\omega - \tfrac{i}{4}\Gamma_{\omega - \tfrac{i}{4}\Gamma_{\omega}-\ldots}},
      \label{Eq:renoSELFconsistent}
\end{equation}
where we suppress the time dependence for clarity.
Such branch shifts can qualitatively modify the long-time structure of
the resummation, including its invertibility properties. In the present
work we restrict the reconstruction to the analytic branch
Eq.~\eqref{eq:renoSELF} that is continuously connected to the
weak-coupling (Davies) limit. This choice preserves continuity with the
microscopic weak-coupling theory and defines the branch used throughout.
The role of the self-consistent renormalizations in
Eq.~\eqref{Eq:renoSELFconsistent} is left for future work.

\subsection{Disentangled Map for the RWA}
\label{Sec:resumationDmapRWA}

For the RWA Hamiltonian, the Davies generator decomposes into population and coherence sectors,
\begin{align}
(L_0)_{\mathrm{pop}}
&=\frac{1}{2}
\begin{pmatrix}
-J_\Delta & 0\\
J_\Delta & 0
\end{pmatrix},
\label{Eq:RWAresumPOD}
\end{align}
and
\begin{align}
(L_0)_{\mathrm{coh}}
&=
\begin{pmatrix}
i\tilde{\Delta}-\tfrac{1}{4}J_\Delta & 0\\
0 & -i\tilde{\Delta}-\tfrac{1}{4}J_\Delta
\end{pmatrix}.
\label{Eq:RWAresumCOD}
\end{align}

Because $[L(t),L_0]=0$, the reconstruction
integral~\eqref{Eq:Csolution} simplifies to
\begin{equation}
\label{Eq:RWAsolRE}
\mathbf{C}(t)=e^{L_0 t}\int_0^t d\tau\,[L(\tau)-L_0].
\end{equation}

We define the complex Bohr frequency
\begin{equation}
z=\Delta-\frac{i}{4}\Gamma_\Delta .
\label{Eq:Bohrzb}
\end{equation}

Using the explicit matrices above, the reconstructed
coherence amplitude
$f(t)=\rho_{12}(t)/\rho_{12}(0)$ becomes
\begin{align}
f(t)
&=\Big(1+\frac{\Gamma_\Delta}{4}t\Big)e^{-izt}
-\frac{1}{4}e^{-iz t}
\int_0^t d\tau\,\Gamma_{u(\tau)}(\tau) .
\label{Eq:Phi_RECO}
\end{align}

The linear prefactor in front of the first exponential in Eq.~\eqref{Eq:Phi_RECO} originates from the disentanglement identity Eq.~\eqref{Eq:Phi_constraint}.
Its cancellation is clarified below.

\subsubsection{Markovian Approximation for the Frequency\label{sec:MAF}}
To obtain a closed analytic description of the dynamics, we consider the asymptotic regime
$t\gg\tau_C$, where $\tau_C\equiv 1/\omega_c$ is the bath correlation
time. 
The condition $t\gg\tau_C$ is typically associated with the
onset of the Markovian limit~\cite{BreuerHeinz-Peter1961-2007TToO}.

In the regime $t\gg\tau_C$ we freeze the frequency
renormalization in Eq.~\eqref{eq:renoSELF} at its asymptotic value
and treat the pole as time independent,
\[
u(t)\rightarrow z .
\]
This is justified because the bath correlation function has
already decayed on the scale $\tau_C$, so the self-energy
contribution entering $u(t)$ has reached its stationary limit.

In contrast, we retain the explicit time dependence of the memory kernel in Eqs.~\eqref{Eq:RWAresumPO} and~\eqref{Eq:RWAresumCO}. The cumulant integral continues to
grow and produces the large logarithmic derivative
$\Gamma_z(t)$ even after the renormalized frequency has saturated.
Consequently the late-time behavior is governed by a stationary pole but
a time-dependent generator. 

The resulting survival amplitude simplifies to
\begin{equation}
f(t)
=\left(1+\frac{1}{4}\Gamma_\Delta t\right)e^{-iz t}
-\frac{1}{4}\Sigma_{z}(t),
\label{Phi_preMT}
\end{equation}
where
\begin{equation}
\Sigma_z(t)=\int_0^t d\tau\,\tau\,C(t-\tau)e^{-iz\tau}
\label{SigmaDe}
\end{equation}
is termed the diagonal kernel. 
It governs the late-time
nonexponential decay of the survival amplitude.

The diagonal kernel admits the contour representation
\begin{equation}
\Sigma_z(t)=
-\frac{1}{2\pi}\int_{-\infty+i\gamma}^{\infty+i\gamma}
d\omega\,
e^{-i\omega t}\frac{\Gamma_\omega}{(z-\omega)^2},
\label{SigmaContou}
\end{equation}
where the contour offset \(i\gamma\) lies above all
singularities of the integrand. This form provides the
starting point for the asymptotic analysis below.

\subsubsection{Wigner--Weisskopf Regime}
To leading order in $\lambda^2$, the contour integral in
Eq.~\eqref{SigmaContou} yields the survival amplitude
\begin{align}
f(t) &=
e^{-iz t} +
\frac{1}{4\pi}\int_0^\infty d\omega\,
\frac{e^{-i\omega t}\,\delta J_\omega}{(z-\omega)^2},
\label{Eq:fCmain}
\end{align}
where \(\delta J_\omega=J_\omega-J_\Delta\).

The pole contribution together with the on--shell part of
the branch cut on \(\omega\in\mathbb{R}^+\) cancels the
linear term \(\Gamma_\Delta t/4\) generated by the
disentanglement theorem, leaving the pure exponential
decay of the van Hove limit. The remaining integral is the
residual branch--cut correction.

\subsubsection{Khalfin Regime}

At long times the pole contribution has fully decayed, and the dynamics
is governed entirely by the continuum component. The remaining amplitude
is therefore determined by the highly oscillatory Fourier integral
in~\eqref{Eq:fCmain}, whose leading behavior is controlled by the
analytic structure of the spectral density.

One obtains
\begin{equation}
    f(t)\simeq 
    \frac{C(t)}{4z^2},
    \label{Eq:KhalfinMTrwa}
\end{equation}
so that the coherence amplitude is directly proportional to the bath
correlation function. This agrees, up to $O(\lambda^2)$ corrections in $z$, with the
Khalfin long-time tail of the exact solution~\eqref{Eq:KhalfinExact}.

\subsection{Exact Solution versus Asymptotic Reduction}

\begin{figure}[h]
\centering
\includegraphics[width=0.45\textwidth]{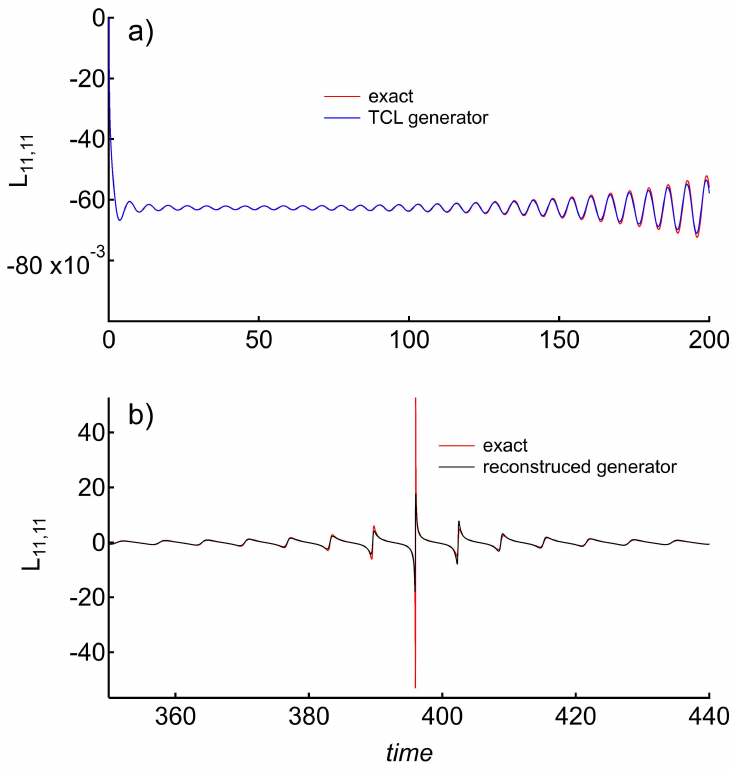}
\caption{\label{Fig:Compare}
\textbf{Numerical validation of the asymptotic reduction procedure.}
(a) Early-time generator element $L_{11,11}(t)$: resummed TCL (blue) agrees with the exact result (red). 
(b) Near $t_P$, destructive interference suppresses $f(t)$ and produces a sharp generator spike. 
The generator reconstructed from the disentangled map (black) reproduces the exact result (red) before, at, and after the spike. 
Parameters: $s=1/3$, $\lambda^2=0.01$, $\omega_c=4$, $\Delta=1$, $T=0$.}
\end{figure}

Figure~\ref{Fig:Compare} verifies the asymptotic reduction procedure by solving
the reconstruction equation~\eqref{Eq:Cdot} numerically, without invoking
the asymptotic approximation $t\gg\tau_C$.

Panel (a) shows the early-time generator matrix element
\(L_{11,11}(t)\) obtained from the resummed TCL expansion
and from the exact solution. The agreement demonstrates
that the TCL generator captures the correct analytic
structure at short times. This regime was termed
``secular inflation'' in Ref.~\cite{Lampert2025}, but that
description is misleading in this time range, since the
growth is physical and reproduced by the exact solution.

Panel~(b) focuses on the time $t_P$ where destructive interference
strongly suppresses the survival amplitude $f(t)$ and the dynamical
map becomes nearly singular. The generator reconstructed from the dynamical map
matches well the exact generator before, at, and after the
spike, except at the very near-singular peaks. The
agreement beyond the spike reflects that the map remains
invertible.

\subsection{From the RWA to the Full Spin--Boson Model}

The RWA provides a first validation of the disentanglement procedure and
identifies the origin of the near-singular behavior of the dynamical map.
Interference between the exponential pole and the algebraic Khalfin tail
strongly suppresses the survival amplitude \(f(t)\), bringing the map close
to singularity. Except under measure-zero phase-matching conditions,
however, the RWA map remains invertible.

As a result, the RWA fails to select a classical coherence of the system at late times. As shown in
Eq.~\eqref{Eq:rhoExact},
\[
\rho_{12}(t)\propto f(t)c_1c_2^\ast.
\]
 Because \(\lvert f(t)\rvert\)
decays algebraically in the Khalfin regime, the initial quantum-phase
encoded in $c_1c_2^\ast$ survives at arbitrarily long times, although with algebraically diminishing
amplitude.  Thus, although the direction of the coherence becomes fixed above the Khalfin transition as the precession ceases, that direction remains determined by the initial coherence. The late-time RWA dynamics therefore remains bijective rather than becoming projective.

The full spin--boson model behaves differently. Counter-rotating terms
activate nonsecular coherence transfer, allowing the dynamics to select a preferred quadrature of the coherence rather than preserve the direction set by the initial state.

\section{Full Spin-Boson Model}
\label{Sec:QubitDynamics}

We now consider a two-level system coupled to a bosonic environment through
the full interaction operator \(A=\sigma_x/2\) in
Eq.~\eqref{Eq:SBM_RWA}. The interaction $A$ contains counter-rotating terms that generate nonsecular transfer between different coherence quadratures, providing the mechanism by which coherence is suppressed along one direction while the orthogonal direction survives.

We now perform the Feynman disentanglement theorem explicitly
for the full spin--boson model. The derivation is highly technical but
necessary to establish the dynamical map used throughout the remainder of
the paper. The physical analysis begins in Sec.~\ref{Sec:Disent}. Readers
primarily interested in the science are encouraged to begin there and
return to the derivation as needed.

Unlike the RWA, the full spin--boson model is not integrable and there is no exact solution. The
identification of the partially resummed generator with the exact singular
time-local generator therefore cannot be established analytically. We
instead expect that the reconstruction validated in the RWA remains
applicable to the full spin--boson model. This will be tested
against numerically exact time-evolving matrix product operator (TEMPO)
simulations.

\subsection{Reference and TCL Generators in the SBM}
\label{Sec:baseline}

The reference generator is given by
Eqs.~\eqref{Eq:RWAresumPOD} and~\eqref{Eq:RWAresumCOD}, with the
renormalized transition frequency
\[
\widetilde{\Delta}
=\Delta+\frac{1}{4}\left(S_\Delta-S_{-\Delta}\right).
\]
This expression contains the correct Lamb shift, in contrast to the
well-known incorrect Lamb shift produced by the rotating-wave
approximation~\cite{Fleming2010}. This expression contains the correct Lamb shift, in contrast to the
well-known incorrect Lamb shift produced by the rotating-wave
approximation~\cite{Fleming2010}. In the reference dynamics, the coherence
undergoes a uniform contraction at rate \(J_\Delta/4\) and rotates at
frequency \(\widetilde{\Delta}\), while the reference channel remains
invertible at every finite time.

The TCL generator matrix elements are obtained by specializing the results
of Appendix~\ref{App:TCLSBM} to the unbiased spin--boson model at zero
temperature. The population and coherence dynamics decouple, so the TCL generator reads
\begin{equation}
L(t)=
\begin{pmatrix}
L_{\mathrm{pop}}(t) & 0\\
0 & L_{\mathrm{coh}}(t)
\end{pmatrix}.
\label{Eq:LSBM_block}
\end{equation}

The generator depends on a time--dependent renormalized transition frequency
\begin{equation}
\Delta'(t)=\Delta-\frac{i}{4}\!\left(\Gamma_{\Delta}(t)-\Gamma_{-\Delta}(t)\right).
\label{Eq:Deltaprimeb}
\end{equation}

For times $t\gg\tau_C$ the frequency approaches its Markovian limit
\begin{equation}
\Delta'(\infty)\equiv a
=\Delta-\frac{i}{4}\bigl(\Gamma_\Delta+\Gamma^\ast_{-\Delta}\bigr)
=\tilde{\Delta}-\frac{i}{4}J_\Delta .
\label{Eq:Bohrz}
\end{equation}

In the following, analytical computations are performed within this
Markov approximation for the frequency. The numerical calculations
instead solve Eq.~\eqref{Eq:Cdot}  directly, without this approximation. The
analytical result is the asymptotic limit of the disentangled map obtained
from the direct numerical solution.

For compactness we introduce the notation
\(
\Gamma_\pm(t)\equiv \Gamma_{\pm\Delta'(t)}(t).
\)
The coherence block then takes the form
\begin{align}
L_{\mathrm{coh}}(t)
&=
\begin{pmatrix}
\ell_{21,21}(t) & \ell_{21,12}(t)\\
\ell_{12,21}(t) & \ell_{12,12}(t)
\end{pmatrix},
\label{Eq:LSBM_coh_compact}
\end{align}
with matrix elements
\begin{align}
\ell_{12,12}(t)
&= -i\Delta-\tfrac14\!\left(\Gamma_{+}(t)+\Gamma_{-}^\ast(t)\right),\\
\ell_{21,12}(t)
&= \tfrac14\!\left(\Gamma_{+}(t)+\Gamma_{-}^{*}(t)\right),\label{Eq:LSBM_coh_entries}\\
\ell_{21,21}(t)&=\ell_{12,12}^\ast(t)\\
\ell_{12,21}(t)
&= \ell_{21,12}^{*}(t).
\label{Eq:LSBM_coh_entries1}
\end{align}

The diagonal elements contain the Bohr frequency \(\Delta\) together with
the dissipative combination \(\Gamma_{+}(t)+\Gamma_{-}^\ast(t)\), which
governs the decay of the coherence amplitudes.
By contrast, the off--diagonal elements \(\ell_{21,12}(t)\) and
\(\ell_{12,21}(t)\) generate nonsecular transfer between the two
coherence components \(\rho_{12}\) and \(\rho_{21}\).
The dissipative parts of the secular and nonsecular sectors have equal
magnitude and opposite sign, whereas only the secular rates contains
the unitary splitting \(-i\Delta\).

The phase of the nonsecular coherence channel determines the classical
quadrature selected by the bath. For a transverse coupling
\[
A=\frac{1}{2}\,\vec{\sigma}\cdot\vec{a},
\]
where \(\vec{a}\) lies in the plane orthogonal to the energy axis, this
phase is fixed by the pointer direction \(\vec{a}\). Thus,
\(A=\sigma_x/2\) produces a real nonsecular coupling, whereas
\(A=\sigma_y/2\) produces a purely imaginary one. The diagonal decay rates
are independent of \(\vec{a}\); only the phase of the nonsecular channel
records which quadrature becomes the phase-selected classical remnant.

The population block has an analogous structure,
\begin{align}
L_{\mathrm{pop}}(t)
&=
\begin{pmatrix}
\ell_{11,11}(t) & \ell_{11,22}(t)\\
\ell_{22,11}(t) & \ell_{22,22}(t)
\end{pmatrix},
\label{Eq:LSBM_pop_compact}
\end{align}
with matrix elements
\begin{align}
\ell_{11,11}(t)
&=- \frac{1}{2}\text{Re}\,\Gamma_{+}(t)\\
\ell_{11,22}(t)&=\frac{1}{2}\text{Re}\,\Gamma_{-}(t)\\
\ell_{22,11}(t)
&= -\ell_{11,11}(t)\\
\ell_{22,22}(t)
&= -\ell_{11,22}(t).
\end{align}

\subsection{Reconstructed Map}

Substituting the generator into the disentanglement formula
Eq.~\eqref{Eq:Csolution} yields the matrix elements of the dynamical map.
The intermediate algebra is given in the Appendix~\ref{App:SBM}. 

The coherence sector contains two qualitatively different components. The
secular component propagates each coherence into itself:
\begin{align}
\Phi_{12,12}(t)
&=\Big(1+\frac{\Gamma_\Delta+\Gamma_{-\Delta}}{4}t\Big)
e^{-i\widetilde{\Delta}t-\frac{J_\Delta}{4}t}
\nonumber\\
&\quad
-\frac{1}{4}
e^{-i\widetilde{\Delta}t-\frac{J_\Delta}{4}t}
\int_0^t d\tau\,
\big[
\Gamma_{\Delta'(\tau)}(\tau)
+\Gamma_{-\Delta'(\tau)}^\ast(\tau)
\big],
\label{Eq:diagF1212}
\end{align}
whereas the nonsecular component transfers one coherence into its conjugate:
\begin{align}
\Phi_{21,12}(t)
&=\frac{1}{4}
e^{i\widetilde{\Delta}t-\frac{J_\Delta}{4}t}
\int_0^t d\tau\,
e^{-2i\widetilde{\Delta}\tau}
\nonumber\\
&\quad\times
\big[
\Gamma_{\Delta'(\tau)}(\tau)
+\Gamma_{-\Delta'(\tau)}^\ast(\tau)
\big].
\label{Eq:offDiag}
\end{align}
The secular component contains both the reference semigroup and its memory
correction. The nonsecular coherence transfer, by contrast, is generated
entirely by the bath memory and has no counterpart in the reference
dynamics. The classicalization mechanism is now taking shape.

The other matrix elements are obtained by complex conjugation,
$\Phi_{21,21}=\Phi_{12,12}^\ast$ and
$\Phi_{12,21}=\Phi_{21,12}^\ast$.
The linear in time prefactor in Eq.~\eqref{Eq:diagF1212} is a non--Markovian
term generated by the disentanglement theorem [see Eq.~\eqref{Eq:Phi_constraint}].
It will be canceled by the integral contribution, restoring the van Hove limit.

The population dynamics can be reconstructed using the same
disentanglement procedure as in the coherence sector. The calculation
is algebraically more cumbersome because the population block couples
the two diagonal components of the density matrix, so the main results
are summarized in Subsubsection~\ref{sub:popu}, while the full derivation is
deferred to Appendix~\ref{Sec:PopuDisen}.

\subsection{Diagonal Coherence Dynamics}

The resulting integrals are evaluated in the Markov approximation for
the frequency argument. The diagonal coherence element becomes
\begin{align}
\Phi_{12,12}(t)
&=
\Big(1+\frac{\Gamma_\Delta+\Gamma_{-\Delta}^\ast}{4}t\Big)
e^{-iat}
-\frac14\,\Sigma_a^{\text{sbm}}(t),
\label{Eq:C1212_mi}
\end{align}
where
\begin{equation}
\Sigma_a^{\text{sbm}}(t)
=\Sigma_a(t)
+e^{-J_\Delta t/2}\,[\Sigma_{-a}(t)]^\ast ,
\label{Eq:SigSB}
\end{equation}
with \(\Sigma_a(t)\) defined in Eq.~\eqref{SigmaDe}. The two terms
represent the emission and absorption contributions, respectively.

As in the RWA, the diagonal kernel admits the contour representation
\begin{equation}
\Sigma_a^{\text{sbm}}(t)=  
-\frac{1}{2\pi}\int_{-\infty+i\gamma}^{\infty+i\gamma}
d\omega\,e^{-i\omega t}
\frac{\Gamma_\omega+\big[\Gamma_{-\omega^\ast+iJ_\Delta/2}\big]^\ast}
{(\omega-a)^2}.
\label{WW_SB}
\end{equation}
where the contour offset \(i\gamma\) lies above all singularities of
the integrand.

\subsubsection{Diagonal Wigner--Weisskopf Regime in the SBM}
As in the RWA, to leading order in \(\lambda^2\) the contour integral
yields
\begin{align}
[\Phi(t)]_{12,12}
&=e^{-iat}
\nonumber\\
&+\frac{1}{4\pi}\!\int_0^\infty d\omega
\left[
\frac{e^{-i\omega t}\,\delta J_\omega}{(a-\omega)^2}
+\frac{e^{i\omega t-J_\Delta t/2}\,J_\omega}{(a^\ast+\omega)^2}
\right],
\label{WW_SBM}
\end{align}
where \(\delta J_\omega=J_\omega-J_\Delta\). The pole contribution
together with the on--shell part of the branch cut on
\(\omega\in\mathbb{R}^+\) cancels the linear term
\((\Gamma_\Delta+\Gamma_{-\Delta}^\ast)t/4\) generated by the
disentanglement theorem, leaving the pure exponential decay of the van
Hove limit. The remaining integral is the residual branch--cut
correction.

\subsubsection{Diagonal Khalfin Regime in the SBM}

At long times \(t\to\infty\), the dominant frequencies scale as
\(\omega\sim 1/t\), so the denominators may be expanded in \(\omega\).
The leading term is then proportional to the bath correlation function,
and the resulting Khalfin tail is
\begin{equation}
[\Phi(t)]_{12,12}
=
\frac{1}{4\Delta^2}\Big[C(t)+e^{-J_\Delta t/2}C(t)^\ast\Big].
\label{Eq:Khalfin1212}
\end{equation}

\subsubsection{Pole--plus--tail interference}
Combining the exponential and algebraic contributions yields the
interpolation formula
\begin{equation}
\Phi_{12,12}(t)
\approx
e^{-iat}
+\frac{1}{4\Delta^2}
\Big[C(t)+e^{-J_\Delta t/2}C(t)^\ast\Big],
\label{Eq:F1212Fullt}
\end{equation}
valid for \(t\gg\tau_C\), where \(a\) is defined in
Eq.~\eqref{Eq:Bohrz}. Here we see the disentanglement in action: the
secular coherence separates into an exponentially decaying pole and an
algebraic memory tail, analogous to the survival amplitude in the RWA.

The absorption contribution in Eq.~\eqref{Eq:F1212Fullt} is subleading
relative to the reference exponential \(e^{-iat}\) and therefore has
little effect on the observable dynamics. Keeping only the leading part
of each contribution, one may write
\begin{equation}
\Phi_{12,12}(t)
\approx
e^{-iat}+\frac{C(t)}{4\Delta^2},
\label{Eq:F1212Fullt1}
\end{equation}
which provides a uniformly accurate interpolation for
\(t\gg\tau_C\). Up to this point, the structure remains
essentially the same as in the RWA model.

\subsection{Off--diagonal coherence dynamics}

The nonsecular component differs qualitatively from the RWA dynamics and
therefore requires separate analysis. It underlies the central results of
this work: anisotropy, emergence of the pointer direction, projective
behavior, loss of invertibility, and classicalization.

Since the off--diagonal map element is \(O(\lambda^2)\),
replacing \(\tilde{\Delta}\) by \(\Delta\) in nonoscillatory prefactors
and denominators changes the result only at \(O(\lambda^4)\). We make
this replacement as convenient in the derivation, while retaining
\(\tilde{\Delta}\) in oscillatory phases and trigonometric functions,
where the Lamb shift contributes to the long-time phase.

In the Markov approximation for
the frequency, Eq.~\eqref{Eq:offDiag} can be rewritten as
\begin{equation}
\Phi_{21,12}(t)
=
\frac{1}{4}\Big[
Z_a(t)
+
e^{-J_\Delta t/2}[Z_{-a}(t)]^\ast
\Big],
\label{Eq:Zsb}
\end{equation}
where we introduced the offdiagonal kernel
\begin{equation}
Z_{x+iy}(t)=
\int_0^t d\tau\, C(t-\tau)
\frac{\sin(x\tau)}{x}e^{y\tau}.
\label{Eq:Zkerne}
\end{equation}

Taking the Laplace transform and inverting along the Bromwich contour
gives
\begin{equation}
\Phi_{21,12}(t)=
-\frac{1}{8\pi}
\int_{-\infty+i\gamma}^{\infty+i\gamma} d\omega\,
e^{-i\omega t}
\frac{\Gamma_\omega+\big[\Gamma_{-\omega^\ast+iJ_\Delta/2}\big]^\ast}
{(\omega+iJ_\Delta/4)^2-\tilde{\Delta}^2}.
\label{Eq:BromZ}
\end{equation}
The result separates naturally into pole and branch--cut contributions.

\subsubsection{Off-diagonal Wigner--Weisskopf regime in SBM}

\paragraph*{Pole contribution.}
Evaluating the residues at
\(\omega_\pm=\pm\widetilde{\Delta}-iJ_\Delta/4\) yields
\begin{align}
\Phi_{21,12}^{\mathrm{pole}}(t)
&=\frac{i}{8\widetilde{\Delta}}\,e^{-J_\Delta t/4}
\Big[
e^{-i\widetilde{\Delta}t}
\Big(\Gamma_a+\Gamma_{-a}^\ast\Big)
\nonumber\\
&\qquad
-
e^{i\widetilde{\Delta}t}
\Big(\Gamma_{-a^\ast}+\Gamma_{a^\ast}^\ast\Big)
\Big].
\label{Eq:Phi2112_pole_final}
\end{align}
In the weak-coupling limit \(a\to\Delta-i0^+\), this reduces to
\begin{equation}
\Phi_{21,12}^{\mathrm{pole}}(t)
=
-\frac{i}{4}
e^{-J_\Delta t/4}
\frac{\cos(\widetilde{\Delta}t)}{\Delta}
\left(\Gamma_\Delta^\ast+\Gamma_{-\Delta}\right),
\label{Eq:F1221polelimit}
\end{equation}
which traces a one-dimensional trajectory in the complex plane, although
its direction does not yet coincide with the pointer axis.

\paragraph*{Branch--cut contribution.}

The emission and absorption continua give
\begin{align}
\Phi_{21,12}^{\rm cut}(t)
&=
-\frac{1}{4\pi}\!\int_{0}^{\infty}\! d\omega\,
\frac{e^{-i\omega t}\,J_\omega}{(\omega+iJ_\Delta/4)^2-\tilde{\Delta}^2}
\nonumber\\
&\quad
-\frac{e^{-J_\Delta t/2}}{4\pi}\!\int_{0}^{\infty}\! d\omega\,
\frac{e^{i\omega t}\,J_\omega}{(\omega+iJ_\Delta/4)^2-\tilde{\Delta}^2}.
\label{Eq:Phi2112_cut_total}
\end{align}

Unlike the diagonal sector, the resolvent has only
first-order poles. The on-shell and off-shell contributions
can nevertheless be determined by contour deformation, as
shown in Subsubsection~\ref{sec:BCZkerenl} of
Appendix~\ref{App:SBM}. One finds the on-shell contribution
\begin{equation}
\Phi_{21,12}^{\rm cut,em}(t)
=
\frac{iJ_\Delta}{4\Delta}
e^{-J_\Delta t/4}e^{-i\tilde{\Delta} t}
\label{Eq:Phi2112_cut_intermediate}
\end{equation}
valid to \(O(\lambda^2)\).
The absorption integral has no pole in the upper half--plane and
therefore produces only subleading off--resonant contributions.

Adding Eq.~\eqref{Eq:Phi2112_cut_intermediate} to the pole term
\eqref{Eq:F1221polelimit} yields the pole plus on--shell contribution
\begin{equation}
\Phi_{21,12}^{\text{pole+shell}}(t)
=
e^{-J_\Delta t/4}X(t),
\label{Eq:Phi2112_combined}
\end{equation}
where $X(t)$ is a {\underline{\it real}} function
\begin{equation}
\begin{split}
X(t)
&=\frac{1}{4\Delta}
\left[J_\Delta\sin(\tilde{\Delta} t)
-(S_\Delta-S_{-\Delta})\cos(\tilde{\Delta} t)\right]\\
&=\frac{1}{4}\frac{\vert\Gamma_\Delta+\Gamma^\ast_{-\Delta}\vert}{\Delta}
\,\sin(\tilde{\Delta} t-\theta)\\
&=X_0\,\sin(\tilde{\Delta} t-\theta).
\end{split}
\label{Eq:Xoft}
\end{equation}
The phase shift is determined by the ratio of the Lamb shift to the
spectral density:
\begin{equation}
\theta
=
\tan^{-1}\!\left(
\frac{S_\Delta-S_{-\Delta}}{J_\Delta}
\right).
\label{Eq:thetaA}
\end{equation}
This phase represents the initial slippage.

Thus, the trajectory of
\(\Phi_{21,12}^{\mathrm{pole+shell}}\) lies on the real axis of the
complex plane, i.e., along the pointer direction. This early-time behavior
coincides with the dynamics generated by the Bloch--Redfield master
equation with time-dependent coefficients.

\begin{remark}
\normalfont
[Memory of slippage and anisotropy in the exponentially relaxing regime]

It can be shown, relatively straightforwardly, that the traditional Redfield
equation, which corresponds to TCL2 with asymptotic coefficients, does not
generate a dynamical map satisfying Eq.~\eqref{Eq:Phi2112_combined}, where
the direction along the real axis and the phase shift $\theta$ encode the pointer axis and the initial
slippage of the dynamics, respectively. These are characteristics retained
by the asymptoyic solution of the Redfield equation with time-dependent coefficients.
The solution of the traditional Redfield equation instead has a zero phase
shift, but a rotation of the anisotropy direction by angle $\theta$ relative
to the pointer axis.

Thus, purely exponential relaxation observed in the disentangled map does
not imply memorylessness,
\begin{equation}
    \Phi_{\mathrm{rec}}(t) \neq \Phi_{\mathrm{Markov}}(t),
\end{equation}
where $\Phi_{\mathrm{rec}}(t)$ denotes the reconstructed dynamical map in the asymptotic limit and
$\Phi_{\mathrm{Markov}}(t)$ the dynamical map generated by the corresponding
TCL2  master equation with asymptotic coefficients. Replacing the time-dependent generator by its
asymptotic coefficients does something more drastic than merely approximating
the subsequent decay rate by pure exponentials: it erases the accumulated transformation produced
during the transient, including both the phase shift and the selected
quadrature of the dynamics.
\end{remark}

\subsubsection{Off-Diagonal Khalfin Tail in the SBM}

The remaining off--shell continuum contribution to the disentangled map is evaluated in
Subsection~\ref{App:SBM2112} of Appendix~\ref{App:SBM}. Adding this contribution to the exponential
part in Eq.~\eqref{Eq:Phi2112_combined} gives
\begin{equation}
\Phi_{21,12}(t)
=
e^{-J_\Delta t/4}X(t)
+
\frac{C(t)}{4\Delta^{2}}.
\label{Eq:Tail2112}
\end{equation}
This expression shares the same asymptotic tail as the diagonal
coherence element in Eq.~\eqref{Eq:F1212Fullt1}.

This reveals a sharp contrast between the Markovian
regime and the late-time Khalfin-tail regime. In the Markovian regime, the nonsecular map element is
suppressed by \(O(\lambda^2)\) relative to the secular term. This
suppression originates from the oscillatory structure of the interaction-picture
generator: averaging over times of order \(T_2\) reduces the nonsecular
contribution and yields the asymmetry characteristic of the Davies
semigroup.

In the Khalfin-tail regime, by contrast, this oscillatory
averaging no longer controls the dynamics. The exponential contribution has already decayed, and the continuum tail dominates both secular and nonsecular elements.
As a result, the secular and nonsecular map elements acquire the same asymptotic tail, and the Markovian asymmetry is lost. 

\subsection{Population dynamics}
\label{sub:popu}

The population dynamics can be reconstructed using the same
disentanglement procedure as in the coherence sector. The derivation is
more cumbersome and is given in Sec.~\ref{Sec:PopuDisen} of
Appendix~\ref{App:SBM}.

The final result simplifies considerably. After cancellation of the
disentanglement-induced linear term, the population map takes the affine
form
\begin{equation}
\Phi_{\mathrm{pop}}(t)
=
\begin{pmatrix}
e^{-J_\Delta t/2} & 0 \\
1-e^{-J_\Delta t/2} & 1
\end{pmatrix}
+
B_\infty
\begin{pmatrix}
1 & 1 \\
-1 & -1
\end{pmatrix},
\label{Eq:PopMapMain}
\end{equation}
to leading order in \(\lambda^2\), \emph{with no contribution from the
Khalfin tail}.

The stationary population is determined by
\begin{equation}
B_\infty
=
-\frac{1}{4}\,
\partial_\omega S_\omega\big|_{-\Delta},
\end{equation}
which is an \(O(\lambda^2)\) correction and coincides with the excited-state
population of the mean-force Gibbs state.

The absence of a Khalfin tail in Eq.~\eqref{Eq:PopMapMain} is remarkable.
Although the same algebra used to derive the coherence dynamics is applied
here, no branch-cut discontinuity survives in the population sector. The
populations in the full spin--boson model relax purely exponentially.

\section{Disentangled Dynamical Map\\ for the SBM~\label{Sec:Disent}}

We now assemble the ingredients derived above. The reduced dynamics
contains three distinct contributions: (i)~Markovian relaxation and
decoherence, described by the Davies semigroup \(e^{L_0 t}\);
(ii)~an exponentially damped nonsecular coherence term, given by
Eq.~\eqref{Eq:Phi2112_combined}; and (iii)~the algebraic Khalfin tail,
which follows from Eqs.~\eqref{Eq:F1212Fullt1} and~\eqref{Eq:Tail2112},
\begin{equation}
[\Phi(t)]_{\mathrm{coh,Khal}}
=
\frac{|C(t)|}{4\Delta^2}
\begin{pmatrix}
e^{-i\phi_c} & e^{i\phi_c} \\
e^{-i\phi_c} & e^{i\phi_c}
\end{pmatrix},
\end{equation}
where $\phi_c=-\pi(s+1)/2$ is the phase of the bath correlation function.

Together they give the weak--coupling spin--boson dynamical map at zero
temperature. \emph{We have thus obtained a quantum dynamical map in which the distinct physical mechanisms are explicitly separated:}
\begin{empheq}[box=\fbox]{equation}
\Phi(t)=
\Phi_{\mathrm{GAD}}(t)
+
e^{-J_\Delta t/4}X(t)\,\mathcal T
+
\Phi_{\text{Khal}}(t).
\label{Eq:MainMap}
\end{empheq}

Equation~\eqref{Eq:MainMap} gives the asymptotic form, valid for
\(t\gg 1/\omega_c\), of the solution
\[
\Phi_{\mathrm{td}}(t)=e^{L_0t}+C(t)
\]
of Eq.~\eqref{Eq:Cdot}. We refer to both \(\Phi_{\mathrm{td}}(t)\) and its
asymptotic form \(\Phi(t)\) as the disentangled map, since both follow from
the Feynman disentangling theorem. Only in the asymptotic regime defined by $t\gg 1/\omega_c$, however,
does the map acquire the simple decomposition in
Eq.~\eqref{Eq:MainMap}, which separates the distinct physical mechanisms.
Quantitative benchmarking against exact reduced dynamics uses the full
time-dependent map \(\Phi_{\mathrm{td}}(t)\), whereas
Eq.~\eqref{Eq:MainMap} exposes the asymptotic building blocks of the
dynamics.

$\Phi_{\mathrm{GAD}}(t)$ is the generalized amplitude--damping
channel with relaxation parameter
\begin{equation}
\gamma\equiv \gamma(t)=1-e^{-J_\Delta t/2},
\end{equation}
excitation probability
\begin{equation}
p=B_\infty,
\end{equation} and unitary phase rotation 
at frequency $\tilde\Delta$. Explicitly, 
\[
\Phi_{\text{GAD}}(t)\rho=
\begin{pmatrix}
(1-\gamma)\rho_{11}+\gamma p
&
\sqrt{1-\gamma}\,e^{-i\tilde\Delta t}\rho_{12}
\\[4pt]
\sqrt{1-\gamma}\,e^{+i\tilde\Delta t}\rho_{21}
&
(1-\gamma)\rho_{22}+\gamma(1-p)
\end{pmatrix}.
\]
Operationally, in the vectorization
$[\rho_{11},\rho_{22},\rho_{21},\rho_{12}]^{\mathsf T}$,
$\Phi_{\text{GAD}}(t)=$
\begin{equation}
\begin{pmatrix}
1-\gamma(1-p) & \gamma p & 0 & 0\\
\gamma(1-p) & 1-\gamma p & 0 & 0\\
0 & 0 & \sqrt{1-\gamma}\,e^{+i\tilde\Delta t} & 0\\
0 & 0 & 0 & \sqrt{1-\gamma}\,e^{-i\tilde\Delta t}
\end{pmatrix}.
\end{equation}

The operator $\mathcal T$ acts exclusively in the coherence sector,
where it implements transposition and generates the real, anisotropic
nonsecular correction. In the vectorization order
$[\rho_{11},\rho_{22},\rho_{21},\rho_{12}]^{\mathsf T}$ used here, its
matrix representation is
\[
\tau=0_2\oplus\sigma_x.
\]

The last term $\Phi_{\text{Khal}}(t)$ represents the Khalfin contribution to the dynamical map,
which we refer to simply as the \emph{Khalfin map}.  In the
vectorization
$[\rho_{11},\rho_{22},\rho_{21},\rho_{12}]^{\mathsf T}$,
define the coherence-sector projector
\begin{equation}
\Pi_{\mathrm{coh}}
=
\operatorname{diag}(0,0,1,1).
\end{equation}
The Khalfin map can then be written as
\begin{equation}
\Phi_{\mathrm{Khal}}(t)
=
\frac{|C(t)|}{2\Delta^2}\,
\Pi_{\mathrm{coh}}\mathcal P_c\Pi_{\mathrm{coh}},
\end{equation}
where
\begin{equation}
\mathcal P_c(\rho)
=
\sum_{\pm}P_\pm U_c^\dagger\rho U_cP_\pm.
\end{equation}
The projector on the right restricts the input to the coherence
sector, while the projector on the left restricts the output to the
same sector. Thus, only the input coherences contribute to the
Khalfin effect, and the resulting contribution has support exclusively
in the output coherence sector. The Khalfin contribution does not affect the population dynamics.

Here
\begin{equation}
P_\pm=\tfrac12(I\pm\sigma_x),
\qquad
U_c=e^{-i\phi_c\sigma_z/2},
\qquad
\phi_c=-\frac{\pi}{2}(s+1).
\end{equation}
Thus, the Khalfin map is a rotation of the state set by the correlation-function phase $\phi_c$, followed by projection onto the pointer axis selected by the interaction Hamiltonian.
The projection is symmetry selective and thereby breaks rotational
covariance about the $z$ axis.

For an initial condition along $\sigma_x$, the asymptotic algebraic time dependence of the coherence determined by the map in Eq.~\ref{Eq:MainMap} agrees with that of Leggett et al~\cite{Leggett1987}.
The Khalfin map in Eq.~\ref{Eq:MainMap} reproduces the coherence map of the unbiased spin--boson model implicit in the Born approximation of DiVincenzo and Loss~\cite{DiVincenzoLoss2005}. However, the absence of a Khalfin tail in the populations in Eq.~\ref{Eq:MainMap} differs from their Born-approximation result. Conversely, the population map in Eq.~\ref{Eq:MainMap} agrees with that obtained in the time-dependent renormalization-group derivation of Lindner and Schoeller~\cite{LindnerSchoeller2018}, while the coherence map differs from their result through the absence of the exponentially decaying prefactor found in Ref.~\cite{LindnerSchoeller2018}.  None of these long-time results are accessible within standard second-order Redfield or Lindblad master equations. The central point of the present construction is that the generator~\eqref{Eq:resumationGen} contains asymptotic information that becomes accessible only after the nonperturbative reconstruction and disentanglement developed here.

In summary, equation~\eqref{Eq:MainMap} is the central result of this work. It provides an explicit quantum dynamical map for the unbiased spin–boson model in the weak–coupling regime, separating the contributions of relaxation, nonsecular coherence mixing, and long–time projection driven by bath memory.
It is completely positive and trace preserving up to corrections of
order $O(\lambda^2)$.
We now examine its consequences.

\subsection{Finite-time loss of invertibility\label{Sec:FTLI}}

The three contributions in Eq.~\eqref{Eq:MainMap}—Markovian relaxation,
nonsecular coherence mixing, and the Khalfin tail—play distinct roles
whose combined action drives the smallest singular value of the
dynamical map to zero at a finite time.

Let $t_P$ denote the smallest time at which
\[
|\Phi_{12,12}(t_P)| = |\Phi_{12,21}(t_P)|.
\]
At this time the dynamical map loses invertibility.

This follows directly from the singular--value decomposition of the
coherence block,
\[
\begin{pmatrix}
\Phi_{21,21}(t) & \Phi_{21,12}(t)\\
\Phi_{12,21}(t) & \Phi_{12,12}(t)
\end{pmatrix},
\]
whose singular values are
\begin{equation}
\sigma_\pm(t)
=
|\,|\Phi_{12,12}(t)| \pm |\Phi_{12,21}(t)|\,|.
\end{equation}
Thus $\sigma_-(t_P)=0$, signaling loss of invertibility.

The matrix elements follow directly from the map~\eqref{Eq:MainMap},
\begin{equation}
\label{Eq:Ind}
\begin{aligned}
\Phi_{12,12}(t)
&=
e^{-J_{\Delta}t/4-i\tilde{\Delta}t}
+
\frac{C(t)}{4\Delta^2},
\\
\Phi_{12,21}(t)
&=
\frac{|\Gamma_\Delta+\Gamma^\ast_{-\Delta}|}{4\Delta}
e^{-J_{\Delta}t/4}
\sin(\tilde{\Delta}t-\theta)
+
\frac{C(t)}{4\Delta^2}.
\end{aligned}
\end{equation}

Both terms exhibit interference between an exponentially
damped contribution and the bath correlation function.
When their amplitudes become comparable, destructive
interference suppresses the amplitude, though incomplete
phase matching prevents exact cancellation in the individual
terms, Eqs.~\eqref{Eq:Ind}, like in the RWA.

\begin{figure}[t]
\centering
\includegraphics[width=\columnwidth]{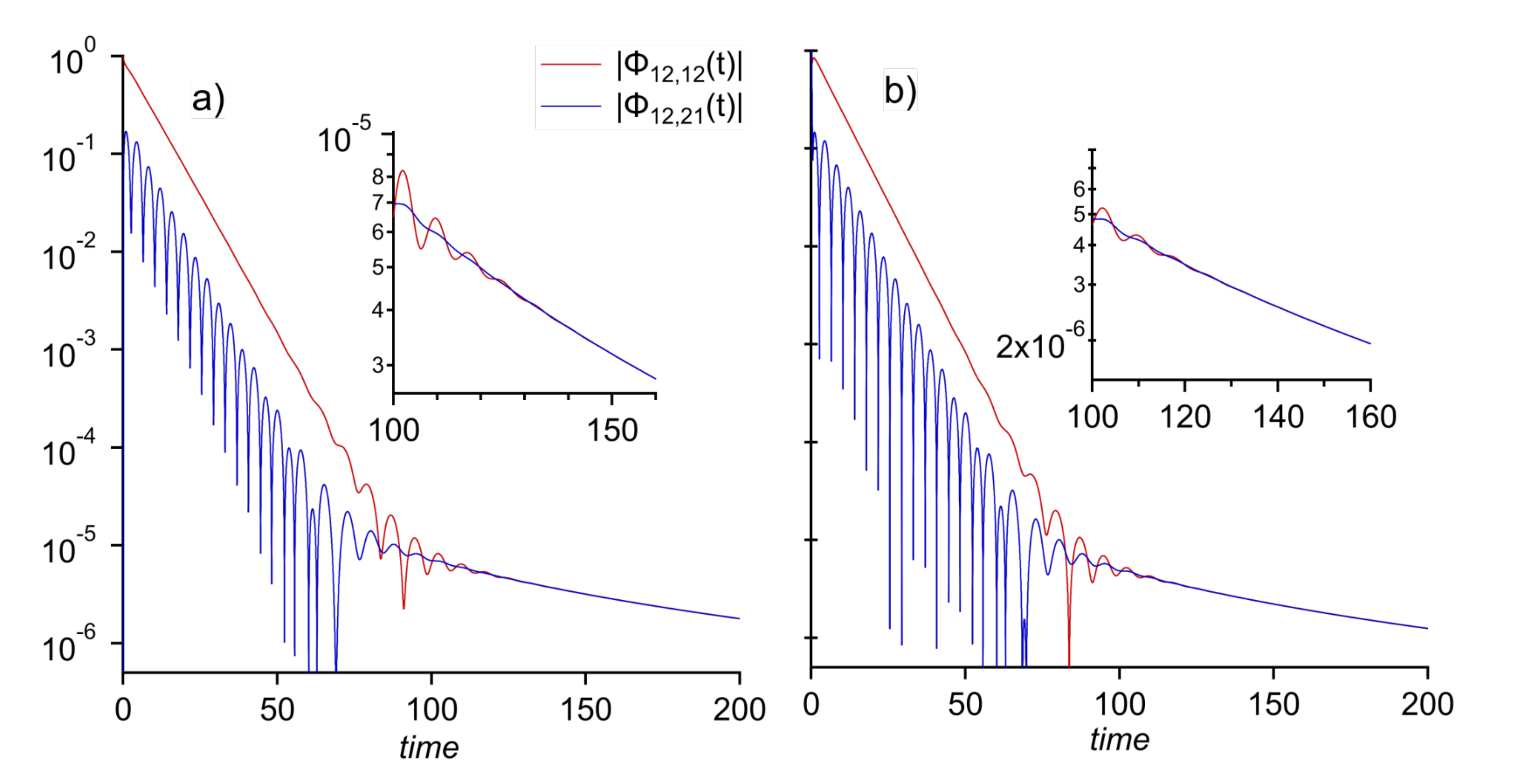}
\caption{
Magnitudes of the secular (red) and nonsecular (blue) matrix elements of
the coherence block. At early times, both decay exponentially, with their
ratio scaling as \(O(\lambda^2)\). In the Khalfin regime, they approach the
same algebraic tail determined by the bath correlation function. Their
pole--tail interference produces isolated crossings at which
\(\sigma_-(t)\), the difference between the two curves, vanishes and the
dynamical map becomes noninvertible. The map regains invertibility between successive crossings.  By contrast, conventional second order master equations exhibit no algebraic tails.  Panel~(a) shows the solution of
Eq.~\eqref{Eq:Cdot}, in the asymptotic limit of the frequency, and panel~(b) is the asymptotic map in
Eq.~\eqref{Eq:MainMap}. The vertical spikes on the logarithmic scale mark
zeros of the matrix elements. The insets
magnify the Khalfin regime, showing the persistence of the exponential decay in the
difference between the secular and nonsecular matrix elements. 
Parameters:
\(s=1\), \(\lambda^2=0.1\), \(\omega_c=10\), \(\Delta=1\), and \(T=0\).
}
\label{Fig:SVD}
\end{figure}

The loss of invertibility ultimately arises because the secular and nonsecular components of the map share identical Khalfin tails but exhibit parametrically different Markovian components. The nonsecular component is \(O(\lambda^2)\) and  enters the interference regime earlier. By the time the diagonal component reaches this regime, the off-diagonal contribution is deep in the Khalfin tail. Destructive interference at the later time then suppresses the diagonal amplitude below the common Khalfin tail, forcing the two contributions to cross. At this crossing, a singular value vanishes and the dynamical map loses invertibility.

Thus, the crossing follows from three ingredients: exponential Markovian
decay; an algebraic Khalfin tail, which can interfere destructively with
the exponential contribution; and the appearance of the same correlation
function in the secular and nonsecular components. The first two were
established independently of the present work by the Davies theory of
Markovian decay and the Khalfin--Peres theory of late-time decay. The
present derivation establishes the third. Once these ingredients are
combined, the origin of the crossing becomes transparent without requiring
the reader to follow every step of the derivation.

In the Khalfin regime, the deviations of the coherence-block matrix elements in Eq.~\eqref{Eq:MainMap} from their common bath-correlation tail remain exponentially suppressed. Consequently, in a coarse-grained sense, the smallest singular value of the coherence block continues to follow a Markovian exponential decay. The algebraic persistence of the individual coherences therefore does not impede the bath's ability to exponentially suppress the initial quantum states.

\section{How accurately can we benchmark every element of a dynamical
map over many orders of magnitude?
\label{Sec:RefBullS}}

Benchmarking the dynamical map~\eqref{Eq:MainMap} over its full time domain requires two complementary tests. At early times, finite-order TCL dynamics provides a controlled perturbative reference. At the long time scales of the Khalfin transition, however, the finite-order TCL dynamics no longer captures the algebraic tail. Numerically exact methods, such as the time-evolving matrix product operator (TEMPO) method~\cite{Strathearn2018} and related approaches~\cite{lacroix2026tensornetworkmethodsnonperturbative}, then provide a potential independent reference, although convergence becomes increasingly difficult in this regime.

We therefore implement two distinct validation tests. First, since we demonstrated in Ref.~\cite{Lampert2025} that the solutions of the master equation~\eqref{Eq:TCLmeq} are more accurate than TCL2 at small time scales relative to $t_P$, we test whether this improvement carries over to the map~\eqref{Eq:MainMap}.

\subsection{Early-Time Validation of the Dynamical Map}
For the parameters considered here, the finite-order TCL expansion was
analyzed in detail in Ref.~\cite{Lampert2025}. Asymptotic stability is first
lost at TCL6, the critical perturbative order for the Ohmic bath, while
secular growth appears at TCL8 and beyond. Thus, TCL4 incorporates the leading finite-order
correction beyond TCL2 while remaining below the critical order at which
asymptotic stability is lost.

We therefore compare the TCL2- and TCL4-generated maps with the disentangled map of  Eq.~\eqref{Eq:MainMap}
 For TCL4, we use the canonical generator introduced
by Crowder et al.~\cite{Crowder}. Using the TCL4 map as a reference, we compute the operator-norm distances of both the TCL2 and disentangled maps from the TCL4 map.

\begin{figure}[t]
\centering
\includegraphics[width=1\linewidth]{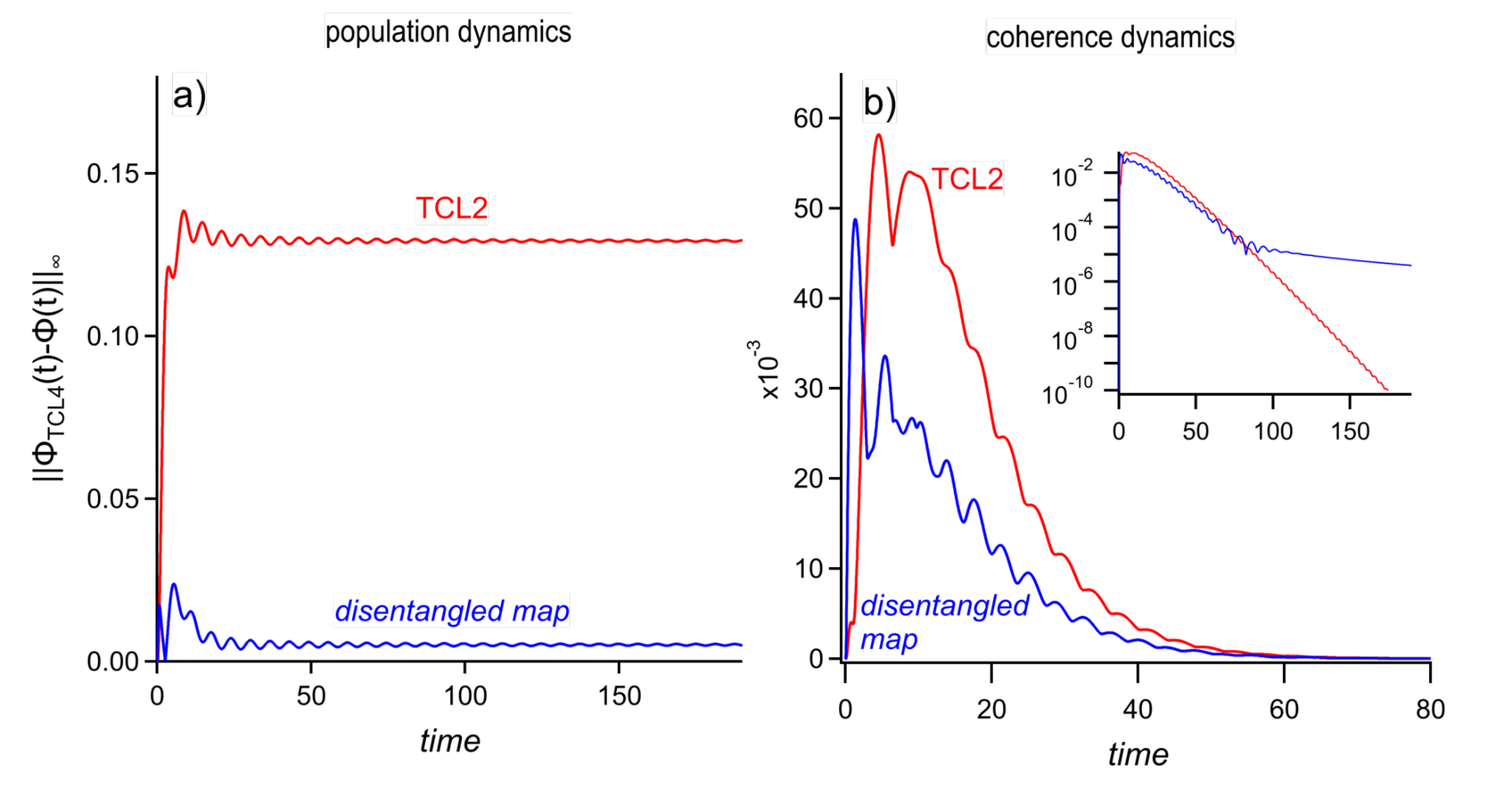}
\caption{
Operator-norm distances
\(\|\Phi_X^{(P,C)}(t)-\Phi_{\mathrm{TCL4}}^{(P,C)}(t)\|_\infty\)
for \(\Delta=1\), \(\omega_c=10\), \(s=1\), \(\lambda^2=0.1\), and
\(T=0\), where \(X\) denotes the TCL2 or disentangled map and
\(\|\cdot\|_\infty\) is the largest singular value. (a) Population block.
(b) Coherence block. The inset shows the coherence-block distances on a
logarithmic scale at longer times.
}
\label{Fig:TCL4benchmark}
\end{figure}

In the unbiased spin--boson model, the population and coherence sectors
decouple~\cite{kaplanek2020hot}, allowing their accuracies to be assessed
independently. Figures~\ref{Fig:TCL4benchmark}(a) and
\ref{Fig:TCL4benchmark}(b) show their respective distances from the TCL4
reference.

For the population block, the disentangled map is substantially closer to
TCL4 than the TCL2 map throughout the interval shown. This is consistent
with the improvement from \(O(\lambda^0)\) to \(O(\lambda^2)\) expected theoretically~\cite{Fleming} and established
for the partially resummed master-equation~\eqref{Eq:TCLmeq} in Ref.~\cite{Lampert2025}. The
known improvement therefore carries over directly to the population block
of the reconstructed map.

The coherence block accuracy also improves, as seen in Fig.~\ref{Fig:TCL4benchmark}(b) although much more modestly. Over much of the
early-time regime, the disentangled map reduces the distance from TCL4 by
approximately a factor of two relative to the TCL2-generated map. 

Near the Khalfin time scale, however, the ordering of the coherence-block distances
reverses: the disentangled map becomes farther from TCL4 than does TCL2.
This does not indicate a deterioration of the disentangled map, because the Khalfin
tail is absent from finite-order TCL dynamics and requires resummation.
TCL4 therefore ceases to be an appropriate reference in this regime.

\subsection{Khalfin Effect and Distinguishability of Initial States}

We now compare the Khalfin regime of Eq.~\eqref{Eq:MainMap} with that of TEMPO, a numerically exact method in the limit of
convergence. A point-by-point quantitative comparison is difficult because the numerical value of the Khalfin time ($t_P$) is unusually sensitive to convergence parameters. For example, we find that a relatively small change in a parameter, such as
$\omega_c$, can produce a disproportionately large shift in the Khalfin
transition time $t_P$  in TEMPO, making a point-by-point quantitative comparison
especially demanding. This sensitivity ultimately reflects the structure of the
Feynman--Vernon influence functional: the Khalfin contribution enters
it as a logarithmic correction to the dominant contribution responsible
for the exponential decay. Consequently, even a small relative error in the
influence functional and the singular value precision cut-off can accumulate over the very long time scale of
the Khalfin crossover and produce a substantial shift in $t_P$.

A qualitative comparison, however, remains possible. Rather than testing the precise location of the transition, we test the two characteristic features predicted by the map~\eqref{Eq:MainMap}: the emergence of a common algebraic tail in the coherence-block matrix elements and the exponential suppression of their difference.

\begin{figure}[t]
\centering
\includegraphics[width=0.95\linewidth]{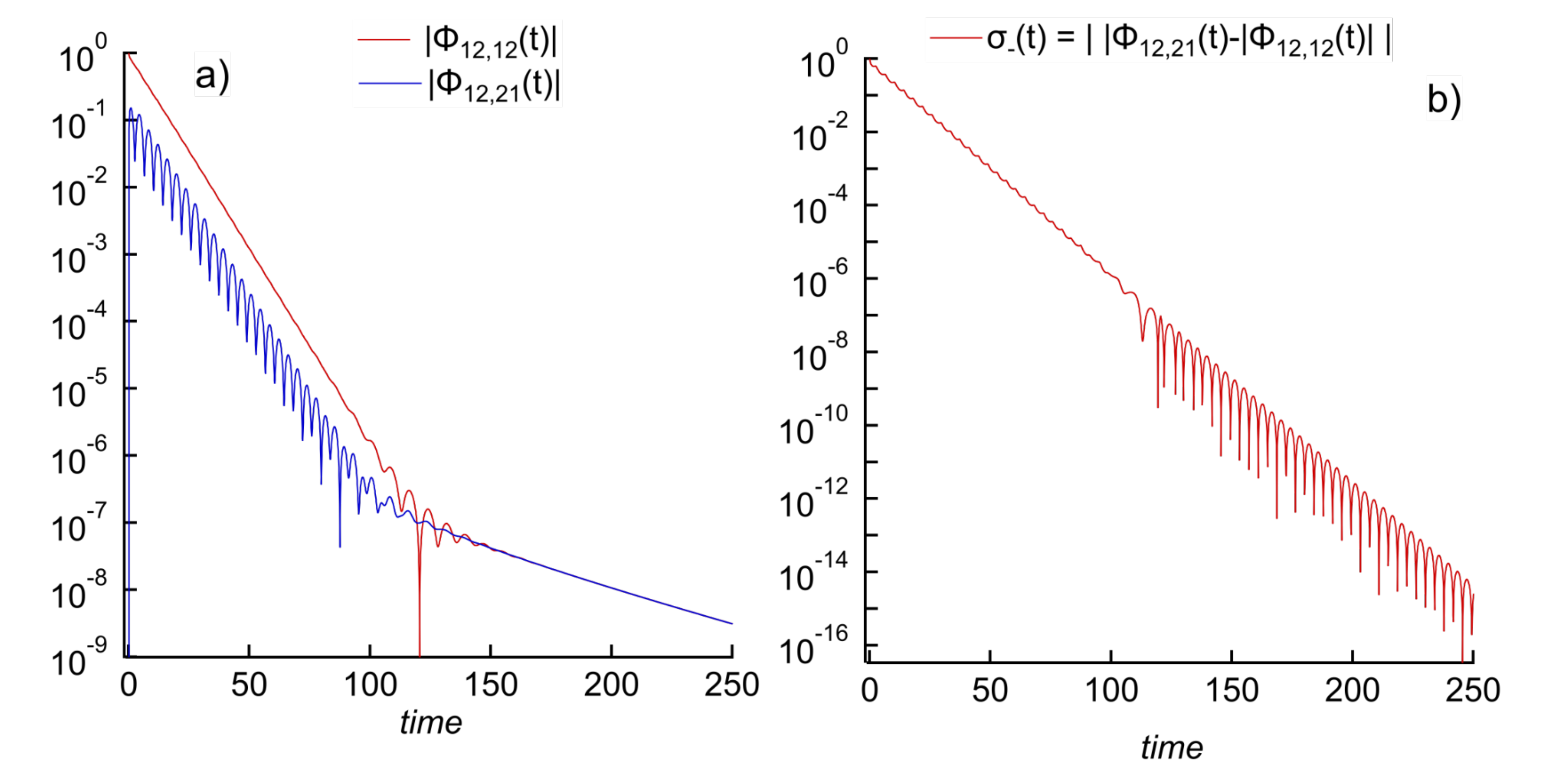}
\caption{
\textbf{Classical coherence without distinguishability of the initial coherence emerges as non-Markovian memory.}
TEMPO results for the coherence block of the dynamical map.
(a) Magnitudes of the secular and nonsecular elements,
\(\lvert\Phi_{12,12}(t)\rvert\) (red) and
\(\lvert\Phi_{12,21}(t)\rvert\) (blue). Both undergo the Khalfin transition at
\(t\gtrsim t_P\), where exponential relaxation gives way to an algebraic
tail. The two coherences then approach the same long-time behavior.
(b) Their difference,
\(\sigma_-(t)=
\bigl|
\lvert\Phi_{12,21}(t)\rvert-
\lvert\Phi_{12,12}(t)\rvert
\bigr|\),
which measures the surviving distinguishability of the
initial superpositions. Although each coherence decays only algebraically
beyond the Khalfin transition, their distinguishability continues to
disappear, repeatedly collapsing to zero and partially recovering with an exponentially
decreasing envelope. The Khalfin tail therefore preserves coherence
without preserving the initial quantum information encoded in it.
}
\label{fig:classicalization}
\end{figure}

 Figure~\ref{fig:classicalization} reproduces with TEMPO the
Khalfin regime, repeated losses and partial recoveries of distinguishability,
and the asymptotic merging of the secular and nonsecular coherences predicted
by the disentangled map in Fig.~\ref{Fig:SVD}. The qualitative agreement between Figs.~\ref{Fig:SVD}
and~\ref{fig:classicalization} therefore validates the reconstruction procedure developed in this work.

Figure~\ref{fig:classicalization}(b) resolves the smaller singular-value dynamics
beyond the transition in TEMPO. We quantify the remaining quantum information by the
smallest singular value \(\sigma_-(t)\) of the coherence block, since
whenever \(\sigma_-(t)=0\), the map becomes noninvertible and
maps two initially distinct superpositions onto the same reduced state.
We also define the relative remaining quantum information as
\(\mathcal P(t)=\sigma_-(t)/\sigma_+(t)\).
At \(t\ll t_P\), \(\mathcal P\approx 1+O(\lambda^2)\), so there is little loss of relative distinguishability at this stage of the evolution.

Information backflow, however, begins well before $t_P$, as manifested by the oscillations of $\sigma_-(t)$ around its exponentially decaying envelope. The Khalfin transition is therefore
not the onset of backflow, but the separatrix at which algebraic memory
becomes dominant in each coherence. Beyond $t_P$, backflow persists while $\sigma_-(t)$
remains exponentially suppressed in a coarse-grained sense. This differs fundamentally from the
rotating-wave approximation, in which the two singular values decay algebraically beyond the Khalfin transition. In the
full spin--boson model, the Khalfin transition thus acquires a new role: it
marks the onset of classicalization: One singular direction becomes exponentially suppressed relative to the other, so that $\mathcal P(t)$ decays exponentially and the coherence block becomes asymptotically rank one.

Beyond $t_P$, the surviving algebraic correlations select the classical quadrature
determined by the phase of the bath correlation function. The disentangled
map makes this separation explicit: the exponential sector suppresses
initial quantum information, while the algebraic Khalfin sector projects onto the
phase-fixed classical coherence.

\section{Conclusion}

\label{Sec:Discussion}

The disentangled map developed here provides a regular and physically transparent description of the spin--boson dynamics through and beyond the singularity of the time-local generator. Its accuracy at intermediate times follows the improvement previously established for the resummed master equation and is confirmed here at the level of the dynamical map, while numerically exact TEMPO results independently validate the quantum-information structure of the Khalfin effect.

The principal physical result is a separation between
memory, quantum information, and classicalization. Algebraically long-lived correlations and information backflow coexist with exponentially vanishing quantum information. The surviving correlations do not protect initial quantum states from decoherence, but instead select the classical quadrature that remains. The simplicity of the disentangled map makes this separation explicit, revealing a physical structure that is obscured in the Feynman--Vernon functional.

This result opens two broader questions. Could primordial field fluctuations in the inflationary universe similarly survive while the information they carry changes from quantum to classical, with the Khalfin transition selecting the classical inhomogeneities that seed the subsequent nonuniform expansion? And, more generally, into what elementary dynamical structures can non-Markovian dynamics be resolved? These questions provide natural directions for future work.

\section*{Acknowledgments}

The author thanks Jiahao Chen for valuable discussions and for extending the resummed TCL generator to the multi-bath regime, and Sirui Chen for helpful discussions. This work
was supported by the David and Lucile Packard Foundation and the School of Physics at the Georgia Institute of Technology
through a seed grant. The author remains fully responsible for the content.

\appendix
\section{Partially Resummed TCL Generator for the Spin--Boson Model}
\label{App:TCLSBM}

This appendix provides an explicit representation of the partially
resummed TCL generator to clarify the structure of
Eqs.~\eqref{Eq:renogen} and \eqref{Eq:renofreqs}, whose physical content
may be obscured by the density of indices in the main text. The
expressions given here are intended both as a guide to interpretation
and as a reference for practical implementation.

The formulation also serves as a starting point for future extensions
of the present framework to finite temperature and biased spin--boson
models, where spectral overlap play a
significant role in the dynamics.

We write the time-local generator in the eigenbasis of the system Hamiltonian
\(
H_S=\frac{\Delta}{2}\sigma_z,
\)
where indices $1,2$ denote the eigenstates of $\sigma_z$.
The system--bath coupling operator is written in this basis as
$A_{nm}=\langle n|A|m\rangle$. We assume $A_{11}=-A_{22}$.

The bath enters through the correlation function
\begin{equation}
C(t)=\langle B(t)B(0)\rangle_\beta,
\end{equation}
and its one-sided Fourier transform
\begin{equation}
\Gamma_{\omega}(t)=\int_0^t d\tau\, e^{i\omega\tau} C(\tau).
\end{equation}
 All correlation functions are evaluated at finite temperature.
We also define the zero-frequency component
\begin{equation}
J_0(t)=\int_0^t d\tau\, C(\tau).
\end{equation}

Excluding the spectral overlap, the interaction renormalizes the bare level splitting $\Delta$ into a
time-dependent complex quantity
\begin{equation}
\Delta'(t)=\Delta-i|A_{12}|^2\big[\Gamma_{\Delta}(t)-\Gamma_{-\Delta}(t)\big].
\label{Eq:Deltaprime}
\end{equation}

\paragraph{Role of the spectral-overlap shift.}
The partial resummation introduces a spectral-overlap
shift of the renormalized splitting, such that the effective
frequencies entering the bath response functions acquire
imaginary contributions of the form
$\Delta'(t)\pm 4iJ_0(t)A_{11}^2$.  We therefore define
\begin{equation}
\Omega(t)=4A_{11}^2J_0(t),
\label{Eq:OmegaDef}
\end{equation}
which approaches the asymptotic decay rate of the
off-diagonal density-matrix elements (the transverse
dephasing rate of the Davies generator) in the long-time limit.

The overlap shift enters the population sector with
$+i\Omega(t)$, whereas the coherence block depends on
$-i\Omega(t)$.  This sign asymmetry has a direct dynamical
consequence: the population dynamics is stabilized
(broadening the effective transition frequencies appearing
in the rates), while the coherence sector is destabilized,
producing large or negative effective rates in the
logarithmic time-local generator. 

The generator preserves trace and Hermiticity of the density matrix. Consequently,
its matrix elements satisfy
\begin{equation}
\sum_n L_{nn,ij}(t)=0 ,
\label{Eq:trace}
\end{equation}
and
\begin{equation}
L_{nm,ij}(t)=L_{mn,ji}^\star(t).
\label{Eq:HermiticityConstraint}
\end{equation}
Therefore only an independent subset of components needs to be specified; all
others follow from Eqs.~\eqref{Eq:trace}--\eqref{Eq:HermiticityConstraint}.

\paragraph{Population sector.}
\begin{align}
L_{11,11}(t)
&=-2|A_{12}|^2\,\mathrm{Re}\,\Gamma_{\Delta'(t)+i\Omega(t)}(t),\\
L_{11,22}(t)
&= 2|A_{12}|^2\,\mathrm{Re}\,\Gamma_{-\Delta'(t)+i\Omega(t)}(t).
\end{align}
All remaining population elements follow from trace preservation,
Eq.~\eqref{Eq:trace}.

\paragraph{Population--coherence couplings.}
\begin{align}
L_{21,11}(t)
&=2A_{21}A_{11}\left[\Gamma_{\Delta'(t)+i\Omega(t)}(t)-iS_0(t)\right],\\
L_{21,22}(t)
&=2A_{21}A_{11}\left[-\Gamma^\star_{-\Delta'(t)+i\Omega(t)}(t)-iS_0(t)\right].
\end{align}
All remaining couplings follow from the Hermiticity condition
Eq.~\eqref{Eq:HermiticityConstraint}.

\paragraph{Coherence sector.}
\begin{align}
L_{21,21}(t)
&= i\Delta-\Omega(t)
-|A_{12}|^2\!\left[
\Gamma^\star_{\Delta'(t)-i\Omega(t)}(t)
\right.\nonumber\\
&\qquad\left.
+\Gamma_{-\Delta'(t)-i\Omega(t)}(t)
\right],\\
L_{12,21}(t)
&= |A_{12}|^2\!\left[
\Gamma^\star_{\Delta'(t)-i\Omega(t)}(t)
+\Gamma_{-\Delta'(t)-i\Omega(t)}(t)
\right].
\end{align}

\paragraph{Coherence-to-population transfer.}
The coherence-to-population transfer term is
\begin{equation}
L_{11,21}(t)=2A_{12}A_{11}J_0(t),
\end{equation}
and trace preservation, Eq.~\eqref{Eq:trace}, implies
$L_{22,21}(t)=-L_{11,21}(t)$.
All other coherence elements are obtained from
Eq.~\eqref{Eq:HermiticityConstraint}.

These expressions completely determine the TCL generator for the qubit spin--boson model and reduce, in the zero-temperature limit, to the unbiased case discussed in the main text.

\section{TCL Generator in the RWA
\label{Sec:nesting}}

\noindent\textit{Roadmap.}
Starting from the Volterra iteration of the RWA survival amplitude, we
(i) expand the solution into iterates $f_n$,
(ii) construct the van Kampen cumulants $K_{2n}$ via the recursion
Eq.~\eqref{Eq:recursive},
(iii) isolate the secular structure using the $\Delta$--identity
Eq.~\eqref{Eq:Lampert}, and
(iv) show (to $O(\lambda^6)$) that the resulting cumulant series matches the
nested--frequency form Eq.~\eqref{Eq:nest}.

\subsection{Iterated solution and cumulant recursion}

Integrating both sides of Eq.~\eqref{Eq:IDeq} from $0$ to $t$ and changing the order of integration yields a 
Volterra integral equation of the second kind,  
\begin{equation}
    f'(t) = 1 - \frac{1}{4} \int_{0}^{t} d\tau \, \Gamma_{\omega}(t - \tau)\, f'(\tau),
    \label{Eq:Volt2}
\end{equation}
with the memory kernel $\Gamma_{\omega}(t)$ defined in Eq.~\eqref{Eq:timedSD},
evaluated at $\omega=\Delta$. To reduce notational clutter, throughout this appendix
$f$ is understood to be in the interaction picture (so we omit the prime).
 We use the shorthand for time ordered integrals as
\[
\idotsint_\leftarrow^{t} dt_{1\ldots n}=\int_0^t dt_1\int_0^{t_1}dt_2\cdots\int_0^{t_{n-1}}dt_n.
\]

Let us write down individual terms of the series obtained by iterating
Eq.~\eqref{Eq:Volt2}. 
Defining $f(t)\equiv f_0+f_1+f_2+f_3+\ldots$, we have
\begin{align}
    f_0 &= 1,\\
    f_1 &= -\frac{1}{4}\int_0^t dt_1\,\Gamma_\omega(t-t_1),\\
    f_2 &= \Big(\!-\frac{1}{4}\Big)^{\!2}
    \iint_\leftarrow^t dt_{12}\,
    \Gamma_\omega(t-t_1)\Gamma_\omega(t_1-t_2),\\
    f_3 &= \Big(\!-\frac{1}{4}\Big)^{\!3}
    \iiint_\leftarrow^t dt_{123}\,
    \Gamma_\omega(t-t_1)\Gamma_\omega(t_1-t_2)\Gamma_\omega(t_2-t_3).
\end{align}
For the derivatives we obtain
\begin{align}
    \dot f_0 &= 0,\\
    \dot f_1 &= -\frac{1}{4}\Gamma_\omega(t),\\
    \dot f_2 &= \Big(\!-\frac{1}{4}\Big)^{\!2}
    \int_0^t dt_{1}\,
    \Gamma_\omega(t-t_1)\Gamma_\omega(t_1),\\
    \dot f_3 &= \Big(\!-\frac{1}{4}\Big)^{\!3}
    \iint_\leftarrow^t dt_{12}\,
    \Gamma_\omega(t-t_1)\Gamma_\omega(t_1-t_2)\Gamma_\omega(t_2).
\end{align}

According to the van Kampen cumulant series, the coherence generator admits the
expansion
\begin{align}
    \frac{\dot f}{f}=K_2+K_4+K_6+\ldots\; .
\end{align}
Multiplying by $f$ and collecting equal powers of $\lambda^2$ yields the
recursive relation
\begin{equation}
\label{Eq:recursive}
    K_{2n} = \dot{f}_n - \sum_{j=1}^{n-1} K_{2n-2j}\, f_j ,
\end{equation}
which is a special case of the general recursion for TCL cumulants
\cite{nestmann2019timeconvolutionless}.  The three lowest orders are
\begin{align}
    K_2 &= \dot f_1,\\
    K_4 &= \dot f_2 - f_1\dot f_1,\\
    K_6 &= \dot f_3 - f_1\dot f_2 - f_2\dot f_1 + f_1^2\dot f_1 .
\end{align}
Substituting the explicit iterates gives, for the RWA model,
\begin{align}
    \label{Eq:X2}
    K_2 &= -\frac{1}{4}\Gamma_\omega(t),\\
    K_4 &= \frac{1}{16}\int_0^t dt_1\,
    \big[\Gamma_\omega(t-t_1)-\Gamma_\omega(t)\big]\Gamma_\omega(t_1),\\
    K_6 &= -\frac{1}{4^3}\iint_\leftarrow^t dt_{12}\,
    \Gamma_\omega(t-t_1)\Gamma_\omega(t_1-t_2)\Gamma_\omega(t_2) \notag\\
    &\quad +\frac{1}{4^3}\int_0^t dt_1\,\Gamma_\omega(t_1)
    \int_0^t dt_2\,\Gamma_\omega(t-t_2)\Gamma_\omega(t_2) \notag\\
    &\quad +\frac{1}{4^3}\Gamma_\omega(t)
    \iint_\leftarrow^t dt_{12}\,
    \Gamma_\omega(t-t_1)\Gamma_\omega(t_1-t_2)
    \notag\\
    \label{Eq:K6dd2}
    &\quad -\frac{1}{4^3}\Gamma_\omega(t)
    \bigg[\int_0^t dt_1\,\Gamma_\omega(t_1)\bigg]^2 .
\end{align}

\subsection{Secular structure from the $\Delta$ identity}

Define
\begin{equation}
    D(t,t_1)=\Delta\Gamma_\omega(t,t_1)\equiv
    \Gamma_\omega(t)-\Gamma_\omega(t_1),
    \qquad
    \Gamma\equiv\Gamma_\omega(t).
\end{equation}
This satisfies the identity \cite{Lampert2025}
\begin{equation}
\label{Eq:Lampert}
   \idotsint_\leftarrow^t dt_{1\ldots n}\,
   D(t,t-t_n)
   = \frac{(-i)^n}{n!}\,
   \frac{\partial^n \Gamma_\omega(t)}{\partial \omega^n}.
\end{equation}
Thus the $\Delta$--structure suppresses polynomial secular terms
($\sim t^n$) in the time--ordered integrals; any residual secular growth is
restricted to derivatives of $\Gamma_\omega(t)$, which become dominant for
algebraically decaying $C(t)$.

Using $\Delta$, the fourth--order cumulant becomes
\begin{align}
\begin{split}
    K_4 &=
    \frac{1}{4^2}\int_0^t dt_1\,
    D(t,t-t_1)\big[D(t,t_1)-\Gamma\big]\\
    &=\frac{i}{4^2}\Gamma\frac{\partial\Gamma}{\partial\omega}
      +\frac{1}{4^2}\int_0^t dt_1\,
    D(t,t-t_1)D(t,t_1)\\
    &\equiv \frac{i}{4^2}\Gamma\frac{\partial\Gamma}{\partial\omega}
      +\frac{1}{4^2}H_\omega(t),
\end{split}
\end{align}
where we introduced
\begin{equation}
 H_\omega(t)=\int_0^t dt_1\,D(t,t-t_1)D(t,t_1).
\end{equation}

Next, converting the double integrals in~\ref{Eq:K6dd2} into time--ordered form via
\begin{align}
\nonumber
\int_0^t dt_1 &X(t_1)\int_0^t dt_2 Y(t_2)=\\
&
\nonumber
\iint_{\leftarrow}^{t} dt_{12}\,[X(t_1)Y(t_2)+X(t_2)Y(t_1)],
\end{align}
we obtain

\begin{align}
\nonumber
K_6
&= -\frac{1}{4^3}\!\!\iint_{\leftarrow}^{t} dt_{12} \Big\{\\
\nonumber
  &\big[\Gamma-D(t,t{-}t_1)\big]\big[\Gamma-D(t,t_1{-}t_2)\big]\big[\Gamma-D(t,t_2)\big]\\
  \nonumber
-& \big[\Gamma-D(t,t_1)\big]\big[\Gamma-D(t,t_2)\big]
      \big[2\Gamma-D(t,t{-}t_2)-D(t,t{-}t_1)\big] \\
      \nonumber
-& \Gamma\,\big[\Gamma-D(t,t{-}t_1)\big]\big[\Gamma-D(t,t_1{-}t_2)\big]\\
      +& 2\Gamma\,\big[\Gamma-D(t,t_1)\big]\big[\Gamma-D(t,t_2)\big]
\Big\}\\
=&K_6^{D}+K_6^{D^2}+K_6^{D^3}.
\end{align}

The cumulant structure removes the zeroth order in $\Delta$, so the leading
nonzero contributions are linear in $\Delta$, with $\Delta^2$ and higher powers
subleading.  Using Eq.~\eqref{Eq:Lampert} and
\[
    \iint_\leftarrow^t dt_{12}\,D(t,t-t_1)
    = \iint_\leftarrow^t dt_{12}\,D(t,t_2),
\]
the linear term yields
\begin{align}
\begin{split}
    K_6^D
    &= \frac{1}{4^3}\Gamma^2
        \iint_\leftarrow^t dt_{12}\,D(t,t_2)
     = -\frac{1}{4^3}\frac{1}{2!}\Gamma^2
        \frac{\partial^2\Gamma}{\partial\omega^2}.
\end{split}
\end{align}

The quadratic contribution is
\begin{align}
\begin{split}
    K_6^{D^2}
    &=-\frac{\Gamma}{4^3}\iint_\leftarrow^t dt_{12}\Big\{
      D(t,t_1-t_2)D(t,t_2)
      -D(t,t_1)D(t,t-t_2)\\
    &\qquad\qquad
      -D(t,t_1)D(t,t-t_1)
      -D(t,t_2)D(t,t-t_2)\Big\}.
\end{split}
\label{Eq:C26}
\end{align}
Using
\begin{equation}
\int_0^t d\tau\,\Gamma_\omega(\tau)
=t\Gamma_\omega(t)+i\frac{\partial\Gamma_\omega(t)}{\partial\omega},
\end{equation}
and
\begin{equation}
\int_0^t dt_1\,\Gamma_\omega(t_1)\Gamma_\omega(t-t_1)
=t\Gamma_\omega^2(t)+2i\Gamma_\omega(t)\frac{\partial \Gamma_\omega(t)}{\partial\omega}
+H_\omega(t),
\end{equation}
one finds that the secular ($\sim t$) parts cancel between the two lines of
Eq.~\eqref{Eq:C26}, leaving only subleading contributions.  The leading cubic
term $K_6^{\Delta^3}$ remains finite and we keep it implicit.

Collecting terms gives
\begin{align}
    K_6
    &=\frac{\Gamma}{4^3}\bigg\{
      \bigg[\frac{\partial\Gamma_\omega(t)}{\partial\omega}\bigg]^2
      -i\frac{\partial H_\omega(t)}{\partial\omega}
    \bigg\}
    +K_6^{\Delta^3}.
\end{align}

\subsection{Matching to nested frequency renormalization}

Summing the cumulants to $O(\lambda^6)$ yields
\begin{align}
    K &= K_2+K_4+K_6 \notag\\
      &= -\frac{\Gamma}{4}
         +\frac{i}{4^2}\Gamma\frac{\partial\Gamma}{\partial\omega}
         +\frac{1}{4^2}H_\omega(t)
         -\frac{\Gamma^2}{2!\,4^3}\frac{\partial^2\Gamma}{\partial\omega^2} \notag\\
      &\quad +\frac{\Gamma}{4^3}\bigg[
         \bigg(\frac{\partial\Gamma_\omega(t)}{\partial\omega}\bigg)^2
         -i\frac{\partial H_\omega(t)}{\partial\omega}
      \bigg]
      +K_6^{\Delta^3}.
\end{align}
One may verify that these terms coincide with the Taylor expansion of the
nested--frequency form through $O(\lambda^6)$:
\begin{align}
\label{Eq:nest}
    K
    = -\tfrac{1}{4}\,
      \Gamma_{\omega - \tfrac{i}{4}\Gamma_{\omega - \tfrac{i}{4}\Gamma_{\omega}}}
      + \tfrac{1}{4^2}\,
      H_{\omega - \tfrac{i}{4}\Gamma_{\omega}}
      + K_6^{D^3} + \ldots .
\end{align}

The first term exhibits frequency renormalizations nested to second order.
Truncating Eq.~\eqref{Eq:nest} at order $O(\lambda^2)$ gives
\begin{equation}
K(t)\approx -\frac{1}{4}\,\Gamma_{u(t)}(t),
\qquad
u(t)=\omega-\frac{i}{4}\Gamma_\omega(t),
\end{equation}
which yields the leading resummed quantities in Eqs.~\eqref{eq:renoSELF} and~\eqref{Eq:genU} of the main text.

If the frequency renormalization is iterated indefinitely, as implied by Eq.~\eqref{Eq:nest}, one obtains the self-consistent equation
\begin{equation}
u(t)=\omega-\frac{i}{4}\Gamma_{u(t)}(t),
\label{Eq:iteratedjumps}
\end{equation}
which formally corresponds, in the limit $t\to\infty$, to the pole in the lower half-plane of the exact solution, Eq.~\eqref{Eq:BCintegral}, if such a solution existed. However, due to the multivalued nature of $\Gamma_\omega$, no consistent solution exists in $\mathbb{C}$. Successive iterations lead to discontinuous changes in the frequency, between positive and negative imaginary parts.

\begin{figure}[t]
\centering
\includegraphics[width=0.9\columnwidth]{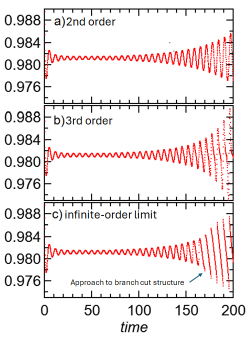}
\caption{
Time evolution of the real part of the renormalized frequency for increasing nesting order of the
frequency renormalization: (a) second order, (b) third order, and (c)
higher-order nesting. As the nesting order increases, late-time
oscillations become progressively enhanced, reflecting the approach to
the branch-cut (Riemann-sheet) structure of the exact solution. The
infinite-order limit corresponds to the nonperturbative continuum
contribution. $s=1/3$, $\lambda^2=0.01$, $\omega_c=4$, $\Delta=1$.}
\label{Fig:Nesting}
\end{figure}

This behavior is illustrated in Fig.~\ref{Fig:Nesting}, which shows the
time evolution of the renormalized frequency (real part) for increasing nesting order. At low
order, the dynamics remains smooth and weakly oscillatory. With each
additional iteration, late-time oscillations become progressively more
pronounced, reflecting the increasing sensitivity to the multivaluedness of
$\Gamma_\omega$. This sequence demonstrates the approach
to the branch-cut, or Riemann-sheet, structure of the exact solution,
which is recovered only in the infinite-order limit.

As a function of time, Eq.~\eqref{Eq:iteratedjumps} admits a unique solution up to a critical time $t_{\mathrm{un}}<t_P$. Beyond this point, the solutions become multivalued and the dynamical map loses invertibility. To remain within the invertible regime, we therefore do not employ the fully nested solution and instead retain only a single iteration in the frequency renormalization.

\section{Disentanglement in the RWA\label{App:RWAdis}}

These appendices present a unified derivation of the dynamical maps used
in the main text. We begin with the RWA model, where the commuting
structure allows the disentanglement formula to be evaluated in closed
form and cleanly exposes the pole/branch--cut decomposition. The
spin--boson appendices then build on this structure. The diagonal
coherence channel follows the same logic with the addition of the
absorption contribution, the off--diagonal channel develops the
pole--plus--shell mechanism that determines the pointer direction, and
the population sector reduces to an affine extension of the Davies
semigroup without a Khalfin tail.

\subsection{Reconstructed Dynamical Map in the RWA Model}
We derive the reduced dynamical map for the rotating--wave (RWA) model
directly from the disentanglement integral~\eqref{Cdef}.


The reduced state obeys the time--local master equation
\begin{equation}
\frac{d}{dt}\rho(t)=L(t)\rho(t).
\end{equation}

Let
\begin{equation}
L_0=\lim_{t\to\infty}L(t)
\end{equation}
denote the asymptotic (Davies) generator. The exact dynamical map may
be written as
\begin{equation}
\rho(t)=\Phi(t)\rho(0),
\qquad 
\Phi(t)=e^{L_0 t}+\mathbf C(t),
\label{Eq:RWAdisentApp}
\end{equation}
with
\begin{equation}
\mathbf C(t)=\int_0^t d\tau\, e^{L_0(t-\tau)}\big[L(\tau)-L_0\big]e^{L_0\tau}.
\label{Cdef}
\end{equation}

\subsection{RWA simplification}

For the RWA Hamiltonian, the TCL generator and the Davies generator are
diagonal in the Bohr operator basis and therefore commute,
\begin{equation}
[L(t),L_0]=0.
\end{equation}
Hence the exponentials in Eq.~\eqref{Cdef} commute with
$L(\tau)-L_0$, and the reconstruction simplifies to
\begin{equation}
\mathbf C(t)=e^{L_0 t}\int_0^t d\tau\,\big[L(\tau)-L_0\big].
\label{Csimple}
\end{equation}

\subsection{Coherence dynamics in the RWA}

Projecting onto the coherence operator $|1\rangle\langle2|$ gives
\begin{equation}
f(t)=\frac{\rho_{12}(t)}{\rho_{12}(0)}=[\Phi(t)]_{12,12}.
\end{equation}

Let the eigenvalue of $L_0$ in this sector be
\begin{equation}
[L_0]_{12,12}=-iz,
\qquad
z=\Delta-\frac{i}{4}\Gamma_\Delta .
\end{equation}

Using Eq.~\eqref{Csimple},
\begin{equation}
[\mathbf C(t)]_{12,12}
=\frac{e^{-iz t}}{4}
\Big[
\Gamma_\Delta t-\int_0^t d\tau\,\Gamma_{u(\tau)}(\tau)
\Big].
\end{equation}

For times $t\gg\tau_C$, the Markov approximation for the frequency
applies,
\[
u(\tau)\simeq z.
\]

Using the bath correlation kernel
\begin{equation}
\Gamma_z(t)=\int_0^t d\tau' \, C(\tau')e^{iz\tau'},
\end{equation}
we obtain
\begin{align}
\int_0^t d\tau\int_0^\tau d\tau' \, C(\tau')e^{iz\tau'}
&=
\int_0^t d\tau' \, C(\tau')e^{iz\tau'}(t-\tau').
\label{Eq:C11}
\end{align}

Define the diagonal kernel
\begin{equation}
\Sigma_z(t)=\int_0^t d\tau\,\tau\,C(t-\tau)e^{-iz\tau}.
\label{SigmaDef}
\end{equation}
Then
\begin{equation}
[\mathbf C(t)]_{12,12}
=\frac{\Gamma_\Delta t}{4}e^{-iz t}-\frac{1}{4}\Sigma_{z}(t),
\label{Eq:C1212app}
\end{equation}
from which Eq.~\eqref{Phi_preMT} in the main text follows.

\subsection{Spectral representation}

The Laplace transform of Eq.~\eqref{SigmaDef} is
\begin{equation}
\tilde\Sigma_z(s)=\frac{\tilde C(s)}{(s+iz)^2},
\end{equation}
where
\begin{equation}
\tilde C(s)=\int_0^\infty dt\, e^{-st} C(t)
\end{equation}
is the ordinary Laplace transform of the bath correlation
function.

Using the standard spectral representation
\begin{equation}
C(t)=\frac{1}{\pi}\int_0^\infty d\omega\,J_\omega e^{-i\omega t},
\end{equation}
one also defines the half-sided Fourier-Laplace transform
\begin{equation}
\Gamma_\omega=J_\omega+iS_\omega
=\int_0^\infty dt\,C(t)e^{i\omega t}.
\end{equation}
The quantity \(\Gamma_\omega\) is simply the Laplace
transform \(\tilde C(s)\) evaluated at \(s=-i\omega\) by
analytic continuation.

We evaluate the inverse transform using the Bromwich contour,
\begin{align}
\Sigma_z(t)
&=
\frac{1}{2\pi i}\int_{-i\infty+\gamma}^{i\infty+\gamma}
ds\,e^{s t}\tilde{\Sigma}_z(s)
\nonumber\\
&=
\frac{1}{2\pi}\int_{-\infty+i\gamma}^{\infty+i\gamma}
d\omega\,e^{-i\omega t}\tilde{\tilde{\Sigma}}_z(\omega),
\end{align}
where the contour offset $i\gamma$ lies above all singularities of the
integrand. Thus
\begin{equation}
\Sigma_z(t)=
-\frac{1}{2\pi}\int_{-\infty+i\gamma}^{\infty+i\gamma}
d\omega\,
e^{-i\omega t}\frac{\Gamma_\omega}{(z-\omega)^2}.
\label{SigmaContour}
\end{equation}
which is Eq.~\eqref{SigmaContou} of the main text.

For $t>0$, the factor $e^{-i\omega t}$ suppresses the contribution from
the large semicircle in the lower half--plane, so the contour may be
deformed downward. The analytic continuation of $\Gamma_\omega$ to the
lower half--plane contains a branch cut on the positive real axis,
while $(z-\omega)^{-2}$ produces a second--order pole at $\omega=z$.
Therefore
\begin{equation}
\Sigma_z(t)
=
\Sigma_z^{\mathrm{pole}}(t)+\Sigma_z^{\mathrm{bc}}(t),
\label{Eq:SigmaSplit}
\end{equation}
with
\begin{equation}
\Sigma_z^{\mathrm{pole}}(t)
=
i\,\mathrm{Res}_{\omega=z}
\left[
e^{-i\omega t}\frac{\Gamma_\omega}{(z-\omega)^2}
\right],
\label{Eq:SigmaPoleDef}
\end{equation}
and
\begin{equation}
\Sigma_z^{\mathrm{bc}}(t)
=
-\frac{1}{2\pi}\int_0^\infty d\omega\,
e^{-i\omega t}
\frac{\Gamma_\omega^+-\Gamma_\omega^-}{(z-\omega)^2}.
\label{Eq:SigmaBCDef}
\end{equation}
Here
\begin{equation}
\Gamma_\omega^\pm
\equiv
\lim_{\epsilon\downarrow0}\Gamma_{\omega\pm i\epsilon}
\end{equation}
denote the boundary values on the two sides of the branch cut.

\subsection{Pole contribution from the Bromwich contour}

We now evaluate the residue term in Eq.~\eqref{Eq:SigmaPoleDef}. Since
the only isolated singularity enclosed by the deformed contour is the
second--order pole at $\omega=z$,
\begin{align}
\mathrm{Res}_{\omega=z}\left[
\frac{\Gamma_\omega e^{-i\omega t}}{(z-\omega)^2}
\right]
&=
\left.\frac{d}{d\omega}\Big(\Gamma_\omega e^{-i\omega t}\Big)\right|_{\omega=z}
\nonumber\\
&=
\left(\partial_\omega\Gamma_\omega-it\,\Gamma_\omega\right)_{\omega=z}
e^{-izt}.
\end{align}
Therefore
\begin{equation}
\Sigma_z^{\mathrm{pole}}(t)
=
e^{-izt}\Big[t\,\Gamma_z+i\,\partial_\omega\Gamma_\omega\big|_z\Big].
\label{SigmaPole}
\end{equation}

Substituting this into Eq.~\eqref{Eq:SigmaSplit} and then into Eqs.~\eqref{Eq:C1212app} and~\eqref{Eq:RWAdisentApp} gives
\begin{align}
[\Phi(t)]_{12,12}
&=
\left(1+\frac{\Gamma_\Delta t}{4}\right)e^{-izt}
-\frac{1}{4}e^{-izt}\Big[t\,\Gamma_z + i\,\partial_\omega\Gamma_\omega\big|_{z}\Big]
\nonumber\\
&\quad
-\frac{1}{4}\Sigma_z^{\mathrm{bc}}(t)
\nonumber\\
&=
\Big(1+\frac{\Gamma_\Delta-\Gamma_z}{4}t
-\frac{i}{4}\partial_\omega\Gamma_\omega\big|_{z}\Big)e^{-izt}
-\frac{1}{4}\Sigma_z^{\mathrm{bc}}(t).
\label{Phi_pole_split}
\end{align}

\subsection{Branch discontinuity and the $2J_\Delta$ identity}

The Sokhotski--Plemelj identity for the half--sided transform implies
\begin{equation}
\Gamma_\omega^+-\Gamma_\omega^-=2J_\omega.
\label{jump}
\end{equation}

The Davies rate is the upper boundary value,
\[
\Gamma_\Delta=\Gamma_\Delta^+,
\]
whereas $z=\Delta-i\Gamma_\Delta/4$ lies in the lower half--plane, so
\[
\Gamma_{z}=\Gamma_\Delta^-+(z-\Delta)\partial_\omega\Gamma_\omega^-|_\Delta+O(\lambda^6),
\]
since $z-\Delta=O(\lambda^2)$ and $\partial_\omega\Gamma_\omega=O(\lambda^2)$. Thus
\begin{equation}
\Gamma_\Delta-\Gamma_{z}
=2J_\Delta-(z-\Delta)\partial_\omega\Gamma_\omega^-|_\Delta+O(\lambda^4).
\label{2Jresult}
\end{equation}
At the time scale $t\sim T_1=2/J_\Delta$, substituting into
Eq.~\eqref{Phi_pole_split} yields the leading contribution
\begin{align}
[\Phi(t)]_{12,12}
\approx \left(1+\frac{J_{\Delta}t}{2}\right)e^{-iz t}
-\frac{1}{4}\Sigma_z^{\mathrm{bc}}(t).
\label{Phi_pole_splitP1}
\end{align}
The pole contribution alone therefore does not restore exponential
decay.

\subsection{Evaluation of the branch--cut integral}

Split
\[
J_\omega=J_\Delta+\delta J_\omega.
\]

The constant part gives the on--shell contribution,
\begin{align}
\Sigma_{\mathrm{bc,shell}}(t)
&=
-\frac{J_\Delta}{\pi}
\int_{-\infty}^{\infty}d\omega\,
\frac{e^{-i\omega t}}{(z-\omega)^2}.
\end{align}

Closing the contour in the lower half--plane, the second--order pole at
$z$ yields
\begin{equation}
\int_{-\infty}^{\infty}d\omega\,
\frac{e^{-i\omega t}}{(z-\omega)^2}
=
-2\pi t\,e^{-iz t}.
\end{equation}
Therefore
\begin{equation}
\Sigma_{\mathrm{bc,shell}}(t)
=
2J_\Delta t\,e^{-iz t}.
\label{I0}
\end{equation}

\subsection{Wigner--Weisskopf Regime in the RWA}

Using Eqs.~\eqref{Eq:SigmaBCDef}, \eqref{Phi_pole_splitP1}, and
\eqref{I0}, we obtain
\begin{align}
[\Phi(t)]_{12,12}
&=
\left(1+\frac{J_\Delta t}{2}\right)e^{-iz t}
-\frac{J_\Delta t}{2}e^{-iz t}
\nonumber\\
&\quad
+\frac{1}{4\pi}\int_0^\infty d\omega\,
\frac{e^{-i\omega t}\delta J_\omega}{(z-\omega)^2}.
\end{align}
The two linear contributions cancel, leaving the exponential as the
leading behavior:
\begin{equation}
[\Phi(t)]_{12,12}
=
e^{-iz t}
+\frac{1}{4\pi}\int_0^\infty d\omega\,
\frac{e^{-i\omega t}\delta J_\omega}{(z-\omega)^2}.
\label{WWfinal}
\end{equation}
This is Eq.~\eqref{Eq:fCmain} of the main text. 

\subsection{Khalfin regime in the RWA}

At late times the exponential term in Eq.~\eqref{WWfinal} may be
neglected, so
\begin{equation}
f(t)\simeq 
\frac{1}{4\pi}\int_0^\infty
d\omega\,
\frac{e^{-i\omega t} J_\omega}{(z-\omega)^2}.
\label{WWfinalRL}
\end{equation}

On the chosen analytic branch, $z$ lies in the lower half--plane, so
the denominator never vanishes for real $\omega\ge 0$.
For standard spectral densities the integrand is absolutely integrable,
and the Riemann--Lebesgue lemma implies $f(t)\to 0$ as $t\to\infty$.

At large times the integral is dominated by low frequencies.
Expanding the denominator around $\omega=0$ gives
\begin{equation}
\frac{1}{(z-\omega)^2}
=
\frac{1}{z^2}
+ O(\omega),
\end{equation}
so that
\begin{equation}
f(t)\simeq 
\frac{1}{4z^2}
\frac{1}{\pi}\int_0^\infty d\omega\, e^{-i\omega t} J(\omega)
= \frac{C(t)}{4z^2}.
\end{equation}
This is Eq.~\eqref{Eq:KhalfinMTrwa} of the main text. 

\section{Disentanglement of the Spin--Boson Model\label{App:SBM}}

Before proceeding, we note that all analytic expressions in this
and the previous appendix were verified against numerical solutions of the
reconstruction equation~(\ref{Eq:Cdot}), providing an independent
check of the Laplace transforms and contour manipulations used
in this paper.

\subsection{Diagonal Coherence Dynamics}

This channel differs from the RWA derivation only by the additional
absorption contribution, so the pole/branch--cut analysis can be reused
almost verbatim.

The disentanglement of the diagonal coherence element in the spin--boson
model proceeds analogously to the RWA Hamiltonian. The explicit
matrix element reads
\begin{align}
\Phi_{12,12}(t)
&= e^{(L_0)_{12,12}t}
+\int_0^t d\tau\,
e^{(L_0)_{12,12}(t-\tau)}
\nonumber\\
&\quad\times
\big[l_{12,12}(\tau)-(L_0)_{12,12}\big]
e^{(L_0)_{12,12}\tau}
\nonumber\\[4pt]
&=
\Big[1+\frac{1}{4}(\Gamma_\Delta+\Gamma_{-\Delta}^\ast)t\Big]
\,e^{-ia t}
\nonumber\\
&\quad
-\frac{1}{4}e^{-ia t}
\int_0^t d\tau\,
\Big[\Gamma_{\Delta'(\tau)}(\tau)
+\big[\Gamma_{-\Delta'(\tau)}(\tau)\big]^\ast\Big],
\label{Eq:C1212_start}
\end{align}
which yields Eq.~\eqref{Eq:diagF1212} of the main text. The two terms
in the integrand represent the emission and absorption contributions,
respectively.

For \(t\gg\tau_C\) we employ the Markov approximation for the frequency,
\[
\Delta'(t)\simeq a=\tilde{\Delta}-iJ_\Delta/4,
\qquad
\Gamma_{\pm}(\tau)\equiv\Gamma_{\pm a}(\tau),
\]
which gives Eq.~\eqref{Eq:C1212_mi} of the main text,
\begin{align}
\Phi_{12,12}(t)
=
\Big(1+\frac{\Gamma_\Delta+\Gamma_{-\Delta}^\ast}{4}t\Big)
e^{-ia t}
-\frac{1}{4}\Sigma_{a}^{\text{sbm}}(t),
\label{Eq:C1212_mid}
\end{align}
where
\begin{equation}
\Sigma_{a}^{\text{sbm}}(t)=
e^{-iat}
\int_0^t d\tau\,
\Big[\Gamma_{a}(\tau)+\big[\Gamma_{-a}(\tau)\big]^\ast\Big].
\end{equation}

Proceeding as in Eq.~\eqref{Eq:C11}, the time integration yields
Eq.~\eqref{Eq:SigSB} of the main text,
\begin{equation}
\Sigma_{a}^{\text{sbm}}(t)
=\Sigma_{a}(t)
+e^{-J_\Delta t/2}\,\big[\Sigma_{-a}(t)\big]^\ast,
\label{Eq:SigSBM}
\end{equation}
with Laplace transform
\begin{equation}
\tilde{\Sigma}_{a}^{\text{sbm}}(s)
=\frac{\tilde{C}(s)+\big[\tilde{C}\!\big(s^\ast+J_\Delta/2\big)\big]^\ast}
{(s+ia)^2}.
\end{equation}
Equivalently, in Fourier--Laplace form with \(s=-i\omega\) and
\(\tilde{C}(s)=\Gamma_\omega\),
\begin{equation}
\tilde{\tilde{\Sigma}}_{a}^{\text{sbm}}(\omega)=-
\frac{\Gamma_\omega+\big[\Gamma_{-\omega^\ast+iJ_\Delta/2}\big]^\ast}
{(\omega-a)^2}.
\label{Eq:FLTdiag}
\end{equation}
The two terms in the numerator represent emission and absorption,
respectively.

The inverse Laplace transform is evaluated using a Bromwich contour,
\begin{equation}
\Sigma_{a}^{\text{sbm}}(t)
=\frac{1}{2\pi}\int_{-\infty+i\gamma}^{\infty+i\gamma} d\omega\,
\tilde{\tilde{\Sigma}}_{a}^{\text{sbm}}(\omega)\,e^{-i\omega t}.
\label{Eq:Brom1212}
\end{equation}
Inserting Eq.~\eqref{Eq:FLTdiag} into Eq.~\eqref{Eq:Brom1212} yields
Eq.~\eqref{WW_SB} of the main text.

The emission contribution is identical to the RWA case, so the
pole--plus--branch--cut analysis of the previous appendix applies after
the substitution \(z\rightarrow a\). In particular, the linear term
\(\Gamma_\Delta t/4\) in Eq.~\eqref{Eq:C1212_start} is canceled by the
emission pole together with its on--shell branch--cut contribution.

Similarly, the pole in the absorption term cancels the non--Markovian
term \((\Gamma_{-\Delta}^\ast t/4)e^{-iat}\), since \(J_{-\Delta}=0\).
This restores the purely exponential Markovian limit.

The remaining branch--cut integral at \(\omega^\ast=iJ_\Delta/2\)
arising from the absorption contribution is evaluated by the
substitution \(\omega=-\omega'+iJ_\Delta/2\), which leads to the final
integral in Eq.~\eqref{WW_SBM} of the main text.

\subsection{Off-diagonal coherence dynamics.\label{App:SBM2112}}

The evaluation proceeds in three steps. First, the coherence kernel is
written in terms of the bath correlation function, leading to the
representation Eq.~\eqref{Eq:Zsbm}. Second, the Laplace transform
allows the dynamics to be expressed as a Bromwich contour integral,
whose analytic structure separates naturally into pole and continuum
contributions. Finally, we evaluate these contributions: the poles and
the on--shell branch--cut component produce the exponential decay and
determine the pointer direction, while the off--shell branch--cut
component generates the non--Markovian continuum correction and the
Khalfin tail.
\subsubsection{Time-domain representation}
The off-diagonal element $\Phi_{21,12}(t)=C_{21,12}(t)$ reads
\begin{align}
\Phi_{21,12}(t)
&=\int_0^t d\tau\,
e^{(L_0)_{21,21}(t-\tau)}
\,l_{21,12}(\tau)\,
e^{(L_0)_{12,12}\tau}
\nonumber\\
&= e^{ia^\ast t}
\int_0^t d\tau\, e^{-2 i \tilde{\Delta}\tau}\, l_{21,12}(\tau).
\label{Eq:F1221_0}
\end{align}
where $l_{21,12}(t)$ is given by Eq.~\eqref{Eq:LSBM_coh_entries}. This yields
\begin{align}
\Phi_{21,12}
&=\Phi_{21,12}^{\text{emi}}
+\Phi_{21,12}^{\text{abs}}
\nonumber\\
&=\frac{1}{4}e^{ia^\ast t}
\int_0^t d\tau\, e^{-2i\tilde{\Delta}\tau}
\nonumber\\
&\qquad\times
\Big[
\Gamma_{\Delta'(\tau)}(\tau)
+
\big(\Gamma_{-\Delta'(\tau)}(\tau)\big)^\ast
\Big].
\end{align}

Employing the Markovian approximation for the frequency
$\Delta'(t)\simeq a$ gives an emission integral
\begin{align}
I(t)
&=\int_0^t d\tau\,e^{-2i\tilde{\Delta}\tau}
\Gamma_{\Delta'}(\tau)
\nonumber\\
&=\int_0^t d\tau\,e^{-2i\tilde{\Delta}\tau}
\int_0^\tau dx\,C(x)e^{ia x}
\end{align}
Interchanging the order of integration and performing the
$\tau$–integration yields
\begin{align}
I(t)
&=\int_0^t dx\,C(t-x)
e^{-i\tilde{\Delta}t+\frac{J_\Delta}{4}(t-x)}
\frac{\sin(\tilde{\Delta}x)}{\tilde{\Delta}} .
\end{align}

Inserting $a^\ast=\tilde{\Delta}+iJ_\Delta/4$, the emission term becomes
\begin{align}
\frac{1}{4}e^{ia^\star t}
I(t)
&=
\frac{1}{4}\int_0^t dx\, C(t-x)
e^{-\frac{J_\Delta}{4}x}
\frac{\sin(\tilde{\Delta}x)}{\tilde{\Delta}}\\
&=\frac{1}{4}Z_a(t).
\end{align}

The net matrix element reads
\begin{equation}
\Phi_{21,12}(t)
=
\frac{1}{4}\Big[
Z_{a}(t)
+
e^{-J_\Delta t/2}
\big[Z_{-a}(t)\big]^\ast
\Big],
\label{Eq:Zsbm}
\end{equation}
where we introduced the kernel
\begin{equation}
\label{Eq:Zkernel}
Z_{x+iy}(t)\equiv
\int_0^t d\tau\, C(t-\tau)
\frac{\sin(x\tau)}{x}e^{y\tau}.
\end{equation}
The last two equations yield 
Eqs.~\eqref{Eq:Zsb} and ~\eqref{Eq:Zkerne},  respectively.

\subsubsection{Laplace representation}

Next we evaluate the Laplace transform of Eq.~\eqref{Eq:Zsbm}.
The Laplace transform of the kernel is
\begin{align}
\tilde{Z}_{x+iy}(s)
&=\int_0^\infty Z_{x+iy}(t) e^{-s t}\nonumber\\
&=\frac{\tilde{C}(s)}{x^2+(s-y)^2}.
\end{align}

Hence,
\begin{equation}
\tilde{\Phi}_{21,12}(s)
=
\frac{1}{4}
\frac{\tilde{C}(s)+\big[\tilde{C}(s^\ast+J_\Delta/2)\big]^\ast}
{(s+J_\Delta/4)^2+\tilde{\Delta}^2}.
\end{equation}

In the Fourier–-Laplace representation $s=-i\omega$ and
$\tilde{C}(s)=\Gamma_\omega$,

\begin{equation}
\tilde{\tilde{\Phi}}_{21,12}(\omega)=-
\frac{1}{4}
\frac{\Gamma_\omega+\big[\Gamma_{-\omega^\ast+iJ_\Delta/2}\big]^\ast}
{(\omega+iJ_\Delta/4)^2-\tilde{\Delta}^2}.
\end{equation}

Integrating along the Bromwich contour gives Eq.~\eqref{Eq:BromZ} of the main text,
\begin{equation}
\Phi_{21,12}(t)=
-\frac{1}{8\pi}
\int_{-\infty+i0^+}^{\infty+i0^+} d\omega\,
e^{-i\omega t}
\frac{\Gamma_\omega+\big[\Gamma_{-\omega^\ast+iJ_\Delta/2}\big]^\ast}
{(\omega+iJ_\Delta/4)^2-\tilde{\Delta}^2}.
\label{Eq:Brom2112}
\end{equation}
The analytic structure of the integrand contains poles at
$\omega=\pm\tilde{\Delta}-iJ_\Delta/4$ together with a branch cut
along the real axis. Accordingly the coherence separates into
\begin{equation}
\Phi_{21,12}(t)
=
\Phi^{\rm pole}(t)
+
\Phi^{\rm cut}(t).
\end{equation}

\subsubsection{Off-diagonal pole contribution}
The pole contribution is obtained from the residues in the lower half
plane at $\omega_\pm=\pm\tilde{\Delta}-iJ_\Delta/4$. After straightforward algebra, this yields Eq.~\eqref{Eq:Phi2112_pole_finalA} of the main text:
\begin{align}
\Phi_{21,12}^{\text{pole}}(t)
&=\frac{i}{8\tilde{\Delta}}\,e^{-J_\Delta t/4}
\Big[
e^{-i\tilde{\Delta}t}
\Big(\Gamma_{a}
+\big(\Gamma_{-a}\big)^\ast\Big)
\nonumber\\
&\qquad
-
e^{+i\tilde{\Delta}t}
\Big(\Gamma_{-a^\star}
+\big(\Gamma_{a^\star}\big)^\ast\Big)
\Big].
\label{Eq:Phi2112_pole_finalAP}
\end{align}
In the weak--coupling limit
\[
a \to \Delta - i0^+ ,
\]
the pole contribution reduces to Eq.~\eqref{Eq:F1221polelimit} of the main text:
\begin{equation}
\Phi_{21,12}^{\text{pole}}(t)
=
-\frac{i}{4}
e^{-J_\Delta t/4}
\frac{\cos(\tilde{\Delta}t)}{\tilde{\Delta}}
\left(\Gamma_\Delta^\ast+\Gamma_{-\Delta}\right).
\label{Eq:F1221polelimitAPpole}
\end{equation}

\subsubsection{Off-diagonal branch--cut contribution\label{sec:BCZkerenl}} 

In the emission component (integrand proportional to
$\big[\Gamma_{-\omega^\ast+iJ_\Delta/2}\big]^\ast$ in
Eq.~\eqref{Eq:Brom2112}), the change of variables $\omega\to -\omega$
puts the branch--cut contribution into the form
\begin{align}
\Phi_{21,12}^{\rm cut}(t)
&=
-\frac{1}{4\pi}\int_{0}^{\infty} d\omega\,
\frac{e^{-i\omega t}\, J_\omega}{(\omega+iJ_\Delta/4)^2-\tilde{\Delta}^2}
\nonumber\\
&\quad
-\frac{e^{-J_\Delta t/2}}{4\pi}\int_{0}^{\infty} d\omega\,
\frac{e^{+i\omega t}\,J_\omega}{(\omega+iJ_\Delta/4)^2-\tilde{\Delta}^2}.
\label{Eq:Phi2112_cut_mem}
\end{align}
This yields Eq.~\eqref{Eq:Phi2112_cut_total} of the main text.

The two terms in Eq.~\eqref{Eq:Phi2112_cut_mem} correspond to the
emission and absorption continua. These contributions behave
differently. The emission branch cut produces an on--shell component
that combines with the pole term and determines the pointer direction,
whereas the absorption continuum contains no such on--shell
contribution and contributes only to the remaining continuum
correction.

We first analyze the emission continuum contribution
\begin{equation}
I(t)\equiv
-\frac{1}{4\pi}\int_{0}^{\infty} d\omega\,
\frac{e^{-i\omega t}\,J_\omega}
{(\omega+iJ_\Delta/4)^2-\tilde{\Delta}^2}.
\end{equation}

To evaluate $I(t)$ we deform the contour into the lower half plane. The integral then separates into an on--shell residue
and a vertical contour contribution,
\begin{equation}
I(t)=I_{\rm shell}(t)+I_{\rm vert}(t).
\end{equation}

The denominator has poles at
\begin{equation}
\omega_\pm=-\frac{iJ_\Delta}{4}\pm \tilde{\Delta}.
\end{equation}
The pole in the fourth quadrant,
\begin{equation}
\omega_p=\tilde{\Delta}-\frac{iJ_\Delta}{4},
\end{equation}
gives the residue
\begin{equation}
I_{\rm shell}(t)
=
\frac{i}{4\tilde{\Delta}}\,J_{\omega_p}\,e^{-i\omega_p t}
=
\frac{i}{4\tilde{\Delta}}\,J_{\omega_p}
\,e^{-i\tilde{\Delta}t}e^{-J_\Delta t/4}.
\end{equation}
This on--shell contribution arises solely from the emission branch cut and yields Eq.~\eqref{Eq:Phi2112_cut_intermediate} of the main text.
It combines with the pole term~\eqref{Eq:F1221polelimitAPpole} to produce
\begin{align}
\Phi_{21,12}^{\text{pole+shell}}(t)
&= e^{-J_\Delta t/4}X(t),
\label{Eq:Phi2112_combinedAP}
\end{align}
where \(X(t)\) is given in Eq.~\eqref{Eq:Xoft} 
of the main text.

The remaining terms arise from the continuum contributions. In
contrast to the emission branch cut, the absorption continuum does not
produce an on--shell component and therefore contributes only through
the remaining vertical contour integral. 
The vertical segment is oriented from $-i\infty$ to $0$. Writing
\begin{equation}
\omega=-iy,
\qquad
y:\infty\to 0,
\qquad
d\omega=-i\,dy,
\end{equation}
one finds
\begin{equation}
\int_{-i\infty}^{0} f(\omega)\,d\omega
=
i\int_{0}^{\infty} f(-iy)\,dy.
\end{equation}
Since
\begin{equation}
(-iy+iJ_\Delta/4)^2-\tilde{\Delta}^2
=
-\Big[\tilde{\Delta}^2+\big(y-J_\Delta/4\big)^2\Big],
\end{equation}
and
\begin{equation}
J_{-iy}
=
2\pi\lambda^2\,\omega_c\left(\frac{-iy}{\omega_c}\right)^s e^{iy/\omega_c}
=
2\pi\lambda^2\,\omega_c^{\,1-s}\,
y^s e^{-i\pi s/2}e^{iy/\omega_c},
\end{equation}
the vertical contribution becomes
\begin{equation}
I_{\rm vert}(t)
=
-\frac{i\lambda^2\omega_c^{\,1-s}e^{-i\pi s/2}}{2}
\int_{0}^{\infty} dy\,
\frac{y^s e^{-yt}e^{iy/\omega_c}}
{\tilde{\Delta}^2+\big(y-J_\Delta/4\big)^2}.
\end{equation}
The exponential factor $e^{-yt}$ restricts the dominant integration
region to $y\sim t^{-1}$. For times
$t\gg \tilde{\Delta}^{-1},\omega_c^{-1}$ this implies
$y\ll \tilde{\Delta},\omega_c$ over the dominant region. The phase factor
$e^{iy/\omega_c}$ and the denominator
$\tilde{\Delta}^2+\big(y-J_\Delta/4\big)^2$ may therefore be replaced,
to leading order, by their endpoint values at $y=0$. This gives
\begin{align}
I_{\rm vert}(t)
&\approx
-\frac{i\lambda^2\omega_c^{\,1-s}e^{-i\pi s/2}}{2(\tilde{\Delta}^2+J_\Delta^2/16)}
\int_{0}^{\infty} dy\, y^s e^{-yt},
\nonumber\\\qquad &
t\gg \tilde{\Delta}^{-1},\omega_c^{-1}.
\end{align}
The remaining integral is elementary,
\begin{equation}
\int_{0}^{\infty} dy\, y^s e^{-yt}
=
\Gamma(s+1)\,t^{-(s+1)},
\end{equation}
and therefore
\begin{align}
I_{\rm vert}(t)
\approx
-\frac{i\lambda^2\omega_c^{\,1-s}e^{-i\pi s/2}}{2}
\frac{\Gamma(s+1)}{\tilde{\Delta}^2+J_\Delta^2/16}\,
t^{-(s+1)}.
\label{Eq:IvertKhalfin}
\end{align}
Thus, inserting Eq.~\eqref{Eq:BCF}, the vertical contribution already exhibits the Khalfin tail for
$t\gg \tilde{\Delta}^{-1},\omega_c^{-1}$.

The absorption branch cut is treated similarly. Since it remains off
shell, it produces no additional on--shell term and contributes only the
analogous continuum piece,
\begin{equation}
\Phi_{21,12}^{\rm abs,cut}(t)
\sim
e^{-J_\Delta t/2}\frac{C(t)^\ast}{4\Delta^2}.
\end{equation}
It is therefore exponentially suppressed at late times, so that the asymptotic tail of the full branch--cut contribution is
\begin{equation}
\Phi_{21,12}^{\rm cut}(t)\sim \frac{C(t)}{4\Delta^2}.
\end{equation}

Collecting all contributions, the off--diagonal coherence dynamics
therefore consists of an exponentially damped pole--plus--shell term
that selects the pointer direction, together with a continuum
correction whose asymptotic form is the Khalfin tail.

\subsection{Population Dynamics\label{Sec:PopuDisen}}

We first derive the disentangled population map to leading order in weak coupling and then
determine its Wigner--Weisskopf asymptotics.

In the population sector the Davies reference generator is
\begin{equation}
(L_0)_{\mathrm{pop}}
=
\frac{J_\Delta}{2}
\begin{pmatrix}
-1 & 0\\
1 & 0
\end{pmatrix},
\end{equation}
with exponential
\begin{equation}
e^{(L_0)_{\mathrm{pop}}t}
=
\begin{pmatrix}
e^{-J_\Delta t/2} & 0\\
1-e^{-J_\Delta t/2} & 1
\end{pmatrix}. 
\label{Eq:daviesPopu}
\end{equation}

The resummed generator reads
\begin{equation}
L_{\mathrm{pop}}(t)
=
\frac12\,\mathrm{Re}
\begin{pmatrix}
-\Gamma_+(t) & \Gamma_-(t)\\
\Gamma_+(t) & -\Gamma_-(t)
\end{pmatrix}.
\end{equation}

Inserting this into the disentanglement formula,
\begin{equation}
C_{\mathrm{pop}}(t)=
\int_0^t d\tau
e^{(L_0)_{\mathrm{pop}}(t-\tau)}
\big[L_{\mathrm{pop}}(\tau)-(L_0)_{\mathrm{pop}}\big]
e^{(L_0)_{\mathrm{pop}}\tau},
\end{equation}
and noting that the exponential matrices are real, the real part may be
taken outside the integral. One obtains
\begin{equation}
C_{\mathrm{pop}}(t)=
\begin{pmatrix}
A(t) & B(t)\\
-A(t) & -B(t)
\end{pmatrix},
\end{equation}
with
\begin{align}
A(t)
&=\frac{J_\Delta t}{2}e^{-J_\Delta t/2}
\nonumber\\
&\quad
-\frac{e^{-J_\Delta t/2}}{2}\,
\mathrm{Re}\!\int_0^t d\tau\,
\Big[
\Gamma_+(\tau)
-\big(e^{J_\Delta\tau/2}-1\big)\Gamma_-(\tau)
\Big]
\\
B(t)
&=
\frac12\,\mathrm{Re}\!\int_0^t d\tau\,
e^{-J_\Delta (t-\tau)/2}\Gamma_-(\tau).
\label{Eq:CofT}
\end{align}

The disentangled population map therefore becomes
\begin{equation}
\label{Eq:PhiRho}
\Phi_{\mathrm{pop}}(t)
=
\begin{pmatrix}
e^{-J_\Delta t/2}+A(t) & B(t)\\
1-e^{-J_\Delta t/2}-A(t) & 1-B(t)
\end{pmatrix}.
\end{equation}

The key point is that the apparent secular term in $A(t)$ is cancelled
by the memory contribution, so that at Wigner--Weisskopf time scales the
population sector reduces to the Davies semigroup plus an
$O(\lambda^2)$ static correction.

To evaluate \(B(t)\), we substitute
\begin{equation}
\Gamma_-(\tau)=\int_0^\tau ds\, C(s)e^{-ias},
\qquad
a=\tilde{\Delta}-iJ_\Delta/4,
\end{equation}
into Eq.~\eqref{Eq:CofT}. Interchanging the order of integration and
performing the elementary $\tau$–integration yields
\begin{equation}
B(t)=\frac{1}{J_\Delta}\,
\mathrm{Re}\!\int_0^t ds\, C(s)e^{-ias}
\Big(1-e^{-J_\Delta (t-s)/2}\Big).
\label{Eq:BpopKernel}
\end{equation}

Separating the two terms gives
\begin{align}
B(t)
&=\frac{1}{J_\Delta}\,\mathrm{Re}\Bigl[
\int_0^t ds\, C(s)e^{-ias}
\nonumber\\
&\qquad
- e^{-J_\Delta t/2}
\int_0^t ds\, C(s)e^{-i(a+iJ_\Delta/2)s}
\Bigr].
\end{align}

Using the kernel introduced above,
\begin{equation}
\Gamma_z(t)\equiv \int_0^t d\tau\, C(\tau)e^{iz\tau},
\end{equation}
this becomes
\begin{equation}
B(t)=\frac{1}{J_\Delta}\,
\mathrm{Re}\!\Big[
\Gamma_{-a}(t)
-
e^{-J_\Delta t/2}\Gamma_{-(a+iJ_\Delta/2)}(t)
\Big].
\label{Eq:BpopGamma}
\end{equation}

To evaluate \(A(t)\), we write
\begin{equation}
A(t)=\frac{J_\Delta t}{2}e^{-J_\Delta t/2}
-\frac{e^{-J_\Delta t/2}}{2}\,\mathrm{Re}\,I_A(t),
\end{equation}
where
\begin{equation}
I_A(t)\equiv
\int_0^t d\tau\,
\Big[
\Gamma_+(\tau)-\big(e^{J_\Delta\tau/2}-1\big)\Gamma_-(\tau)
\Big].
\end{equation}

Using
\begin{equation}
\Gamma_{\pm}(\tau)=\int_0^\tau ds\,C(s)e^{\pm ias},
\qquad
a=\tilde{\Delta}-iJ_\Delta/4,
\end{equation}
and exchanging the order of integration, the two contributions may be
expressed in terms of the kernels
\begin{equation}
\Gamma_z(t)\equiv \int_0^t d\tau\,C(\tau)e^{iz\tau},
\qquad
\Sigma_z(t)\equiv
\int_0^t d\tau\,\tau\,C(t-\tau)e^{-iz\tau}.
\end{equation}

After straightforward algebra one obtains
\begin{align}
I_A(t)
&=
e^{iat}\Sigma_a(t)
-\frac{2}{J_\Delta}e^{J_\Delta t/2}\Gamma_{-a}(t)
\nonumber\\
&\quad
+\frac{2}{J_\Delta}\Gamma_{-(a+iJ_\Delta/2)}(t)
+e^{-iat}\Sigma_{-a}(t).
\end{align}

Using Eq.~\eqref{Eq:BpopGamma}, the final expression for \(A(t)\) becomes
\begin{align}
A(t)
&=
\frac{J_\Delta t}{2}e^{-J_\Delta t/2}
+B(t)
\nonumber\\
&\quad
-\frac{e^{-J_\Delta t/2}}{2}\,
\mathrm{Re}\!\Big[
e^{iat}\Sigma_a(t)+e^{-iat}\Sigma_{-a}(t)
\Big],
\label{Eq:ApopFinal}
\end{align}
where
\begin{equation}
a=\tilde{\Delta}-iJ_\Delta/4.
\end{equation}

Eqs.~\eqref{Eq:PhiRho},~\eqref{Eq:BpopGamma}, and~\eqref{Eq:ApopFinal}
give the disentangled population map prior to taking asymptotic limits.

\subsection{Wigner--Weisskopf asymptotics of the population sector}

At the Wigner--Weisskopf time scale \(t\sim O(\lambda^{-2})\), only
\(O(\lambda^0)\) terms multiplying the reference exponential are
retained. As in the coherence sector, the explicit linear term
\((J_\Delta t/2)e^{-J_\Delta t/2}\) in Eq.~\eqref{Eq:ApopFinal} is cancelled by the combined contribution of the pole and the on--shell
component of the emission branch cut (see, e.g., Eq.~\eqref{I0}). The intermediate
steps follow the same procedure as in the RWA Laplace--transform
analysis and are therefore not repeated here.

We now determine the asymptotic form of the population map by examining
$A(t)$ in Eq.~\eqref{Eq:ApopFinal}.
As in the coherence sector, the kernels $\Sigma_z(t)$ are decomposed
into pole and branch--cut contributions,
\begin{equation}
\Sigma_z(t)=\Sigma_z^{\mathrm{pole}}(t)+\Sigma_z^{\mathrm{bc}}(t),
\end{equation}
where
\begin{equation}
\Sigma_z^{\mathrm{pole}}(t)
=
e^{-izt}\Big[t\,\Gamma_z+i\,\partial_\omega\Gamma_\omega\big|_{z}\Big],
\label{Eq:SigLinear}
\end{equation}
which is identical to Eq.~\eqref{SigmaPole}.

After the cancellation of the linear terms in Eqs.~\eqref{Eq:ApopFinal} and~\ref{Eq:SigLinear}, the remaining terms are exponentially decaying and
$O(\lambda^2)$. The reference semigroup in Eq.~\eqref{Eq:daviesPopu}
therefore dominates, giving to leading order
\begin{equation}
A(t)=B(t).
\label{Eq:ApopAsymFinal}
\end{equation}

The corresponding population dynamics therefore takes the form of an
affine map on the population vector,
\begin{align}
\Phi_{\mathrm{pop}}(t)
&=
\begin{pmatrix}
e^{-J_\Delta t/2} & 0\\
1-e^{-J_\Delta t/2} & 1
\end{pmatrix}
+
B(t)
\begin{pmatrix}
1 & 1\\
-1 & -1
\end{pmatrix}.
\label{Eq:PopMapAsym}
\end{align}

At long times the second term in Eq.~\eqref{Eq:BpopGamma} vanishes,
since its internal growth is overcompensated by the external factor
$e^{-J_\Delta t/2}$. Therefore
\begin{equation}
B_\infty=\frac{1}{J_\Delta}\,\text{Re}\,\Gamma_{-a}.
\end{equation}
Expanding $\Gamma_{-a}$ about $-\Delta$ and using
$\text{Re}\,\Gamma_{-\Delta}=0$ at zero temperature, one obtains
\begin{equation}
B_\infty
=
-\frac{1}{4}\,\partial_\omega S_\omega\big|_{-\Delta},
\end{equation}
which is precisely the $O(\lambda^2)$ stationary excited--state
population, in agreement with the mean--force Gibbs state~\cite{cresser2021weak}.

For $t\to\infty$, the oscillatory part of $\Gamma_{-a}(t)$ is governed by the pole contribution, so $B(t)$ approaches $B_\infty$ exponentially fast. No branch--cut contribution survives in the population. Hence, Eq.~\eqref{Eq:PopMapAsym} reduces to the population map in Eq.~\eqref{Eq:PopMapMain} of the main text.

Together, these appendices show that the same disentanglement framework
underlies all sectors of the dynamics, while the different analytic
structures of the corresponding kernels determine whether the late-time
behavior is purely exponential, affine, or Khalfin-tailed.

\bibliographystyle{apsrev4-2}

\begin{thebibliography}{33}%
\makeatletter
\providecommand \@ifxundefined [1]{%
 \@ifx{#1\undefined}
}%
\providecommand \@ifnum [1]{%
 \ifnum #1\expandafter \@firstoftwo
 \else \expandafter \@secondoftwo
 \fi
}%
\providecommand \@ifx [1]{%
 \ifx #1\expandafter \@firstoftwo
 \else \expandafter \@secondoftwo
 \fi
}%
\providecommand \natexlab [1]{#1}%
\providecommand \enquote  [1]{``#1''}%
\providecommand \bibnamefont  [1]{#1}%
\providecommand \bibfnamefont [1]{#1}%
\providecommand \citenamefont [1]{#1}%
\providecommand \href@noop [0]{\@secondoftwo}%
\providecommand \href [0]{\begingroup \@sanitize@url \@href}%
\providecommand \@href[1]{\@@startlink{#1}\@@href}%
\providecommand \@@href[1]{\endgroup#1\@@endlink}%
\providecommand \@sanitize@url [0]{\catcode `\\12\catcode `\$12\catcode `\&12\catcode `\#12\catcode `\^12\catcode `\_12\catcode `\%12\relax}%
\providecommand \@@startlink[1]{}%
\providecommand \@@endlink[0]{}%
\providecommand \url  [0]{\begingroup\@sanitize@url \@url }%
\providecommand \@url [1]{\endgroup\@href {#1}{\urlprefix }}%
\providecommand \urlprefix  [0]{URL }%
\providecommand \Eprint [0]{\href }%
\providecommand \doibase [0]{https://doi.org/}%
\providecommand \selectlanguage [0]{\@gobble}%
\providecommand \bibinfo  [0]{\@secondoftwo}%
\providecommand \bibfield  [0]{\@secondoftwo}%
\providecommand \translation [1]{[#1]}%
\providecommand \BibitemOpen [0]{}%
\providecommand \bibitemStop [0]{}%
\providecommand \bibitemNoStop [0]{.\EOS\space}%
\providecommand \EOS [0]{\spacefactor3000\relax}%
\providecommand \BibitemShut  [1]{\csname bibitem#1\endcsname}%
\let\auto@bib@innerbib\@empty
\bibitem [{\citenamefont {Breuer}\ and\ \citenamefont {Petruccione}(2002)}]{BreuerPetruccione2002}%
  \BibitemOpen
  \bibfield  {author} {\bibinfo {author} {\bibfnamefont {H.-P.}\ \bibnamefont {Breuer}}\ and\ \bibinfo {author} {\bibfnamefont {F.}~\bibnamefont {Petruccione}},\ }\href@noop {} {\emph {\bibinfo {title} {The Theory of Open Quantum Systems}}}\ (\bibinfo  {publisher} {Oxford University Press},\ \bibinfo {address} {Oxford},\ \bibinfo {year} {2002})\BibitemShut {NoStop}%
\bibitem [{\citenamefont {Chaturvedi}\ and\ \citenamefont {Shibata}(1979)}]{ChaturvediShibata1979}%
  \BibitemOpen
  \bibfield  {author} {\bibinfo {author} {\bibfnamefont {S.}~\bibnamefont {Chaturvedi}}\ and\ \bibinfo {author} {\bibfnamefont {F.}~\bibnamefont {Shibata}},\ }\href@noop {} {\bibfield  {journal} {\bibinfo  {journal} {Zeitschrift für Physik B Condensed Matter}\ }\textbf {\bibinfo {volume} {35}},\ \bibinfo {pages} {297} (\bibinfo {year} {1979})}\BibitemShut {NoStop}%
\bibitem [{\citenamefont {{Van Kampen}}(1974{\natexlab{a}})}]{VANKAMPEN1}%
  \BibitemOpen
  \bibfield  {author} {\bibinfo {author} {\bibfnamefont {N.}~\bibnamefont {{Van Kampen}}},\ }\href {https://doi.org/https://doi.org/10.1016/0031-8914(74)90121-9} {\bibfield  {journal} {\bibinfo  {journal} {Physica}\ }\textbf {\bibinfo {volume} {74}},\ \bibinfo {pages} {215} (\bibinfo {year} {1974}{\natexlab{a}})}\BibitemShut {NoStop}%
\bibitem [{\citenamefont {{Van Kampen}}(1974{\natexlab{b}})}]{VANKAMPEN2}%
  \BibitemOpen
  \bibfield  {author} {\bibinfo {author} {\bibfnamefont {N.}~\bibnamefont {{Van Kampen}}},\ }\href {https://doi.org/https://doi.org/10.1016/0031-8914(74)90122-0} {\bibfield  {journal} {\bibinfo  {journal} {Physica}\ }\textbf {\bibinfo {volume} {74}},\ \bibinfo {pages} {239} (\bibinfo {year} {1974}{\natexlab{b}})}\BibitemShut {NoStop}%
\bibitem [{\citenamefont {Crowder}\ \emph {et~al.}(2024)\citenamefont {Crowder}, \citenamefont {Lampert}, \citenamefont {Manchanda}, \citenamefont {Shoffeitt}, \citenamefont {Gadamsetty}, \citenamefont {Pei}, \citenamefont {Chaudhary},\ and\ \citenamefont {Davidovi\ifmmode~\acute{c}\else \'{c}\fi{}}}]{Crowder}%
  \BibitemOpen
  \bibfield  {author} {\bibinfo {author} {\bibfnamefont {E.}~\bibnamefont {Crowder}}, \bibinfo {author} {\bibfnamefont {L.}~\bibnamefont {Lampert}}, \bibinfo {author} {\bibfnamefont {G.}~\bibnamefont {Manchanda}}, \bibinfo {author} {\bibfnamefont {B.}~\bibnamefont {Shoffeitt}}, \bibinfo {author} {\bibfnamefont {S.}~\bibnamefont {Gadamsetty}}, \bibinfo {author} {\bibfnamefont {Y.}~\bibnamefont {Pei}}, \bibinfo {author} {\bibfnamefont {S.}~\bibnamefont {Chaudhary}},\ and\ \bibinfo {author} {\bibfnamefont {D.}~\bibnamefont {Davidovi\ifmmode~\acute{c}\else \'{c}\fi{}}},\ }\href {https://doi.org/10.1103/PhysRevA.109.052205} {\bibfield  {journal} {\bibinfo  {journal} {Phys. Rev. A}\ }\textbf {\bibinfo {volume} {109}},\ \bibinfo {pages} {052205} (\bibinfo {year} {2024})}\BibitemShut {NoStop}%
\bibitem [{\citenamefont {Lampert}\ \emph {et~al.}(2025)\citenamefont {Lampert}, \citenamefont {Gadamsetty}, \citenamefont {Chaudhary}, \citenamefont {Pei}, \citenamefont {Chen}, \citenamefont {Crowder},\ and\ \citenamefont {Davidovi{\'c}}}]{Lampert2025}%
  \BibitemOpen
  \bibfield  {author} {\bibinfo {author} {\bibfnamefont {L.}~\bibnamefont {Lampert}}, \bibinfo {author} {\bibfnamefont {S.}~\bibnamefont {Gadamsetty}}, \bibinfo {author} {\bibfnamefont {S.}~\bibnamefont {Chaudhary}}, \bibinfo {author} {\bibfnamefont {Y.}~\bibnamefont {Pei}}, \bibinfo {author} {\bibfnamefont {J.}~\bibnamefont {Chen}}, \bibinfo {author} {\bibfnamefont {E.}~\bibnamefont {Crowder}},\ and\ \bibinfo {author} {\bibfnamefont {D.}~\bibnamefont {Davidovi{\'c}}},\ }\href@noop {} {\bibfield  {journal} {\bibinfo  {journal} {Physical Review A}\ }\textbf {\bibinfo {volume} {111}},\ \bibinfo {pages} {042214} (\bibinfo {year} {2025})}\BibitemShut {NoStop}%
\bibitem [{\citenamefont {Ángel Rivas}\ and\ \citenamefont {Huelga}(2012)}]{RivasHuelga2012}%
  \BibitemOpen
  \bibfield  {author} {\bibinfo {author} {\bibnamefont {Ángel Rivas}}\ and\ \bibinfo {author} {\bibfnamefont {S.~F.}\ \bibnamefont {Huelga}},\ }\href {https://doi.org/10.1007/978-3-642-23354-8} {\emph {\bibinfo {title} {Open Quantum Systems: An Introduction}}},\ SpringerBriefs in Physics\ (\bibinfo  {publisher} {Springer},\ \bibinfo {address} {Heidelberg},\ \bibinfo {year} {2012})\BibitemShut {NoStop}%
\bibitem [{\citenamefont {Hou}\ \emph {et~al.}(2012)\citenamefont {Hou}, \citenamefont {Yi}, \citenamefont {Yu},\ and\ \citenamefont {Oh}}]{hou2012singularity}%
  \BibitemOpen
  \bibfield  {author} {\bibinfo {author} {\bibfnamefont {S.}~\bibnamefont {Hou}}, \bibinfo {author} {\bibfnamefont {X.}~\bibnamefont {Yi}}, \bibinfo {author} {\bibfnamefont {S.}~\bibnamefont {Yu}},\ and\ \bibinfo {author} {\bibfnamefont {C.}~\bibnamefont {Oh}},\ }\href@noop {} {\bibfield  {journal} {\bibinfo  {journal} {Physical Review A—Atomic, Molecular, and Optical Physics}\ }\textbf {\bibinfo {volume} {86}},\ \bibinfo {pages} {012101} (\bibinfo {year} {2012})}\BibitemShut {NoStop}%
\bibitem [{\citenamefont {Chru\ifmmode \acute{s}\else \'{s}\fi{}ci\ifmmode~\acute{n}\else \'{n}\fi{}ski}\ and\ \citenamefont {Kossakowski}(2010)}]{Chru2010}%
  \BibitemOpen
  \bibfield  {author} {\bibinfo {author} {\bibfnamefont {D.}~\bibnamefont {Chru\ifmmode \acute{s}\else \'{s}\fi{}ci\ifmmode~\acute{n}\else \'{n}\fi{}ski}}\ and\ \bibinfo {author} {\bibfnamefont {A.}~\bibnamefont {Kossakowski}},\ }\href {https://doi.org/10.1103/PhysRevLett.104.070406} {\bibfield  {journal} {\bibinfo  {journal} {Phys. Rev. Lett.}\ }\textbf {\bibinfo {volume} {104}},\ \bibinfo {pages} {070406} (\bibinfo {year} {2010})}\BibitemShut {NoStop}%
\bibitem [{\citenamefont {Hegde}\ \emph {et~al.}(2021)\citenamefont {Hegde}, \citenamefont {Athulya}, \citenamefont {Pathak}, \citenamefont {Piilo},\ and\ \citenamefont {Shaji}}]{hegde2021open}%
  \BibitemOpen
  \bibfield  {author} {\bibinfo {author} {\bibfnamefont {A.~S.}\ \bibnamefont {Hegde}}, \bibinfo {author} {\bibfnamefont {K.}~\bibnamefont {Athulya}}, \bibinfo {author} {\bibfnamefont {V.}~\bibnamefont {Pathak}}, \bibinfo {author} {\bibfnamefont {J.}~\bibnamefont {Piilo}},\ and\ \bibinfo {author} {\bibfnamefont {A.}~\bibnamefont {Shaji}},\ }\href@noop {} {\bibfield  {journal} {\bibinfo  {journal} {Physical Review A}\ }\textbf {\bibinfo {volume} {104}},\ \bibinfo {pages} {062403} (\bibinfo {year} {2021})}\BibitemShut {NoStop}%
\bibitem [{\citenamefont {Jagadish}\ \emph {et~al.}(2023)\citenamefont {Jagadish}, \citenamefont {Srikanth},\ and\ \citenamefont {Petruccione}}]{jagadish2023noninvertibility}%
  \BibitemOpen
  \bibfield  {author} {\bibinfo {author} {\bibfnamefont {V.}~\bibnamefont {Jagadish}}, \bibinfo {author} {\bibfnamefont {R.}~\bibnamefont {Srikanth}},\ and\ \bibinfo {author} {\bibfnamefont {F.}~\bibnamefont {Petruccione}},\ }\href@noop {} {\bibfield  {journal} {\bibinfo  {journal} {Physical Review A}\ }\textbf {\bibinfo {volume} {108}},\ \bibinfo {pages} {042202} (\bibinfo {year} {2023})}\BibitemShut {NoStop}%
\bibitem [{\citenamefont {Chruściński}\ and\ \citenamefont {Mukhamedov}(2026)}]{chru2026}%
  \BibitemOpen
  \bibfield  {author} {\bibinfo {author} {\bibfnamefont {D.}~\bibnamefont {Chruściński}}\ and\ \bibinfo {author} {\bibfnamefont {F.}~\bibnamefont {Mukhamedov}},\ }\href {https://arxiv.org/abs/2604.20335} {\bibinfo {title} {Interpolating between positive, schwarz, and completely positive evolution for d-level systems}} (\bibinfo {year} {2026}),\ \Eprint {https://arxiv.org/abs/2604.20335} {arXiv:2604.20335 [quant-ph]} \BibitemShut {NoStop}%
\bibitem [{\citenamefont {Strachan}\ \emph {et~al.}(2024)\citenamefont {Strachan}, \citenamefont {Purkayastha},\ and\ \citenamefont {Clark}}]{strachan2024extracting}%
  \BibitemOpen
  \bibfield  {author} {\bibinfo {author} {\bibfnamefont {D.~J.}\ \bibnamefont {Strachan}}, \bibinfo {author} {\bibfnamefont {A.}~\bibnamefont {Purkayastha}},\ and\ \bibinfo {author} {\bibfnamefont {S.~R.}\ \bibnamefont {Clark}},\ }\href@noop {} {\bibfield  {journal} {\bibinfo  {journal} {The Journal of Chemical Physics}\ }\textbf {\bibinfo {volume} {161}} (\bibinfo {year} {2024})}\BibitemShut {NoStop}%
\bibitem [{\citenamefont {Davies}(1974)}]{davies1974}%
  \BibitemOpen
  \bibfield  {author} {\bibinfo {author} {\bibfnamefont {E.~B.}\ \bibnamefont {Davies}},\ }\href {https://projecteuclid.org:443/euclid.cmp/1103860160} {\bibfield  {journal} {\bibinfo  {journal} {Comm. Math. Phys.}\ }\textbf {\bibinfo {volume} {39}},\ \bibinfo {pages} {91} (\bibinfo {year} {1974})}\BibitemShut {NoStop}%
\bibitem [{\citenamefont {Khalfin}(1958)}]{Khalfin1958}%
  \BibitemOpen
  \bibfield  {author} {\bibinfo {author} {\bibfnamefont {L.~A.}\ \bibnamefont {Khalfin}},\ }\href@noop {} {\bibfield  {journal} {\bibinfo  {journal} {Sov. Phys. JETP}\ }\textbf {\bibinfo {volume} {6}},\ \bibinfo {pages} {1053} (\bibinfo {year} {1958})}\BibitemShut {NoStop}%
\bibitem [{\citenamefont {Leggett}\ \emph {et~al.}(1987)\citenamefont {Leggett}, \citenamefont {Chakravarty}, \citenamefont {Dorsey}, \citenamefont {Fisher}, \citenamefont {Garg},\ and\ \citenamefont {Zwerger}}]{Leggett1987}%
  \BibitemOpen
  \bibfield  {author} {\bibinfo {author} {\bibfnamefont {A.~J.}\ \bibnamefont {Leggett}}, \bibinfo {author} {\bibfnamefont {S.}~\bibnamefont {Chakravarty}}, \bibinfo {author} {\bibfnamefont {A.~T.}\ \bibnamefont {Dorsey}}, \bibinfo {author} {\bibfnamefont {M.~P.~A.}\ \bibnamefont {Fisher}}, \bibinfo {author} {\bibfnamefont {A.}~\bibnamefont {Garg}},\ and\ \bibinfo {author} {\bibfnamefont {W.}~\bibnamefont {Zwerger}},\ }\href {https://doi.org/10.1103/RevModPhys.59.1} {\bibfield  {journal} {\bibinfo  {journal} {Rev. Mod. Phys.}\ }\textbf {\bibinfo {volume} {59}},\ \bibinfo {pages} {1} (\bibinfo {year} {1987})}\BibitemShut {NoStop}%
\bibitem [{\citenamefont {Kaplanek}\ and\ \citenamefont {Burgess}(2020)}]{kaplanek2020hot}%
  \BibitemOpen
  \bibfield  {author} {\bibinfo {author} {\bibfnamefont {G.}~\bibnamefont {Kaplanek}}\ and\ \bibinfo {author} {\bibfnamefont {C.}~\bibnamefont {Burgess}},\ }\href@noop {} {\bibfield  {journal} {\bibinfo  {journal} {Journal of High Energy Physics}\ }\textbf {\bibinfo {volume} {2020}},\ \bibinfo {pages} {1} (\bibinfo {year} {2020})}\BibitemShut {NoStop}%
\bibitem [{\citenamefont {Brahma}\ \emph {et~al.}(2025)\citenamefont {Brahma}, \citenamefont {Calder{\'o}n-Figueroa},\ and\ \citenamefont {Luo}}]{brahma2025time}%
  \BibitemOpen
  \bibfield  {author} {\bibinfo {author} {\bibfnamefont {S.}~\bibnamefont {Brahma}}, \bibinfo {author} {\bibfnamefont {J.}~\bibnamefont {Calder{\'o}n-Figueroa}},\ and\ \bibinfo {author} {\bibfnamefont {X.}~\bibnamefont {Luo}},\ }\href@noop {} {\bibfield  {journal} {\bibinfo  {journal} {Journal of Cosmology and Astroparticle Physics}\ }\textbf {\bibinfo {volume} {2025}}\bibinfo  {number} { (08)},\ \bibinfo {pages} {019}}\BibitemShut {NoStop}%
\bibitem [{\citenamefont {D'Abbruzzo}\ \emph {et~al.}(2025)\citenamefont {D'Abbruzzo}, \citenamefont {Giovannetti},\ and\ \citenamefont {Cavina}}]{DAbbruzzoPRL}%
  \BibitemOpen
\bibfield  {number} {  }\bibfield  {author} {\bibinfo {author} {\bibfnamefont {A.}~\bibnamefont {D'Abbruzzo}}, \bibinfo {author} {\bibfnamefont {V.}~\bibnamefont {Giovannetti}},\ and\ \bibinfo {author} {\bibfnamefont {V.}~\bibnamefont {Cavina}},\ }\href {https://doi.org/10.1103/cb7c-5f66} {\bibfield  {journal} {\bibinfo  {journal} {Phys. Rev. Lett.}\ }\textbf {\bibinfo {volume} {135}},\ \bibinfo {pages} {240401} (\bibinfo {year} {2025})}\BibitemShut {NoStop}%
\bibitem [{\citenamefont {Davidović}(2020)}]{Davidovic2020}%
  \BibitemOpen
  \bibfield  {author} {\bibinfo {author} {\bibfnamefont {D.}~\bibnamefont {Davidović}},\ }\href {https://doi.org/10.22331/q-2020-09-21-326} {\bibfield  {journal} {\bibinfo  {journal} {Quantum}\ }\textbf {\bibinfo {volume} {4}},\ \bibinfo {pages} {326} (\bibinfo {year} {2020})}\BibitemShut {NoStop}%
\bibitem [{\citenamefont {De~Roeck}\ and\ \citenamefont {Kupiainen}(2013)}]{de2013approach}%
  \BibitemOpen
  \bibfield  {author} {\bibinfo {author} {\bibfnamefont {W.}~\bibnamefont {De~Roeck}}\ and\ \bibinfo {author} {\bibfnamefont {A.}~\bibnamefont {Kupiainen}},\ }in\ \href@noop {} {\emph {\bibinfo {booktitle} {Annales Henri Poincar{\'e}}}},\ Vol.~\bibinfo {volume} {14}\ (\bibinfo {organization} {Springer},\ \bibinfo {year} {2013})\ pp.\ \bibinfo {pages} {253--311}\BibitemShut {NoStop}%
\bibitem [{\citenamefont {Merkli}(2022)}]{merkli2022dynamics}%
  \BibitemOpen
  \bibfield  {author} {\bibinfo {author} {\bibfnamefont {M.}~\bibnamefont {Merkli}},\ }\href@noop {} {\bibfield  {journal} {\bibinfo  {journal} {Quantum}\ }\textbf {\bibinfo {volume} {6}},\ \bibinfo {pages} {615} (\bibinfo {year} {2022})}\BibitemShut {NoStop}%
\bibitem [{\citenamefont {Zhang}\ \emph {et~al.}(2012)\citenamefont {Zhang}, \citenamefont {Lo}, \citenamefont {Xiong}, \citenamefont {Tu},\ and\ \citenamefont {Nori}}]{zhang2012general}%
  \BibitemOpen
  \bibfield  {author} {\bibinfo {author} {\bibfnamefont {W.-M.}\ \bibnamefont {Zhang}}, \bibinfo {author} {\bibfnamefont {P.-Y.}\ \bibnamefont {Lo}}, \bibinfo {author} {\bibfnamefont {H.-N.}\ \bibnamefont {Xiong}}, \bibinfo {author} {\bibfnamefont {M.~W.-Y.}\ \bibnamefont {Tu}},\ and\ \bibinfo {author} {\bibfnamefont {F.}~\bibnamefont {Nori}},\ }\href@noop {} {\bibfield  {journal} {\bibinfo  {journal} {Physical review letters}\ }\textbf {\bibinfo {volume} {109}},\ \bibinfo {pages} {170402} (\bibinfo {year} {2012})}\BibitemShut {NoStop}%
\bibitem [{\citenamefont {Peres}(1980)}]{Peres1980}%
  \BibitemOpen
  \bibfield  {author} {\bibinfo {author} {\bibfnamefont {A.}~\bibnamefont {Peres}},\ }\href {https://doi.org/10.1016/0003-4916(80)90288-2} {\bibfield  {journal} {\bibinfo  {journal} {Annals of Physics}\ }\textbf {\bibinfo {volume} {129}},\ \bibinfo {pages} {33} (\bibinfo {year} {1980})}\BibitemShut {NoStop}%
\bibitem [{\citenamefont {Breuer}\ and\ \citenamefont {Petruccione}(2007)}]{BreuerHeinz-Peter1961-2007TToO}%
  \BibitemOpen
  \bibfield  {author} {\bibinfo {author} {\bibfnamefont {H.-P.}\ \bibnamefont {Breuer}}\ and\ \bibinfo {author} {\bibfnamefont {F.}~\bibnamefont {Petruccione}},\ }\href {https://oxford.universitypressscholarship.com/view/10.1093/acprof:oso/9780199213900.001.0001/acprof-9780199213900} {\emph {\bibinfo {title} {The theory of open quantum systems}}}\ (\bibinfo  {publisher} {Oxford University Press},\ \bibinfo {address} {Oxford},\ \bibinfo {year} {2007})\BibitemShut {NoStop}%
\bibitem [{\citenamefont {Fleming}\ \emph {et~al.}(2010)\citenamefont {Fleming}, \citenamefont {Cummings}, \citenamefont {Anastopoulos},\ and\ \citenamefont {Hu}}]{Fleming2010}%
  \BibitemOpen
  \bibfield  {author} {\bibinfo {author} {\bibfnamefont {C.}~\bibnamefont {Fleming}}, \bibinfo {author} {\bibfnamefont {N.~I.}\ \bibnamefont {Cummings}}, \bibinfo {author} {\bibfnamefont {C.}~\bibnamefont {Anastopoulos}},\ and\ \bibinfo {author} {\bibfnamefont {B.~L.}\ \bibnamefont {Hu}},\ }\href {https://doi.org/10.1088/1751-8113/43/40/405304} {\bibfield  {journal} {\bibinfo  {journal} {Journal of Physics A: Mathematical and Theoretical}\ }\textbf {\bibinfo {volume} {43}},\ \bibinfo {pages} {405304} (\bibinfo {year} {2010})}\BibitemShut {NoStop}%
\bibitem [{\citenamefont {DiVincenzo}\ and\ \citenamefont {Loss}(2005)}]{DiVincenzoLoss2005}%
  \BibitemOpen
  \bibfield  {author} {\bibinfo {author} {\bibfnamefont {D.~P.}\ \bibnamefont {DiVincenzo}}\ and\ \bibinfo {author} {\bibfnamefont {D.}~\bibnamefont {Loss}},\ }\href {https://doi.org/10.1103/PhysRevB.71.035318} {\bibfield  {journal} {\bibinfo  {journal} {Physical Review B}\ }\textbf {\bibinfo {volume} {71}},\ \bibinfo {pages} {035318} (\bibinfo {year} {2005})}\BibitemShut {NoStop}%
\bibitem [{\citenamefont {Lindner}\ and\ \citenamefont {Schoeller}(2018)}]{LindnerSchoeller2018}%
  \BibitemOpen
  \bibfield  {author} {\bibinfo {author} {\bibfnamefont {C.~J.}\ \bibnamefont {Lindner}}\ and\ \bibinfo {author} {\bibfnamefont {H.}~\bibnamefont {Schoeller}},\ }\href {https://doi.org/10.1103/PhysRevB.98.115425} {\bibfield  {journal} {\bibinfo  {journal} {Physical Review B}\ }\textbf {\bibinfo {volume} {98}},\ \bibinfo {pages} {115425} (\bibinfo {year} {2018})},\ \Eprint {https://arxiv.org/abs/1802.09846} {arXiv:1802.09846 [cond-mat.stat-mech]} \BibitemShut {NoStop}%
\bibitem [{\citenamefont {Strathearn}\ \emph {et~al.}(2018)\citenamefont {Strathearn}, \citenamefont {Kirton}, \citenamefont {Kilda}, \citenamefont {Keeling},\ and\ \citenamefont {Lovett}}]{Strathearn2018}%
  \BibitemOpen
  \bibfield  {author} {\bibinfo {author} {\bibfnamefont {A.}~\bibnamefont {Strathearn}}, \bibinfo {author} {\bibfnamefont {P.}~\bibnamefont {Kirton}}, \bibinfo {author} {\bibfnamefont {D.}~\bibnamefont {Kilda}}, \bibinfo {author} {\bibfnamefont {J.}~\bibnamefont {Keeling}},\ and\ \bibinfo {author} {\bibfnamefont {B.~W.}\ \bibnamefont {Lovett}},\ }\href {https://doi.org/10.1038/s41467-018-05617-3} {\bibfield  {journal} {\bibinfo  {journal} {Nat. Commun.}\ }\textbf {\bibinfo {volume} {9}},\ \bibinfo {pages} {3322} (\bibinfo {year} {2018})}\BibitemShut {NoStop}%
\bibitem [{\citenamefont {Lacroix}\ \emph {et~al.}(2026)\citenamefont {Lacroix}, \citenamefont {Burgess}, \citenamefont {Lorenzoni}, \citenamefont {Wiercinski}, \citenamefont {Damezin}, \citenamefont {Lim}, \citenamefont {Tamascelli}, \citenamefont {Chin}, \citenamefont {Cygorek}, \citenamefont {Lovett}, \citenamefont {Keeling}, \citenamefont {Huelga}, \citenamefont {Plenio},\ and\ \citenamefont {Gauger}}]{lacroix2026tensornetworkmethodsnonperturbative}%
  \BibitemOpen
  \bibfield  {author} {\bibinfo {author} {\bibfnamefont {T.}~\bibnamefont {Lacroix}}, \bibinfo {author} {\bibfnamefont {A.}~\bibnamefont {Burgess}}, \bibinfo {author} {\bibfnamefont {N.}~\bibnamefont {Lorenzoni}}, \bibinfo {author} {\bibfnamefont {J.}~\bibnamefont {Wiercinski}}, \bibinfo {author} {\bibfnamefont {K.}~\bibnamefont {Damezin}}, \bibinfo {author} {\bibfnamefont {J.}~\bibnamefont {Lim}}, \bibinfo {author} {\bibfnamefont {D.}~\bibnamefont {Tamascelli}}, \bibinfo {author} {\bibfnamefont {A.~W.}\ \bibnamefont {Chin}}, \bibinfo {author} {\bibfnamefont {M.}~\bibnamefont {Cygorek}}, \bibinfo {author} {\bibfnamefont {B.~W.}\ \bibnamefont {Lovett}}, \bibinfo {author} {\bibfnamefont {J.}~\bibnamefont {Keeling}}, \bibinfo {author} {\bibfnamefont {S.~F.}\ \bibnamefont {Huelga}}, \bibinfo {author} {\bibfnamefont {M.~B.}\ \bibnamefont {Plenio}},\ and\ \bibinfo {author} {\bibfnamefont {E.~M.}\ \bibnamefont {Gauger}},\ }\href {https://arxiv.org/abs/2608.09850} {\bibinfo {title} {Tensor network methods for
  non-perturbative dynamics of open quantum systems}} (\bibinfo {year} {2026}),\ \Eprint {https://arxiv.org/abs/2608.09850} {arXiv:2608.09850 [quant-ph]} \BibitemShut {NoStop}%
\bibitem [{\citenamefont {Fleming}\ and\ \citenamefont {Cummings}(2011)}]{Fleming}%
  \BibitemOpen
  \bibfield  {author} {\bibinfo {author} {\bibfnamefont {C.~H.}\ \bibnamefont {Fleming}}\ and\ \bibinfo {author} {\bibfnamefont {N.~I.}\ \bibnamefont {Cummings}},\ }\href {https://doi.org/10.1103/PhysRevE.83.031117} {\bibfield  {journal} {\bibinfo  {journal} {Phys. Rev. E}\ }\textbf {\bibinfo {volume} {83}},\ \bibinfo {pages} {031117} (\bibinfo {year} {2011})}\BibitemShut {NoStop}%
\bibitem [{\citenamefont {Nestmann}\ and\ \citenamefont {Timm}(2019)}]{nestmann2019timeconvolutionless}%
  \BibitemOpen
  \bibfield  {author} {\bibinfo {author} {\bibfnamefont {K.}~\bibnamefont {Nestmann}}\ and\ \bibinfo {author} {\bibfnamefont {C.}~\bibnamefont {Timm}},\ }\href@noop {} {\bibinfo {title} {Time-convolutionless master equation: Perturbative expansions to arbitrary order and application to quantum dots}} (\bibinfo {year} {2019}),\ \Eprint {https://arxiv.org/abs/1903.05132} {arXiv:1903.05132 [cond-mat.mes-hall]} \BibitemShut {NoStop}%
\bibitem [{\citenamefont {Cresser}\ and\ \citenamefont {Anders}(2021)}]{cresser2021weak}%
  \BibitemOpen
  \bibfield  {author} {\bibinfo {author} {\bibfnamefont {J.}~\bibnamefont {Cresser}}\ and\ \bibinfo {author} {\bibfnamefont {J.}~\bibnamefont {Anders}},\ }\href@noop {} {\bibfield  {journal} {\bibinfo  {journal} {Physical Review Letters}\ }\textbf {\bibinfo {volume} {127}},\ \bibinfo {pages} {250601} (\bibinfo {year} {2021})}\BibitemShut {NoStop}%
\end{thebibliography}
%

\end{document}